\documentclass[longauth]{aa}

\usepackage{lscape}
\usepackage{txfonts}
\usepackage[]{graphicx} 
\usepackage{array}
\usepackage{rotating}
\usepackage{color}
\usepackage{url}
\usepackage{natbib}
\usepackage{float}
\usepackage{textcomp,gensymb}
\usepackage[utf8]{inputenc}
\usepackage{amsmath} 

\makeatother

\def \aj {AJ} 
\def \mnras {MNRAS} 
\def \pasp {PASP} 
\def \apj {ApJ} 
\def \apjs {ApJS} 
 
\def \apjl {ApJL} 
\def \aap {A\&A} 
\def \nat {Nature}

\begin{document} 

\title{Luminous red novae: Stellar mergers or giant eruptions?\thanks{Tables A.1-A.6 are only available in electronic form
at the CDS via anonymous ftp to cdsarc.u-strasbg.fr (130.79.128.5) or via http://cdsweb.u-strasbg.fr/cgi-bin/qcat?J/A+A/}}

\author{A. Pastorello\inst{1}\thanks{E-mail: andrea.pastorello@inaf.it} 
          \and
E. Mason\inst{2} 
          \and
S. Taubenberger\inst{3,4} 
          \and
M. Fraser\inst{5,6} 
          \and
G. Cortini\inst{7}
          \and
L.~Tomasella\inst{1} 
          \and
M. T. Botticella\inst{8}  
          \and
N.~Elias-Rosa\inst{9,10}  
          \and
R. Kotak\inst{11} 
          \and
S.~J.~Smartt\inst{12}
          \and
S.~Benetti\inst{1} 
          \and
E.~Cappellaro\inst{1} 
          \and
M.~Turatto\inst{1}  
          \and
L.~Tartaglia\inst{13}  
          \and
S.~G.~Djorgovski\inst{14}
          \and
A.~J.~Drake\inst{14} 
          \and
M.~Berton\inst{15,16}  
          \and
F.~Briganti\inst{17} 
          \and
J.~Brimacombe\inst{18} 
          \and
F.~Bufano\inst{19} 
          \and
Y.-Z.~Cai\inst{1,20} 
          \and
S.~Chen\inst{20,21,22}  
          \and
E.~J.~Christensen\inst{23}  
          \and
F.~Ciabattari\inst{24} 
          \and
E.~Congiu\inst{25,26}
          \and
A.~Dimai\inst{27}\thanks{Deceased 7 March, 2019}
          \and
C.~Inserra\inst{28}  
          \and
E.~Kankare\inst{11,12}
          \and
L.~Magill\inst{29} 
          \and
K.~Maguire\inst{30} 
          \and
F.~Martinelli\inst{17}
          \and
A.~Morales-Garoffolo\inst{31}  
          \and
P.~Ochner\inst{1,20} 
          \and
G.~Pignata\inst{32,33}  
          \and
A.~Reguitti\inst{32}
          \and
J.~Sollerman\inst{13} 
          \and
S.~Spiro\inst{1}  
          \and
G.~Terreran\inst{34} 
          \and
D.~E.~Wright\inst{35}} 

\institute{
 INAF - Osservatorio Astronomico di Padova, Vicolo dell'Osservatorio 5, I-35122 Padova, Italy
          \and
 INAF - Osservatorio Astronomico di Trieste, Via G.B. Tiepolo 11, I-34143 Trieste, Italy
          \and
 Max-Planck-Institut f\"ur Astrophysik, Karl-Schwarzschild-Str. 1, D-85748 Garching, Germany 
          \and
 European Organisation for Astronomical Research in the Southern Hemisphere (ESO), Karl-Schwarzschild-Str. 2, D-85748 Garching bei M\"unchen, Germany
          \and
 School of Physics, O'Brien Centre for Science North, University College Dublin, Belfield, Dublin 4, Ireland
          \and
 Royal Society - Science Foundation Ireland University Research Fellow
          \and
 Osservatorio Astronomico di Monte Maggiore, Via Montemaggiore 3, I-47016 Predappio, Forl\`i-Cesena, Italy
          \and
 INAF - Osservatorio Astronomico di Capodimonte, Salita Moiariello 16, I-80131 Napoli, Italy 
          \and
 Institute of Space Sciences (ICE, CSIC), Campus UAB, Cam\'{i} de Can Magrans s/n, 08193 Cerdanyola del Vall\`es (Barcelona), Spain
          \and
 Institut d’Estudis Espacials de Catalunya (IEEC), c/Gran Capit\`a 2-4, Edif. Nexus 201, 08034 Barcelona, Spain
          \and
 Department of Physics and Astronomy, University of Turku, Vesilinnantie 5, FI-20014  Turku, Finland
          \and
 Astrophysics Research Centre, School of Mathematics and Physics, Queen's University Belfast, Belfast BT7 1NN, United Kingdom
          \and
 The Oskar Klein Centre, Department of Astronomy, Stockholm University, AlbaNova, SE-10691 Stockholm, Sweden 
          \and
 California Institute of Technology, 1200 E. California Blvd, Pasadena, CA 91125, USA  
          \and
Finnish Centre for Astronomy with ESO (FINCA), University of Turku, Quantum, Vesilinnantie 5, FI-20014, Finland
          \and
Aalto University Mets{\"a}hovi Radio Observatory, Mets{\"a}hovintie 114, FI-02540 Kylm{\"a}l{\"a}, Finland          
          \and
 Lajatico Astronomical Centre, Via dei Mulini a Vento, 56030 Lajatico, Pisa, Italy
          \and
 Coral Towers Observatory, Unit 38 Coral Towers, 255 Esplanade, Cairns 4870, Australia
          \and
 INAF - Osservatorio Astrofisico di Catania, Via S. Sofia 78, I-95123 Catania Italy
          \and
 Dipartimento di Fisica e Astronomia, Universit\`a di Padova, Vicolo dell'Osservatorio 3, I-35122 Padova, Italy
          \and
 INFN - Sezione di Padova, Via Marzolo 8, 35131, Padova
          \and
 Center for Astrophysics, Guangzhou University, Guangzhou 510006, China
          \and
 Lunar and Planetary Lab, Department of Planetary Sciences, University of Arizona, Tucson, AZ 85721, USA
          \and
 Osservatorio Astronomico di Monte Agliale, Via Cune Motrone, I-55023 Borgo a Mozzano, Lucca, Italy
          \and
 Las Campanas Observatory - Carnagie Institution of Washington, Colina el Pino, Casilla 601, La Serena, Chile
          \and
  INAF - Osservatorio Astronomico di Brera, via E. Bianchi 46, 23807 Merate (LC), Italy
          \and
 Osservatorio Astronomico del Col Drusci\'e,  I-32043 Cortina d'Ampezzo, Italy
          \and
School of Physics $\&$ Astronomy, Cardiff University, Queens Buildings, The Parade, Cardiff CF24 3AA, UK
          \and
 Gemini Observatory, Southern Operations Center, c/o AURA, Casilla 603, La Serena, Chile 
          \and
 School of Physics, Trinity College Dublin, The University of Dublin, Dublin 2, Ireland
          \and
 Department of Applied Physics, University of C\'adiz, Campus of Puerto Real, 11510 C\'adiz, Spain.
          \and
 Departamento de Ciencias Fisicas, Universidad Andres Bello, Avda. Republica 252, Santiago, Chile 
          \and
 Millennium Institute of Astrophysics (MAS), Nuncio Monse\"nor S\'otero Snz 100, Providencia, Santiago, Chile
          \and
 Center for Interdisciplinary Exploration and Research in Astrophysics (CIERA) and Department of Physics and Astronomy, Northwestern University, Evanston, IL 60208, USA
          \and
 Minnesota Institute for Astrophysics, University of Minnesota, Minneapolis, MN 55455, USA}

    \date{Received Month dd, 20yy; accepted Month dd, 20yy}


 
\abstract{We present extensive datasets for a class of intermediate-luminosity optical transients known as ``luminous red novae'' (LRNe).
They show double-peaked light curves, with an initial rapid luminosity rise to a blue peak 
(at $-13$ to $-15$ mag), which is followed by a longer-duration red peak that sometimes is attenuated, resembling a plateau. 
The progenitors of three of them (NGC4490-2011OT1, M101-2015OT1, and SNhunt248), likely relatively massive blue to yellow stars, were also observed in a 
pre-eruptive stage when their luminosity was slowly increasing. 
Early spectra obtained during the first peak show a blue continuum with superposed prominent narrow Balmer lines, with P~Cygni profiles. 
Lines of Fe II are also clearly observed, mostly in emission. During the second peak, the spectral continuum becomes much redder, 
H$\alpha$ is barely detected, and a forest of narrow metal lines is observed in absorption. Very late-time spectra 
($\sim$6 months after blue peak) show an extremely red spectral continuum, peaking in the infrared (IR) domain. 
H$\alpha$ is detected in pure emission at such late phases, along with broad absorption bands due to molecular overtones (such as TiO, VO). 
We discuss a few alternative scenarios for LRNe. Although  major instabilities of single massive stars cannot be definitely ruled out, 
we favour a common envelope ejection in a close binary system, with possibly a final coalescence of the two stars. The similarity between 
LRNe and the outburst observed a few months before the explosion of the Type IIn SN 2011ht is also discussed. } 
 

\maketitle 

\section{Introduction} 
 
A growing number of transients were discovered in the past few years showing intrinsic luminosities
intermediate between those of classical novae and traditional supernova (SN) types. These objects are labelled as ``Intermediate-Luminosity Optical  
Transients'' \citep[e.g.,][]{ber09,sok12} or ``gap transients'' \citep{kas12,pasto19a}. 
Although some gap transients can be genuine albeit weak SN explosions \citep[see discussions in][and references therein]{hor11,koc12},
most of them are non-terminal instabilities of massive stars that are approaching the final stages of their life, or even outbursts due to binary interaction. 
We have observed super-Eddington eruptions of massive Luminous Blue Variable (LBV) stars, but also LBV-like eruptions of  moderate-mass 
stars \citep[see, e.g.,][for reviews on this subject]{hum94,smi11}. In both cases, these transients may mimic the spectro-photometric 
evolution of weak Type IIn supernovae (SNe) and, for this reason, they frequently gained SN designations. 
As they have not undergone core-collapse and destruction, they are labelled as ``SN impostors'' \citep{van00}. 
  
We are currently observing an increasing variety of SN impostors: some of them have a single-peaked (SN-like) light curve \citep[e.g., SN~2007sv and 
PSN~J09132750+7627410;][]{tar15,tar16}. Other impostors experience a very slow evolution to maximum light lasting years to decades \citep[UGC 2773-2009OT1,][]{smi16a}. 
Finally, an erratic photometric variability is observed for many years in other objects, ranging from the relatively faint SN~2002kg 
\citep[$M_R \approx -10$ mag,][]{weis05,mau06,hum17} to the much more luminous SN~2000ch \citep[$M_R \approx -13$ mag;][]{wag04,pasto10} and the 2009-2012 
outbursts of SN~2009ip \citep[with $M_R$ occasionally exceeding $-14$ mag,][]{pasto13}. 

Recent publications analysed a few gap transients brighter than  $M_V\sim-13$ mag, and characterized by double or even triple-peaked light curves. 
From their photometric evolution, they are reminiscent of eruptive variables such as V1309~Sco \citep{mas10,tyl11}, V838~Mon \citep{mun02,bon03,gor04}, 
V4332~Sgr \citep{mar99}, M31RV \citep{bon03,bos04}, OGLE-2002-BLG-360 \citep{tyl13}, and M31-2015OT1  \citep{kur15,wil15}. These events, typically
fainter than $M_V\sim-10$ mag, are collectively dubbed ``red novae'' (RNe).\footnote{Although the acronym RNe is frequently used for ``recurrent novae'',
it will be adopted in this paper for discriminating dimmer red novae from their more luminous extra-galactic counterparts.} 
They are possibly the observational outcome of stellar coalescences \citep[e.g.,][]{koc14,pej14}.
However, some objects are much more luminous than RNe: SHhunt248 \citep{mau15,kan15}, M101-2015OT1 \citep{bla16,gor16},
NGC4490-2012OT1 \citep{smi16b}, and SN~2017jfs \citep{pasto19}. They have been proposed to be scaled-up versions of RNe, hence are labelled ``luminous red novae'' (LRNe).
These latter have been proposed to result from merging events involving massive binaries \citep{smi16b,mau17,bar17}.

Only a handful of appealing merger candidates from non-degenerate stars have been discovered so far \citep[][and references therein]{mcl18},
and ongoing searches for putative future stellar mergers are in progress  \citep[e.g.,][]{pie17}. 
Constraining the physical and observational parameters of these rare objects and constructing templates is an essential task, 
also accounting for the expected burst of new discoveries with the next generation of optical and IR 
instruments, such as the Large Synoptic Survey Telescope \citep{LSST09} and
 the Wide Field Infrared Survey Telescope \citep[WFIRST;][]{spe15}.

\begin{figure*} 
\centering
{\includegraphics[width=7.6cm,angle=0]{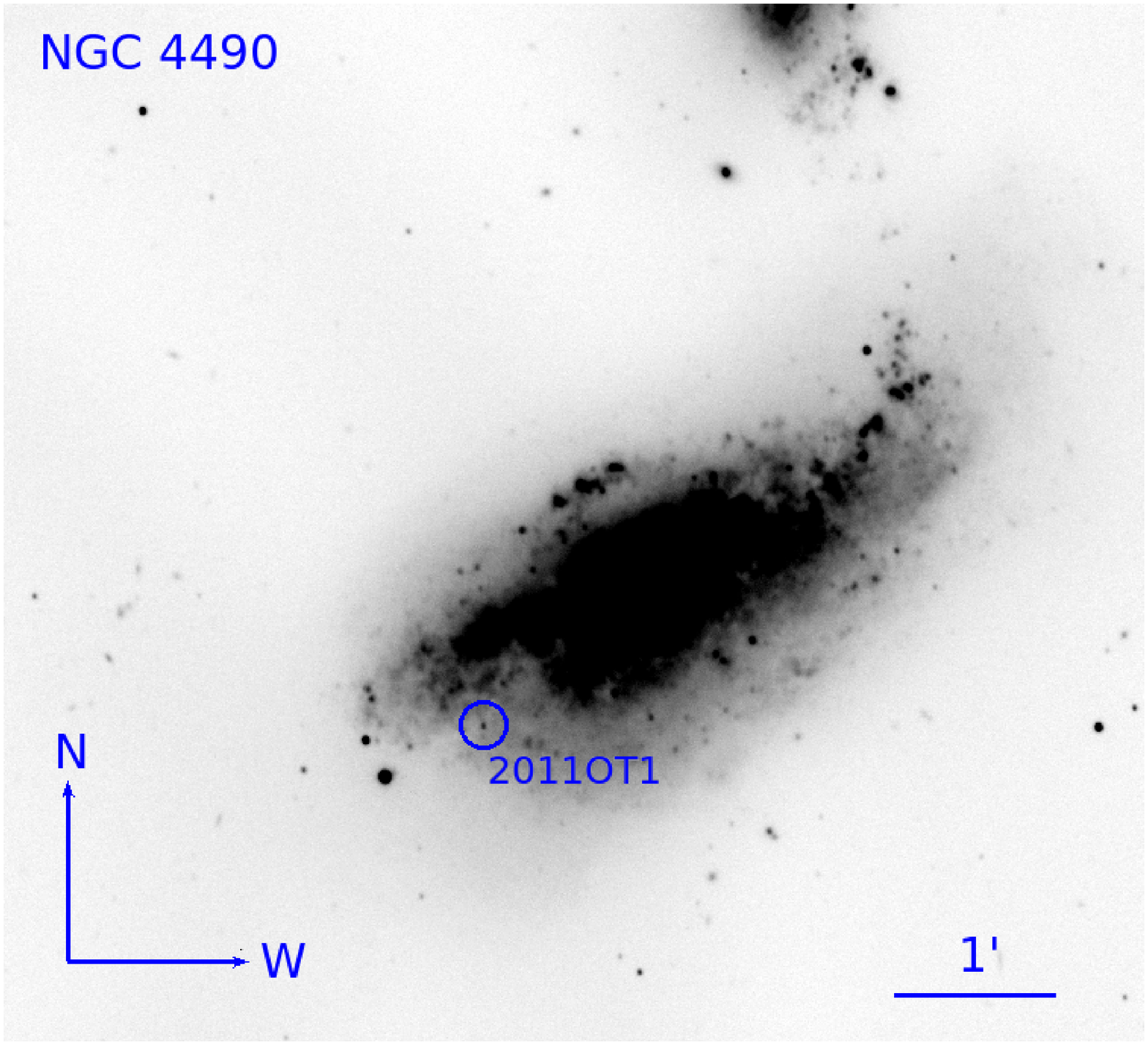}  
\includegraphics[width=7.6cm,angle=0]{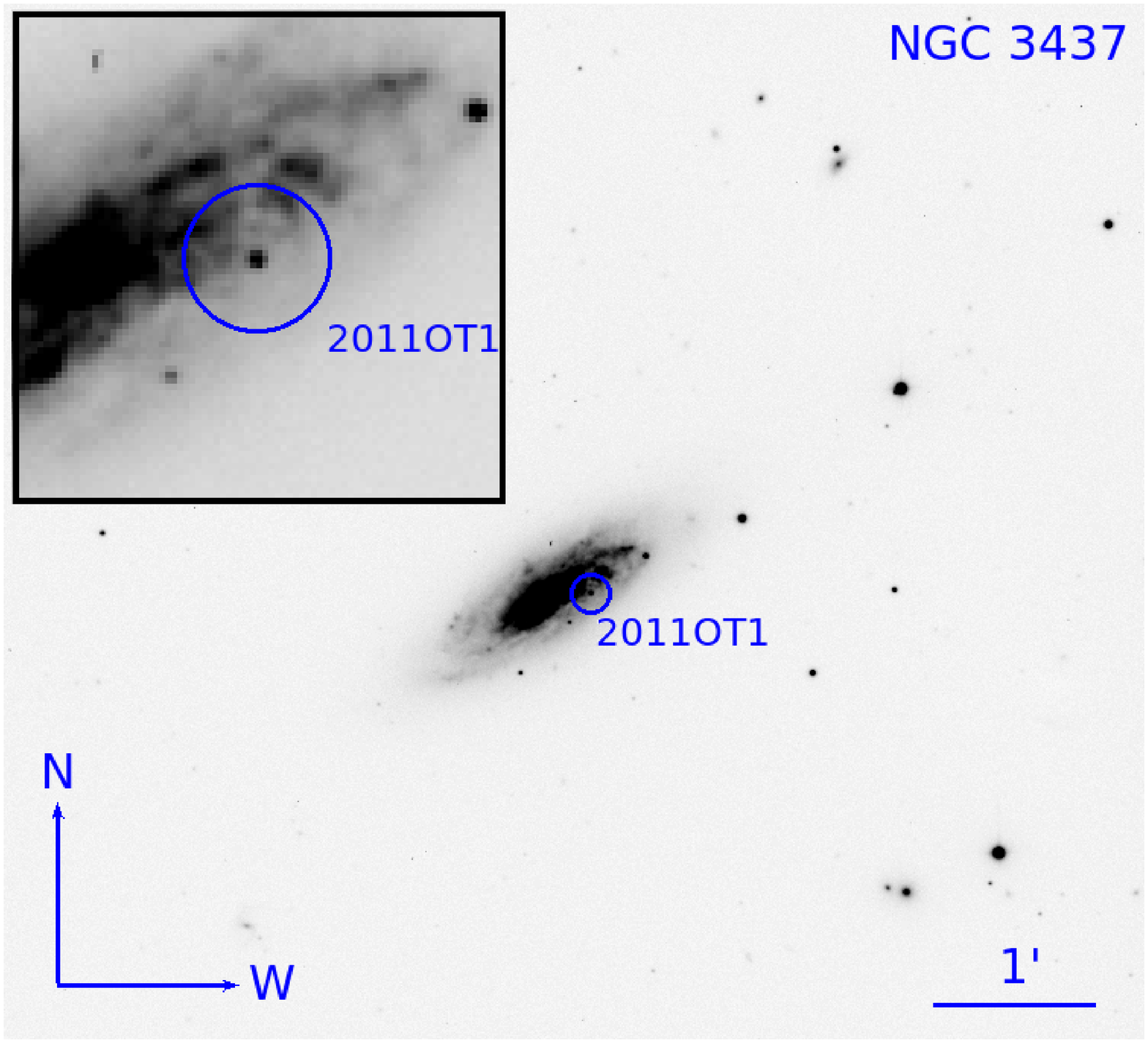} 
\includegraphics[width=7.5cm,angle=0]{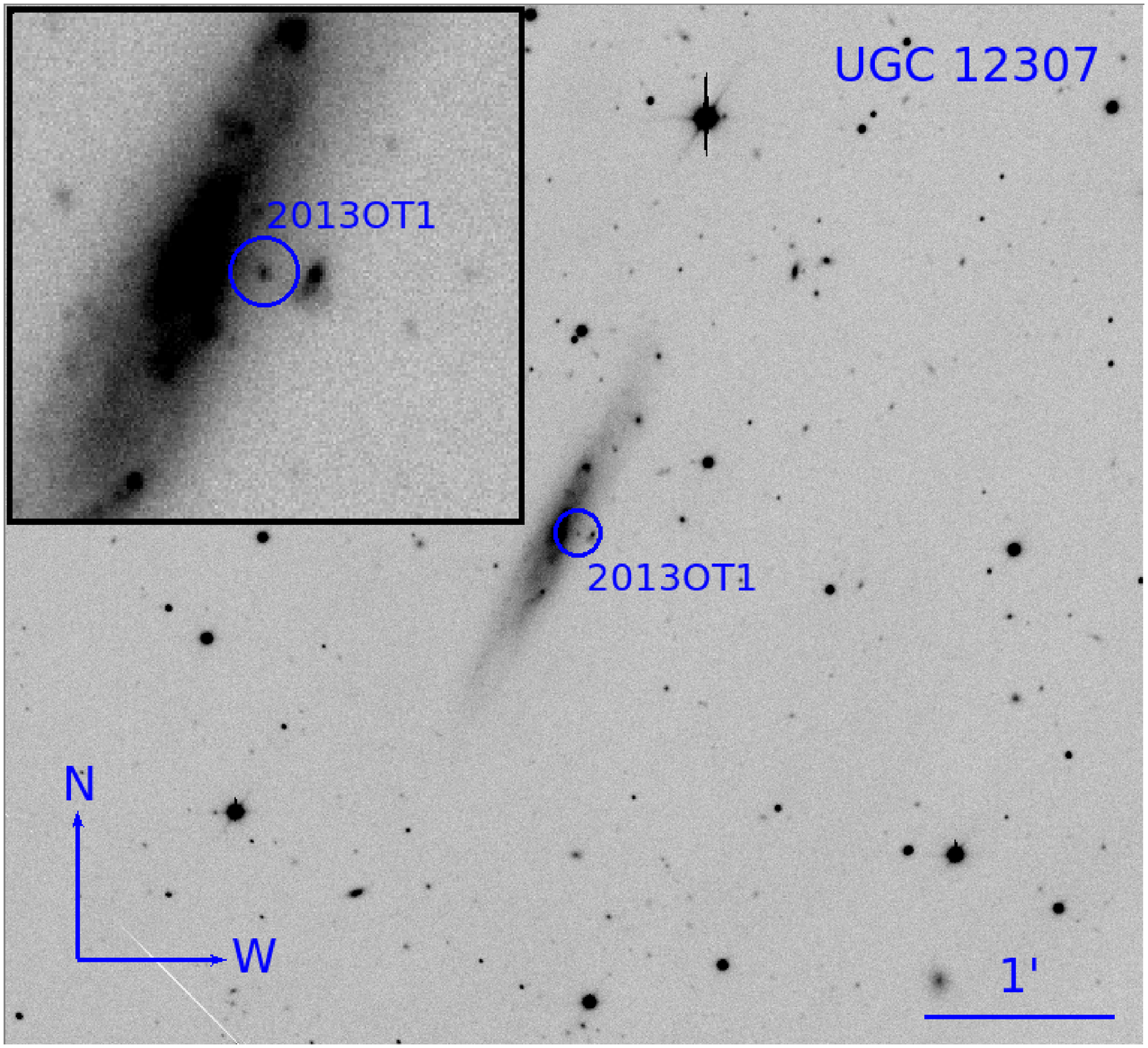}  
\includegraphics[width=7.5cm,angle=0]{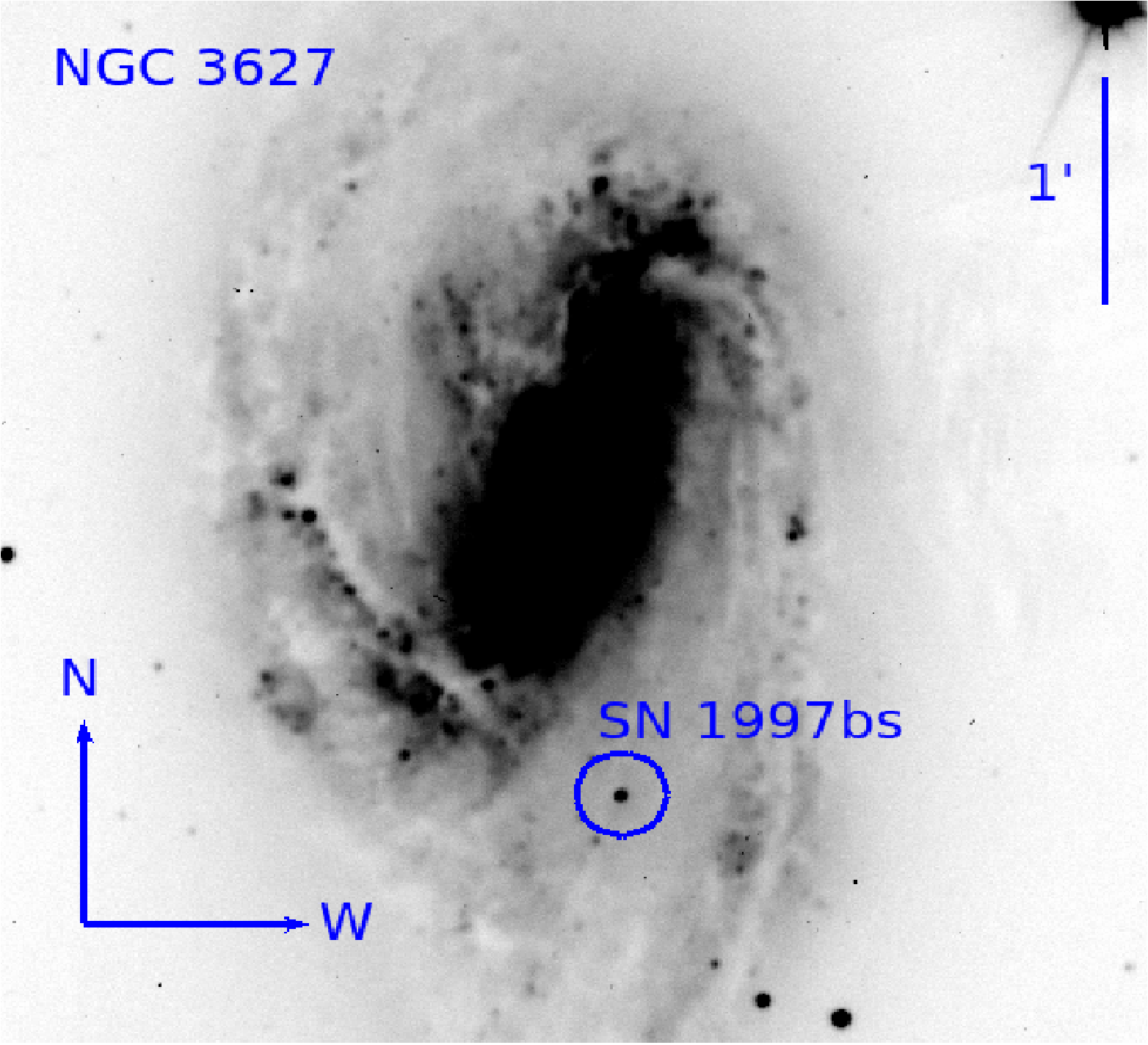} 
\includegraphics[width=7.6cm,angle=0]{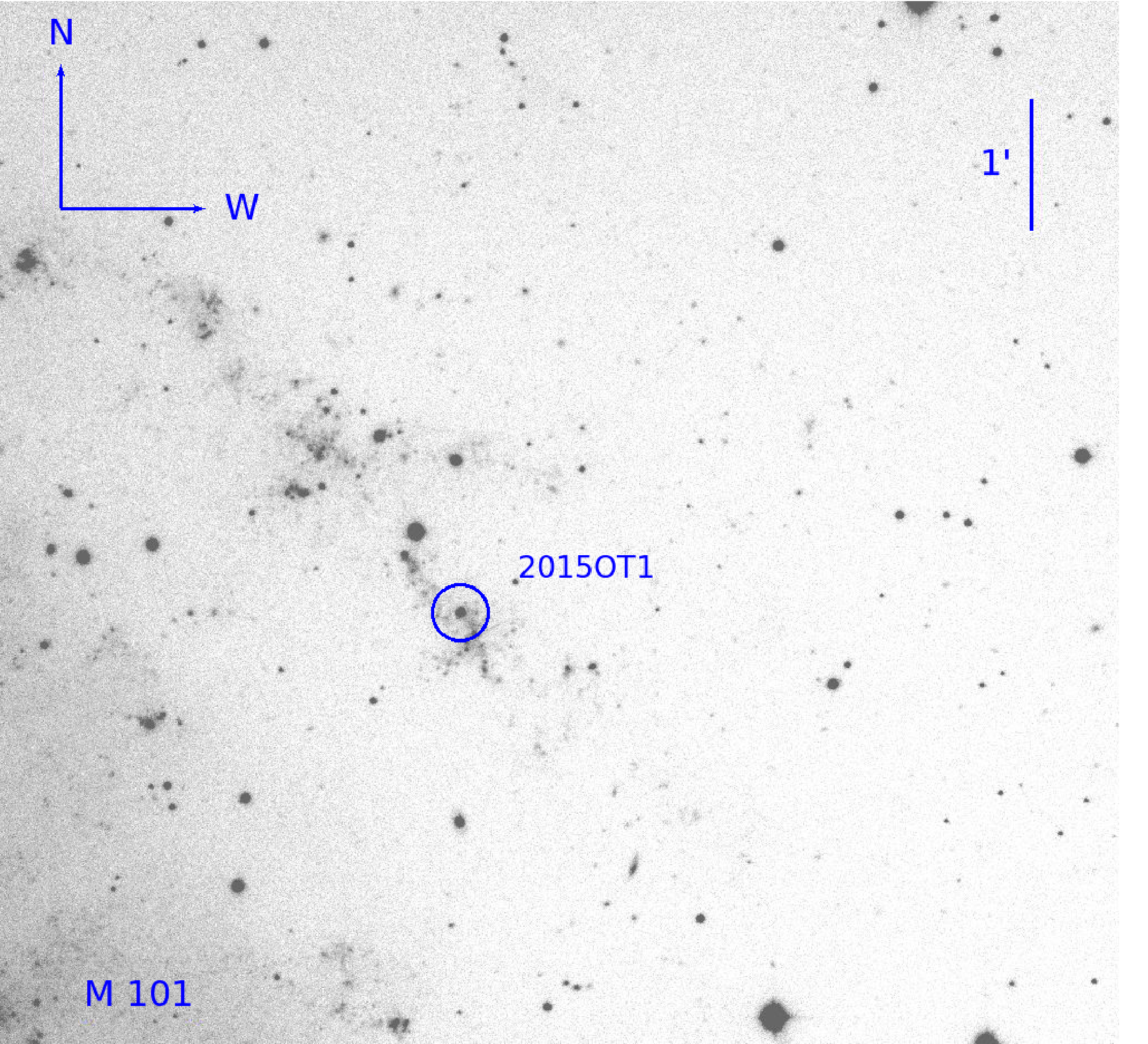} 
\includegraphics[width=7.6cm,angle=0]{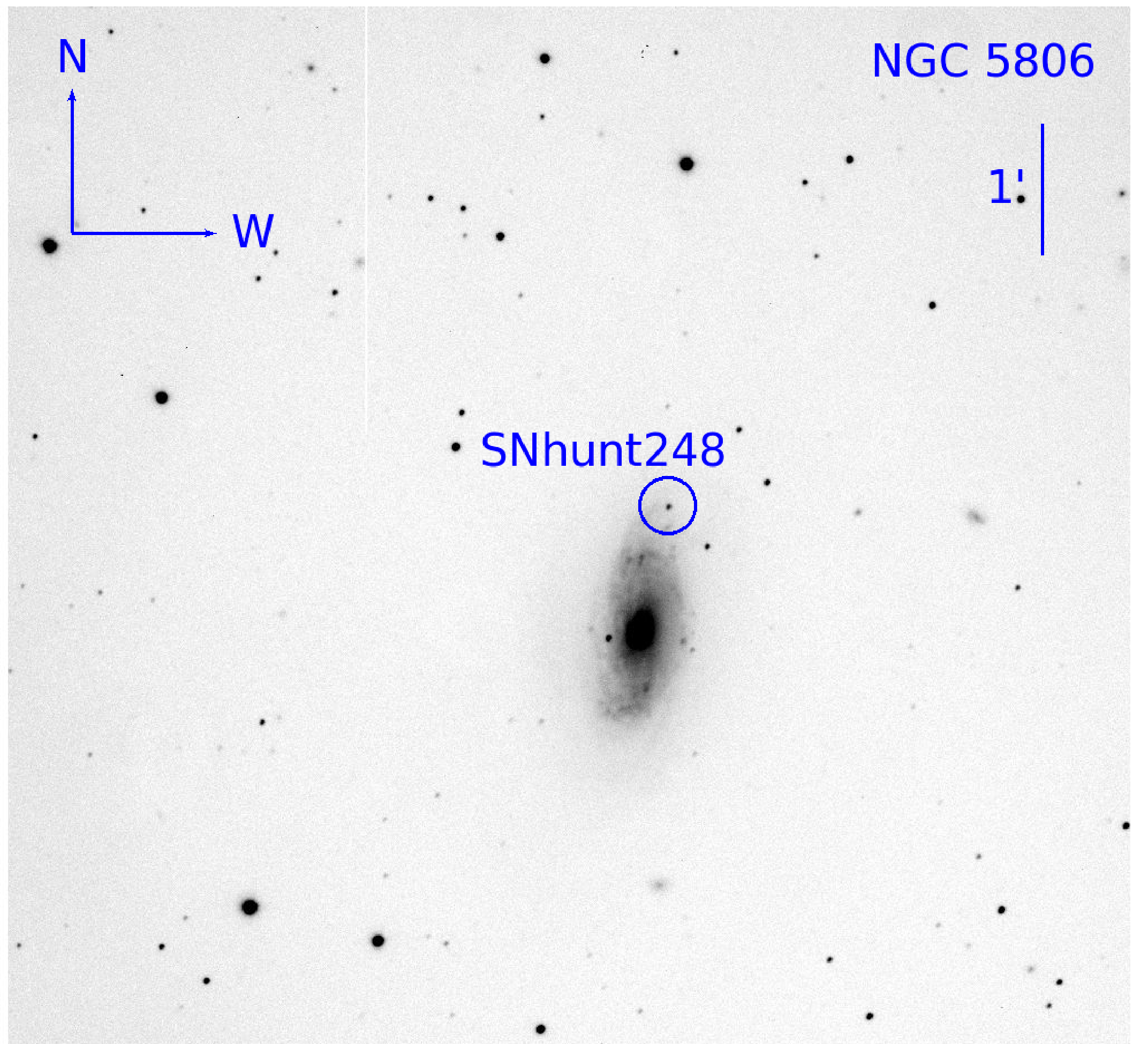}} 
 \caption{Top-left: NGC4490-2011OT1 in its host galaxy. The $R$-band image was obtained on 2012 March 24 with the 2.2m telescope of the Calar Alto Observatory, equipped with CAFOS. 
Top-right: NGC3437-2011OT1 and its host galaxy. The $R$-band image was obtained on 2011 January 15 with the 2.2m telescope of the Calar Alto Observatory, with CAFOS. A blow-up showing the transient 
is in the upper-left inset. Middle-left: UGC12307-2013OT1 and its host galaxy. The $R$-band image was obtained on 2013 October 11 with the 3.58m Telescopio Nazionale Galileo, equipped with LRS.  
A blow-up with the transient is shown in the upper-left inset. Middle-right: SN 1997bs in NGC~3627. The $R$-band image was obtained on 1997 April 27 with the 0.91m Dutch Telescope  
at ESO - La Silla. Bottom-left: M101-2015OT1 in the outskirts of its host galaxy. The Sloan-$r$ band image was obtained on 2015 February 18 with the 1.82m Asiago Copernico Telescope 
plus AFOSC. Bottom-right: SNhunt248 in NGC 5806. The AFOSC image in the Sloan-$r$ band was obtained on 2014 July 18.
\label{maps}} 
\end{figure*}

In this paper, we will analyse an extended sample of LRNe. For a few objects (NGC4490-2011OT1, NGC3437-2011OT1, 
and UGC12307-2013OT1), we will provide the most estensive sets of data available in the literature. We also include 
M101-2015OT1 \citep{bla16,gor16} and SNhunt248 \citep{mau15,kan15} in our sample. For both of them, we will provide
new photometric measurements that complement the existing datasets. We will also discuss the case of SN~1997bs 
\citep{van00}. This object is usually interpreted as an LBV outburst, but possibly shares some similarity with LRNe
(see Sect. \ref{photo_lc}). While we are aware that the nature of some objects discussed in this 
paper is still debated (e.g., SN~1997bs, M85-2006OT1, or SN~2011ht), our main goal is to provide observational criteria 
to constrain the LRN class. On the basis of these arguments, we will try to link or unlink controversial 
cases to this family of transients.
 
\begin{figure*} 
\centering
 \includegraphics[width=18.0cm,angle=0]{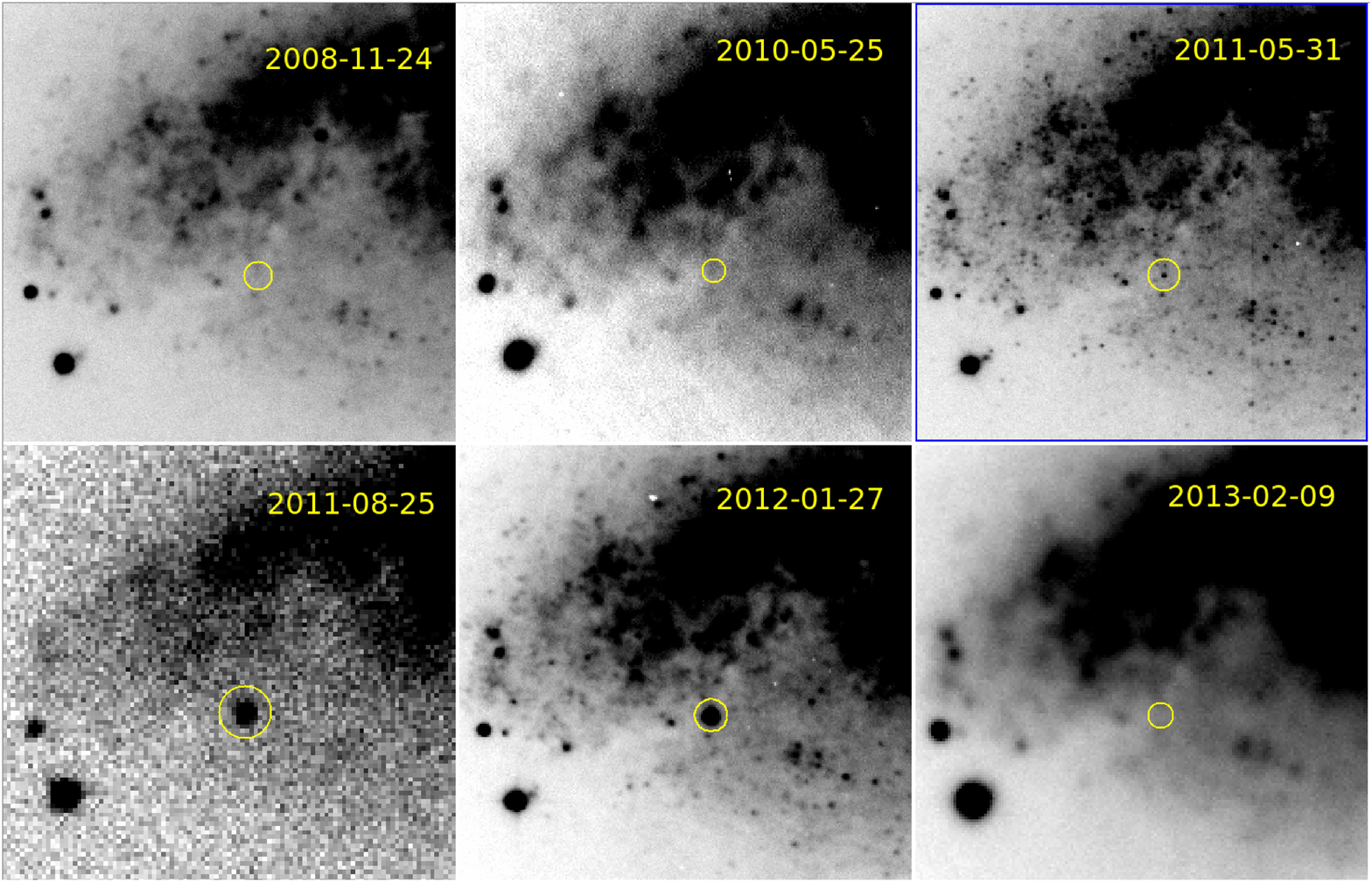}  
 \caption{$R$-band images of the location of NGC4490-2011OT1 at six representative epochs. The field of view is $1.5\arcmin$.   \label{zoom_NGC4490}} 
\end{figure*}

\section{The sample} \label{sample} 
 
The discovery of NGC4490-2011OT1 (also known as PSN J12304185+4137498) was announced by \citet{cor11}. 
The discovery epoch is 2011 August 16.83 UT.\footnote{UT dates will be used throughout this paper.} 
The coordinates of the object are: $\alpha=12^{h}30^{m}41\fs84$ and $\delta=+41\degr37\arcmin49\farcs7$ (equinox J2000.0), which is $57\arcsec$ east and $37\arcsec$ south of 
the centre of NGC~4490 (Fig. \ref{maps}, top-left panel). This galaxy, classified as type SBcd by Hyperleda,\footnote{\sl http://leda.univ-lyon1.fr/} 
belongs to an interacting pair with NGC~4485 known as Arp~260. NGC~4490 also hosted the peculiar Type II-P SN~1982F \cite[likely similar to SN~1987A, see][]{tsv88,pasto12} 
and the Type~IIb SN~2008ax \citep[][]{pasto08,tsv09,rom09,tau11b,cho11}.\\  
Soon after the discovery of  NGC4490-2011OT1, \citet{mag11} spectroscopically classified it as an SN impostor. \citet{fra11} analysed pre-SN 
HST archival images, and found a candidate progenitor with an absolute magnitude of $F606W=-6.2$ mag (neglecting possible reddening 
corrections), which is fainter than the expected luminosity of a massive LBV. More recently, \citet{smi16b} provided follow-up data 
of NGC4490-2011OT1, along with a revised discussion on the candidate progenitor, concluding that the latter was likely an intermediate-mass yellow supergiant
or (adopting a higher reddening correction) a moderate luminosity LBV. In agreement with \citet{pasto08} and \citet{smi16b}, 
we adopt $d=9.6$ Mpc (hence, distance modulus $\mu=29.91\pm0.29$ mag) as the distance to NGC~4490, which is the 
average value of several estimates obtained with different methods and available in the NASA/IPAC Extragalactic Database (NED)\footnote{\sl https://ned.ipac.caltech.edu/}. 
The Galactic reddening, provided by \citet{sch11}, is $E(B-V)=0.02$ mag, while the host galaxy 
extinction is more uncertain. As remarked by \citet{smi16b}, the total extinction may vary from $A_V=0$ to $A_V=2$ mag.
In this paper, following  \citet{smi16b}, we will adopt for NGC4490-2011OT1 their most likely value, $E(B-V)=0.32$ mag. 
Multi-epoch images of the site of NGC4490-2011OT1 are shown in Fig. \ref{zoom_NGC4490}.

NGC3437-2011OT1 (also named PSN J10523453+2256052 and SNhunt31) is a discovery of the Catalina Real-time Transient 
Survey (CRTS),\footnote{{\sl http://crts.caltech.edu/}; CRTS data can be obtained using the CRTS Data Release 3 (CSDR3) web service interface: 
{\sl http://crts.iucaa.in/CRTS/}.} made on 2011 January 10.41 \citep{how11}. The object has coordinates $\alpha=10^{h}52^{m}34\fs53$ and  
$\delta=+22\degr56\arcmin05\farcs2$ (equinox J2000.0), and is located  $16\arcsec$ west and $2\arcsec$ north of the core of the SABc-type host 
galaxy NGC~3437 (Fig. \ref{maps}, top-right panel). That galaxy hosted also the Type Ic SN~2004bm \citep{arm04,fol04}. \\
Independent classifications of NGC3437-2011OT1 as a SN impostor were provided by \citet{vin11} and \citet{tau11b} on the basis 
of its faint absolute magnitude and the presence of narrow emission lines of H superposed on a blue spectral continuum. 
Weak Fe~II emission lines were also detected by \citet{tau11b}. 
Adopting the HyperLeda recessional velocity corrected for the Local Group infall onto the Virgo Cluster ($v_{Vir}=1351$ km s$^{-1}$), 
and a standard cosmology (with Hubble Constant $H_0=73$ km s$^{-1}$ Mpc$^{-1}$, with $\Omega_\Lambda=0.73$ and $\Omega_M=0.27$), we obtain a luminosity distance  
$d=18.6$ Mpc, hence $\mu=31.35\pm0.15$ mag. We note, however, that the above distance is significantly lower than that inferred using other methods (e.g., 
Tully-Fisher and sosie galaxies, reported by NED), which in fact provide a median  
$d=23.9$ Mpc ($\mu=31.89\pm0.24$ mag). 
In this paper, we adopt the recent Tully-Fisher estimate corrected for selection biases of \citet{sor14} and scaled to $H_0=73$ km s$^{-1}$ Mpc$^{-1}$, 
hence $d$ = 20.9 Mpc ($\mu=31.60\pm0.43$ mag). 
As there is no evidence for significant host galaxy dust absorption from the spectra, a total reddening $E(B-V)=0.02$ mag is assumed \citep{sch11}.

UGC12307-2013OT1 (PSN J23011153+1243218) was discovered on 2013 July 17.06 by F. Ciabattari, E. Mazzoni and S. Donati of the Italian Supernova Search Project  
(ISSP).\footnote{\sl http://italiansupernovae.org/} Information on the discovery was posted on the ``Transient Object Followup Reports'' pages of the Central Bureau for 
Astronomical telegrams (CBAT). The transient was observed  in the irregular galaxy UGC~12307, at the following coordinates: $\alpha=23^{h}01^{m}11\fs53$ and $\delta=+12\degr43\arcmin21\farcs8$ 
(equinox J2000.0). The object's position is $6\arcsec$ east of  the nucleus of the host galaxy (Fig. \ref{maps}, middle-left panel). The unfiltered discovery magnitude was about 18.3 mag. 
From the HyperLeda Virgo-corrected recessional velocity  ($v_{Vir}=2881$ km s$^{-1}$), we obtain a luminosity distance $d=39.7$ Mpc, hence $\mu=32.99\pm0.15$ mag. UGC12307-2013OT1 
is significantly affected by line-of-sight Galactic reddening, which is $E(B-V)=0.22$ mag according to the tabulated values of \citet{sch11}. There is no clear evidence of additional 
host galaxy reddening in the spectra of the transient. 
 
For completeness, in this paper we will discuss three additional LRNe whose data are available in the literature:  
SN 1997bs\footnote{M66-1997OT1, according to the naming code adopted in this paper.} \citep[averaging different Cepheid distance
estimates available in NED yields $d=9.20$ Mpc, hence $\mu=29.82\pm0.07$ mag; adopted extinction $E(B-V)=0.21$ mag;][see Fig. \ref{maps}, middle-right panel]{van00};
M101-2015OT1 \citep[a.k.a. PSN J14021678+5426205; from the Cepheid distance $d=6.43$ Mpc, $\mu=29.04\pm0.19$ mag; adopted extinction $E(B-V)=0.01$ mag;][Fig. \ref{maps}, bottom-left panel]{sha11,bla16}; and
SNhunt248\footnote{This object should be named as NGC5806-2014OT1, adopting the naming code of this paper. However, we will maintain the SNhunt designation for
this object, in agreement with other published papers.} \citep[PSN J14595947+0154262; from the Tully-Fisher distance $d=22.50$ Mpc, we obtain $\mu=31.76\pm0.36$ mag; 
adopted extinction $E(B-V)=0.04$ mag;][Fig. \ref{maps}, bottom-right panel]{tul09,kan15,mau15}. For all of them, we also present previously unpublished photometric data. 
A table with a summary of the host galaxy parameters for LRNe is given in Sect. \ref{nature}.
 
Some similarity with LRNe can also be found with the 2010 outburst observed in UGC~5460 \citep[][hereafter, UGC5460-2010OT1]{fra13}. 
Such a faint transient was followed a few months later by the explosion of the Type~IIn SN~2011ht \citep{rom12,hum12,mau13}. 
The implications will be discussed in Sect. \ref{11ht}.
 
\section[]{Photometric data} \label{photo_redu} 
 
The light curves presented in this paper were obtained through follow-up campaigns started after the discovery announcements of each transient, 
but were also complemented by unfiltered data collected from amateur astronomers, scaled to the best matching broadband photometry
according to the quantum efficiency curves of the CCDs used in these observations. Additional historical data were collected through the analysis of images 
available in the public archives. Basic information on the instrumental configurations used in the monitoring campaigns are listed in the footnotes of the
photometry Tables A.1-A.6.
 
Photometric images were first processed by applying overscan, bias and flat-field corrections, using specific tasks in the  
IRAF\footnote{\sl http://iraf.noao.edu/} environment. Multiple exposures were median-combined to increase the signal-to-noise ratio (S/N). 
 The source instrumental magnitudes and their subsequent photometric calibrations were measured using a dedicated pipeline \citep[SNOoPY,][]{cap14}.
The magnitudes of individual ILOTs were obtained through the point spread function (PSF) fitting technique. The procedure allows us to subtract the sky background using a  
low-order polynomial fit (which - in most cases - was a second-order polynomial). A PSF template was computed using the profiles of a limited 
number (5 to 10) of isolated stars in the field of the transient. The fitted object is then removed from the original frames, and the residuals  
at the location of the object are inspected to validate the fit.  

For images obtained with Johnson-Cousins filters, the photometric calibration of instrumental data was based on zeropoints and colour terms obtained through 
observations of standard photometric fields from the \citet{lan92} catalogue performed in photometric nights. When Landolt standards were  
not available (or the observations of the LRNe were taken in poor transparency conditions), archival instrumental zeropoints and colour terms were used. 
In order to improve the photometric calibration, we fixed a secondary sequence of stars in the fields of the LRNe. 
In most cases, photometric sequences were already available in the literature.\footnote{We used reference star magnitudes available in the literature for the 
following objects: SN~1997bs \protect\citep[from][]{van00}, NGC4490-2011OT1 \protect\citep[from][]{tau11a}, SNhunt248  \protect\citep[from][]{kan15}, and 
M101-2015OT1 \protect\citep[from][]{bla16}.} When not available in the literature, Johnson-Cousins magnitudes of secondary sequence stars were derived from 
broadband Sloan magnitudes released by the Sloan Digital Sky Survey (SDSS)\footnote{\sl http://sdss.org},  
following the prescriptions of \citet{chr08}.
These reference stars enabled us to compute zeropoint corrections  for each night, to  improve the photometric calibration. The Sloan-band 
photometry of M101-2015OT1 was directly calibrated using the SDSS catalogue.
 
Photometric errors were computed through artificial star experiments, by placing fake stars of known magnitude near to the position of the LRNe, and 
measuring their magnitudes via PSF-fitting photometry. The dispersion of individual artificial star measurements was combined in quadrature with the PSF-fit 
and the zeropoint calibration errors, giving the final uncertainty of the photometric data. The resulting magnitudes of the transients considered in this paper 
are given in Tables A.1-A.6, available at the CDS, which contain the 
date of the observation, the JD, the optical  magnitudes and errors, and  a numeric code for the instrumental configuration. 

\begin{figure*} 
\centering
{\includegraphics[width=7.6cm,angle=0]{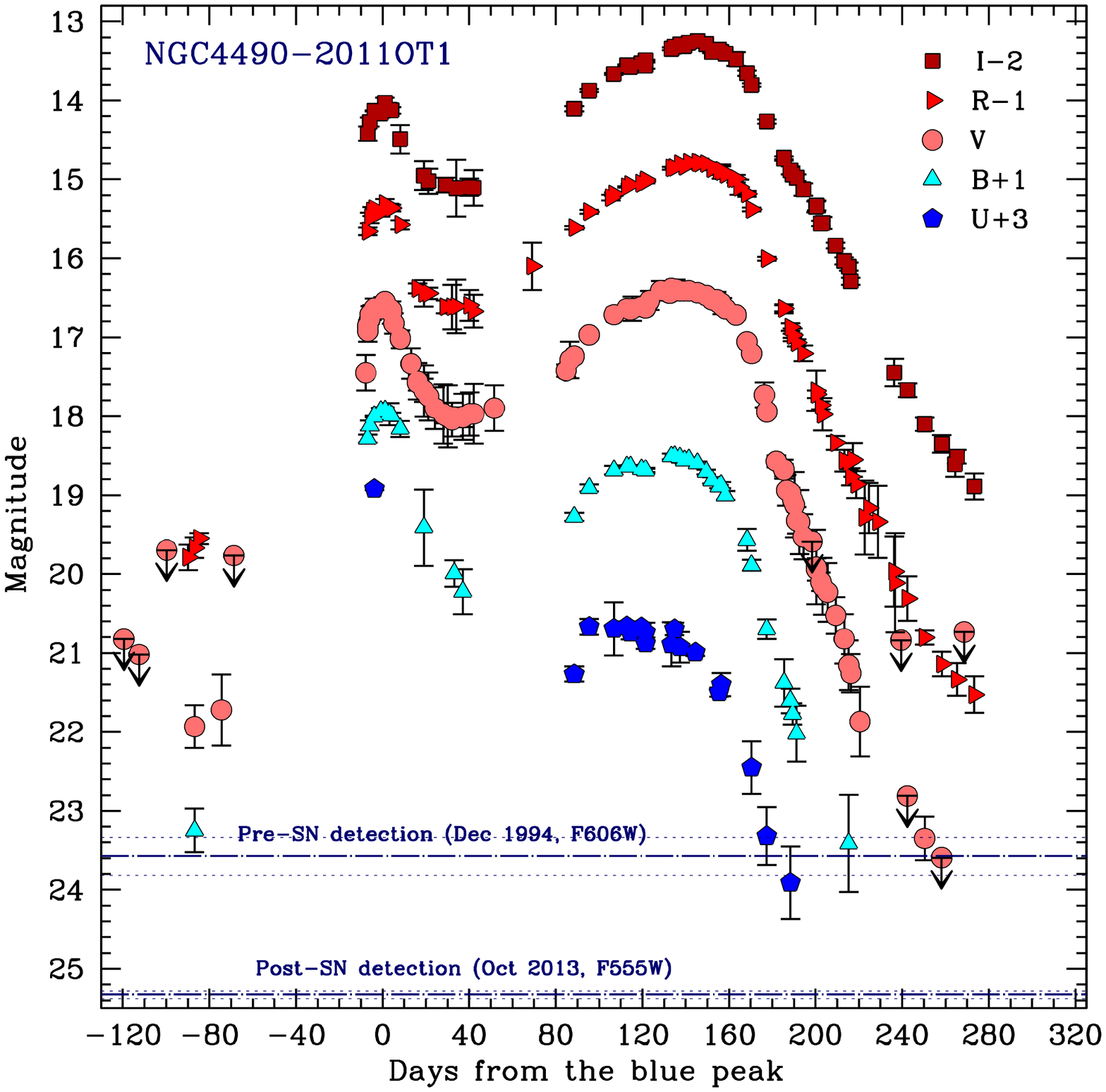}  
\includegraphics[width=7.6cm,angle=0]{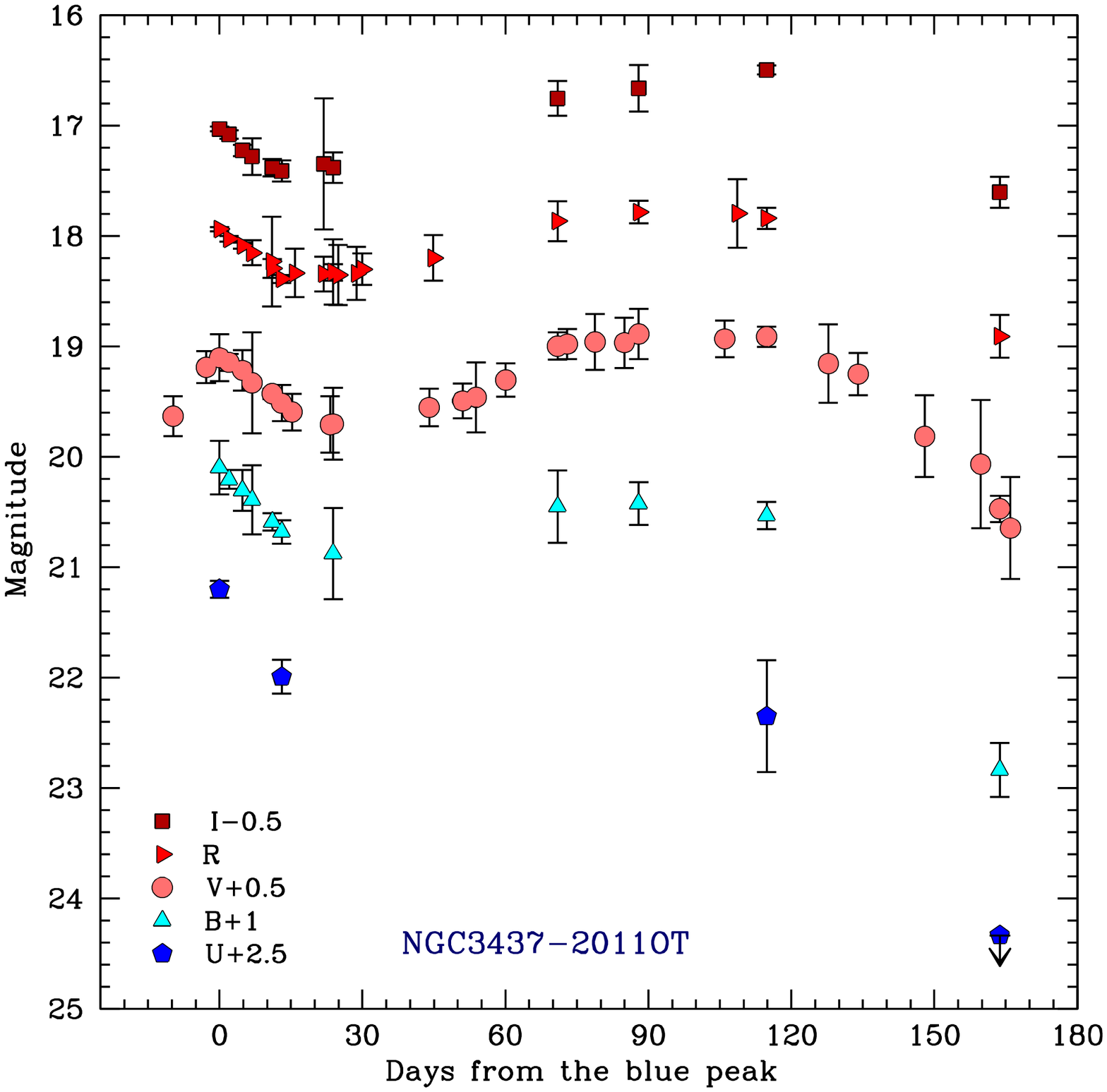}  
\includegraphics[width=7.6cm,angle=0]{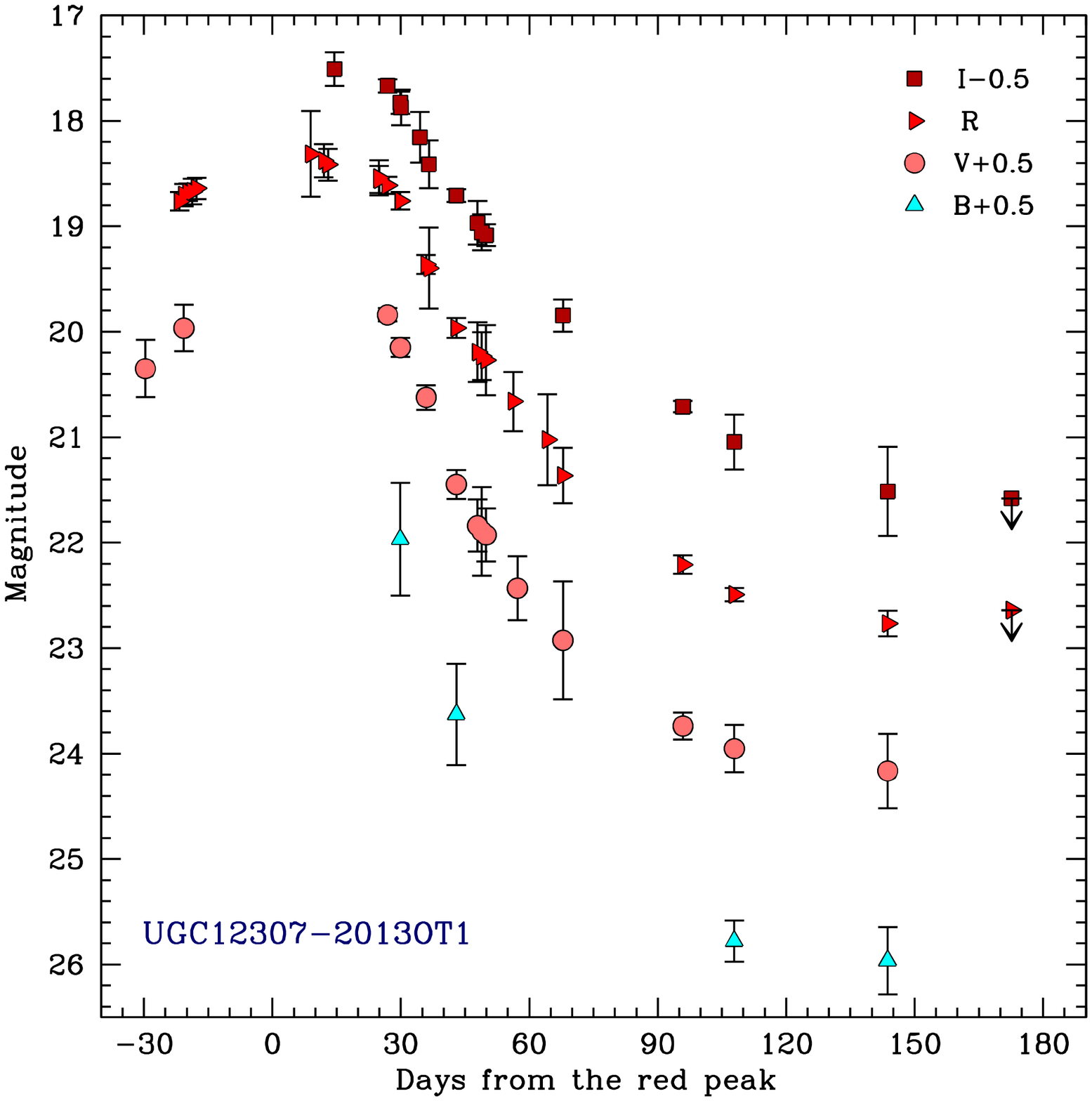}  
\includegraphics[width=7.6cm,angle=0]{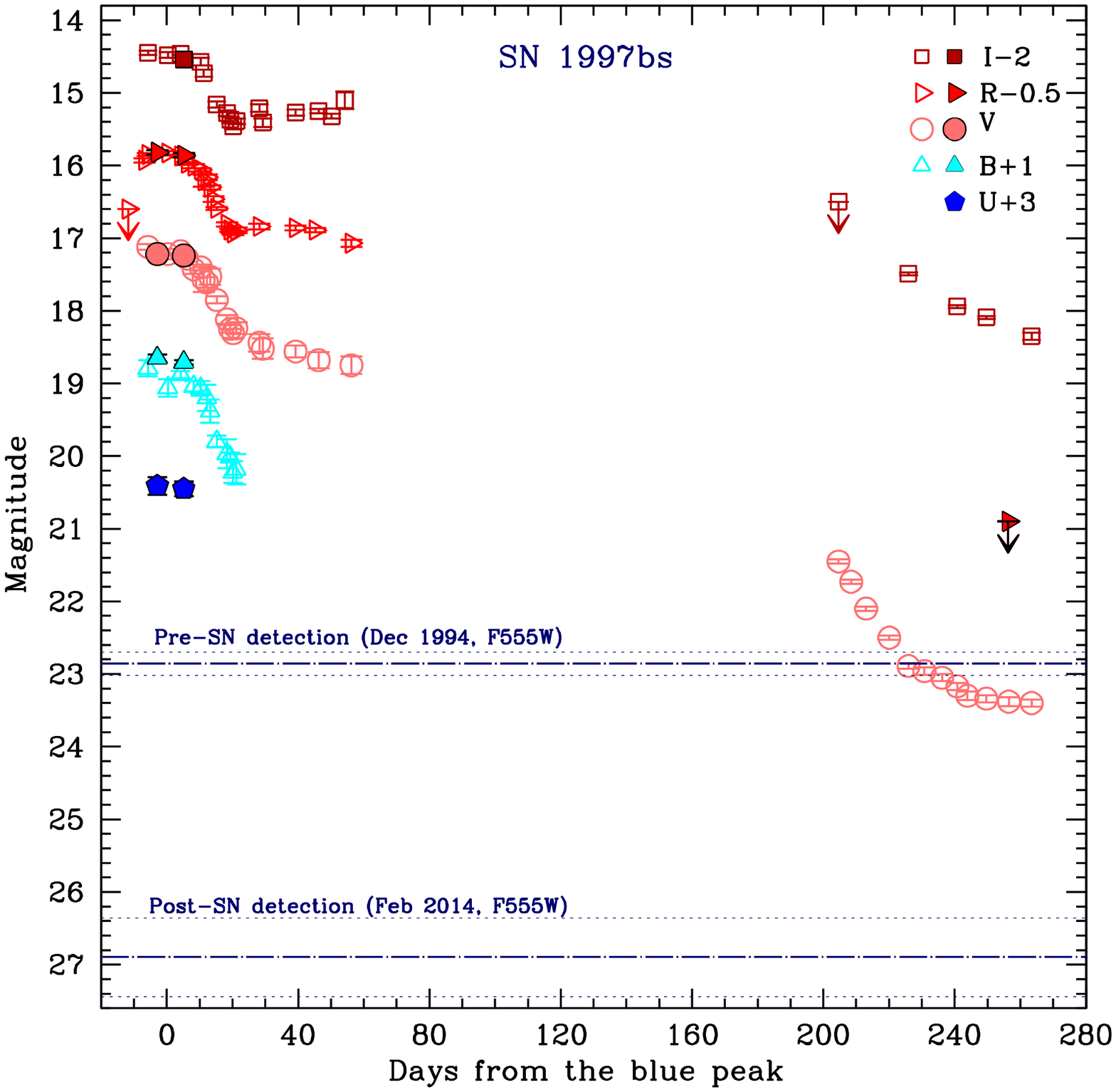}
\includegraphics[width=7.6cm,angle=0]{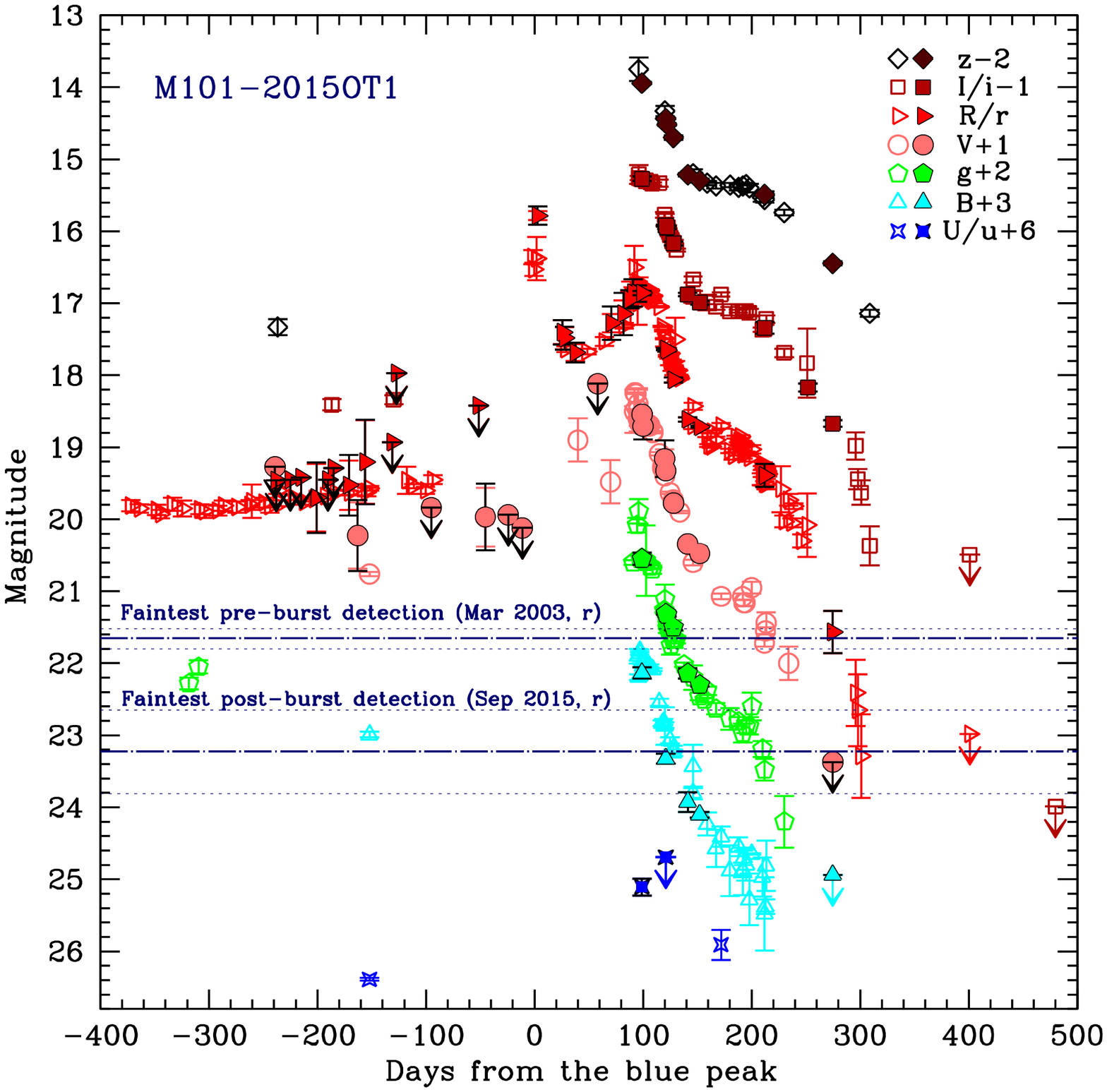}
\includegraphics[width=7.6cm,angle=0]{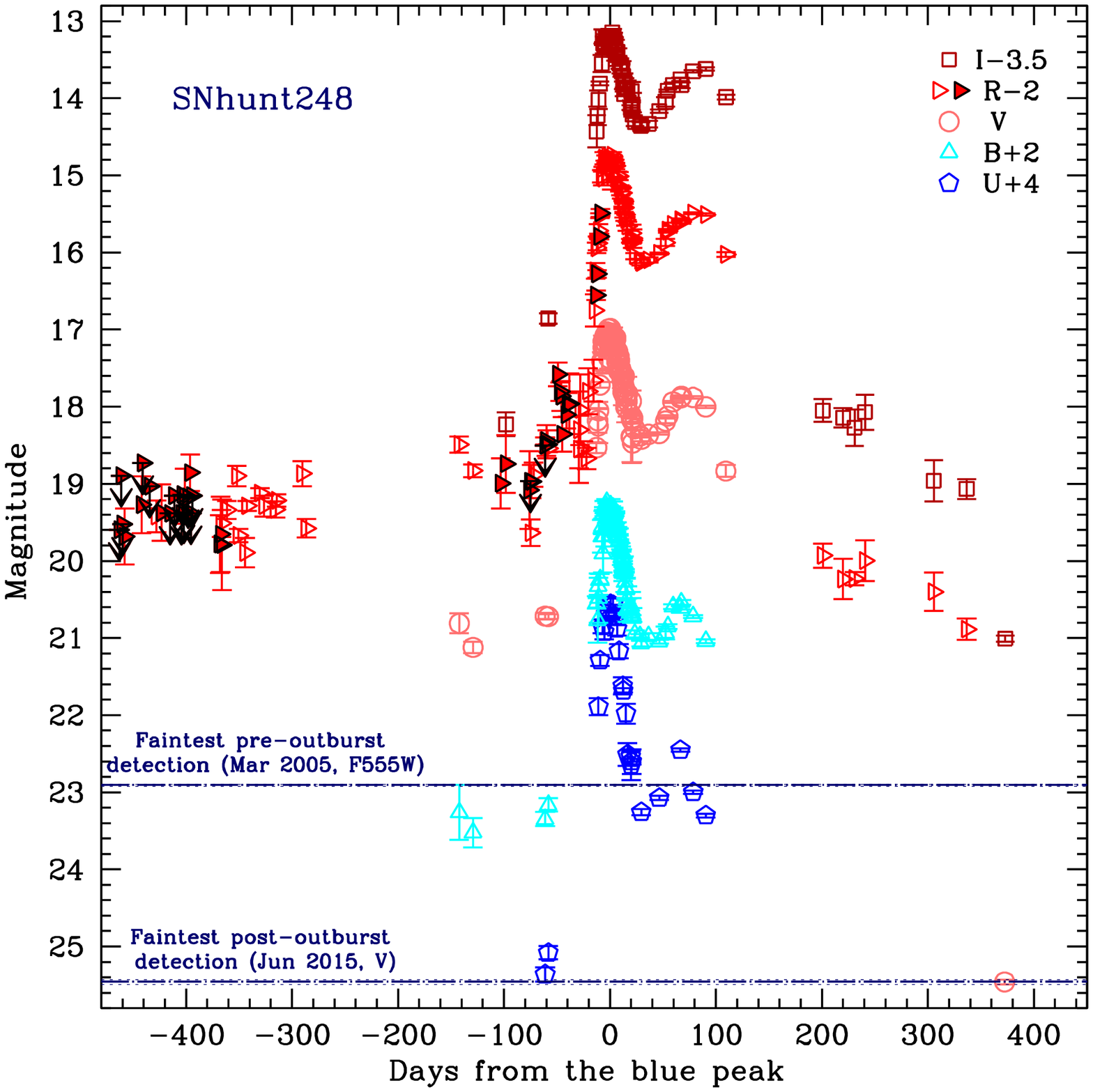}} 
  \caption{Multi-band  light curves of NGC4490-2011OT1 (top-left), NGC3437-2011OT1 (top-right), UGC12307-2013OT1 (middle-left), 
SN~1997bs (middle-right), M101-2015OT1 (bottom-left), and SNhunt248 (bottom-right). Empty symbols in the SN~1997bs, M101-2015OT1 and SNhunt248 
light curves are data from the literature (see text). Only the most significant detection limits are shown. All data are calibrated in the Vega system.
 \label{lc}} 
\end{figure*} 
 
\subsection{Individual light curves} \label{photo_lc} 
  
 The multi-band light curves of the six objects are shown in Fig. \ref{lc}. For most transients, stringent  limits obtained before the first detection 
are not available, and this affects our accuracy in dating the outbursts. The main parameters of the light curves, estimated through low-order polynomial fits, 
are listed in Table \ref{lc_par}.  For comparison, we have also included the parameters for the well-monitored LRN AT~2017jfs \citep{pasto19}.\footnote{An finder chart showing the field of AT~2017jfs is reported in Fig. \ref{finder_17jfs}.}

The object most extensively followed is NGC4490-2011OT1. It was discovered during the fast rise to the first peak, reached on Julian Day (JD) $=2\,455\,798.2\pm0.9$.  
At this epoch, $V=16.55\pm0.05$ mag, while the colours are $B-V\approx0.35$ mag, $V-I\approx0.5$ mag and $R-I\approx0.23$ mag.  
The host galaxy reddening contribution is likely significant but very uncertain, as discussed by \citet{smi16b}, and only the reddening correction due to the Milky Way 
dust is applied to the  colours estimated in this section.
For about one month after this early maximum, the light curve declines in all bands, reaching a minimum at $V=18.05\pm0.22$  mag ($V-R\approx0.4$ mag) on JD = 2\,455\,830.3. 
Then, the luminosity rises again to a second maximum, reached on JD $=2\,455\,937.7\pm3.2$ (i.e., about 140 d after the first $V$-band peak).  
The magnitude at the second peak is $V=16.36\pm0.01$ mag, with the following observed colours: $B-V\approx1.15$  mag, $V-I\approx1.1$  mag, and  
$R-I\approx0.5$ mag. Hence, the spectral energy distribution (SED) at the first maximum peaks at shorter wavelengths than at the second maximum. 
Hereafter, we will label the first and the second peaks as {\it blue} and {\it red} peaks, respectively.  
After the red peak, the light curves decline almost linearly in all bands. We monitored the luminosity decline for further $\sim$4 months, and the object became even redder: 
on JD = 2\,456\,048.6, $V=23.35\pm0.28$ mag, $V-I=3.2\pm0.3$ mag and $R-I=1.7\pm0.1$ mag. In our last detection (JD = 2\,456\,071.5, hence 
272 d after the blue peak), the object has a magnitude $R=22.53\pm0.23$ mag,  
and $R-I=1.6\pm0.3$ mag. The red colours of NGC4490-2011OT1 at late times along with spectroscopic clues (see Sect. \ref{spectra_evol}) suggest that some  
circumstellar dust is forming. We note that in our latest detection, NGC4490-2011OT1 is still about 1 mag brighter than the progenitor's detection reported  
by \cite{fra11} and \citet{smi16b}. 
 In fact, the object was detected by \citet{smi16b} with the Hubble Space Telescope (HST) at even later stages. On JD = 2\,456\,595.7 (over two years after the blue peak),
the object has magnitude $F555W$ (nearly $V$) $=25.33\pm0.05$ mag, and the following approximate colours: $F438W-F555W$ (nearly $B-V$) = 0.1 $\pm$ 0.1 mag  
and $F555W-F814W$ ($\sim V-I$) $=1.9\pm0.1$ mag. This very late source is fainter than our latest detection, 
but  the colours are  bluer  than soon after the red peak.
 This suggests that additional sources are powering the luminosity  
of NGC4490-2011OT1 at very late phases in the Smith et al. images, possibly radiation from shell-shell collisions or light echoes. Incidentally, light echoes 
were observed after the eruption of  V838 Mon \citep{mun02,gor02,kim02,cra03}. 
We will discuss the implications of this analogy with V838 Mon in Sect. \ref{nature}.

\begin{table*} 
\centering
  \caption[]{Light curve parameters for the six LRNe discussed in this paper. We report the JD epochs and the magnitudes of the first (blue) peak 
in columns 2 and 3, those for the second (red) maximum in columns 4 and 5, the time span between the two peaks in the different bands in column 6, and the 
magnitude difference between the blue and red peaks in column 7. The data of AT~2019jfs \protect\citep{pasto19} are also reported (the Sloan-band peak
magnitudes are in the AB system).   \label{lc_par}}
  \begin{tabular}{ccccccc} 
  \hline  \hline 
Filter & JD blue max       &  Mag blue max  &    JD red max       &  Mag red max & $\Delta$d & $\Delta$mag (blue-red) \\  \hline 
   \multicolumn{7}{c}{NGC4490-2011OT1} \\ \hline 
$B$ & 2\,455\,798.7 $\pm$ 0.6 &   16.92 $\pm$ 0.01  & 2\,455\,930.6 $\pm$ 3.5 &  17.50 $\pm$ 0.01 & 131.9 $\pm$ 3.6 & $-$0.58 $\pm$ 0.01 \\ 
$V$ & 2\,455\,798.2 $\pm$ 0.9 &   16.55 $\pm$ 0.05  & 2\,455\,937.7 $\pm$ 3.2 &  16.36 $\pm$ 0.01 & 139.5 $\pm$ 3.3 & $+$0.19 $\pm$ 0.05 \\ 
$R$ & 2\,455\,798.8 $\pm$ 0.6 &   16.29 $\pm$ 0.03  & 2\,455\,941.3 $\pm$ 3.0 &  15.76 $\pm$ 0.01 & 142.5 $\pm$ 3.1 & $+$0.53 $\pm$ 0.03 \\ 
$I$ & 2\,455\,799.0 $\pm$ 1.0 &   16.04 $\pm$ 0.04  & 2\,455\,941.5 $\pm$ 2.7 &  15.27 $\pm$ 0.01 & 142.5 $\pm$ 2.9 & $+$0.77 $\pm$ 0.04 \\ \hline 
   \multicolumn{7}{c}{NGC3437-2011OT1} \\ \hline 
$B$ & $\sim$  2\,455\,574.7   &   19.10 $\pm$ 0.24  & 2\,455\,665.9 $\pm$ 14.2 & 19.32 $\pm$ 0.10 & $\sim$ 91.2   & $\sim -$0.22  \\ 
$V$ & $\sim$  2\,455\,574.7   &   18.60 $\pm$ 0.21  & 2\,455\,671.6 $\pm$ 3.6 &  18.33 $\pm$ 0.02 & $\sim$ 96.7   & $\sim +$0.27  \\ 
$R$ & $\sim$  2\,455\,574.7   &   17.94 $\pm$ 0.02  & 2\,455\,675.3 $\pm$ 6.2 &  17.76 $\pm$ 0.03 & $\sim$ 100.6  & $\sim +$0.18  \\ 
$I$ & $\sim$  2\,455\,574.7   &   17.53 $\pm$ 0.02  & 2\,455\,686.0 $\pm$ 9.5 &  17.00 $\pm$ 0.04 & $\sim$ 111.3  & $\sim +$0.53  \\ \hline 
   \multicolumn{7}{c}{UGC12307-2013OT1} \\ \hline 
$V$ & --                  &   --                & 2\,456\,481.6 $\pm$ 8.0 &  18.64 $\pm$ 0.39 & -- & -- \\  
$R$ & --                  &   --                & 2\,456\,481.9 $\pm$ 5.2 &  18.18 $\pm$ 0.10 & -- & -- \\ 
$I$ & --                  &   --                & 2\,456\,499.4 $\pm$ 8.2 &  17.93 $\pm$ 0.11 & -- & -- \\ \hline 
   \multicolumn{7}{c}{SN 1997bs} \\ \hline 
$B$ & 2\,450\,559.3 $\pm$ 1.7 &  17.70 $\pm$ 0.06 & --                  &   --                & -- & -- \\  
$V$ & 2\,450\,560.5 $\pm$ 2.4 &  17.13 $\pm$ 0.06 & --                  &   --                & -- & -- \\  
$R$ & 2\,450\,562.1 $\pm$ 2.7 &  16.82 $\pm$ 0.03 & --                  &   --                & -- & -- \\ 
$I$ & 2\,450\,562.4 $\pm$ 2.4 &  16.44 $\pm$ 0.03 & --                  &   --                & -- & -- \\ \hline 
   \multicolumn{7}{c}{M101-2015OT1} \\ \hline 
$B$ & --                    &   --                & 2\,457\,065.5 $\pm$ 5.2 &  18.84 $\pm$ 0.05 & --  & --  \\ 
$V$ & $\sim$  2\,456\,972.5     & $\sim$  16.36       & 2\,457\,070.8 $\pm$ 5.8 &  17.61 $\pm$ 0.03 & $\sim$ 98.3 & $\sim -$1.25 \\ 
$R$ & $\sim$  2\,456\,972.5     & $\sim$  16.36       & 2\,457\,071.4 $\pm$ 6.5 &  16.74 $\pm$ 0.05 & $\sim$ 98.9 & $\sim -$0.38 \\ \hline
   \multicolumn{7}{c}{SNhunt248} \\ \hline 
$B$ & 2\,456\,828.5 $\pm$ 2.2   &   17.33 $\pm$ 0.01  & 2\,456\,897.4 $\pm$ 2.6 &  18.59 $\pm$ 0.05 & 68.9 $\pm$ 3.4 & $-$1.26 $\pm$ 0.05 \\ 
$V$ & 2\,456\,827.9 $\pm$ 1.7   &   17.03 $\pm$ 0.01  & 2\,456\,904.1 $\pm$ 1.8 &  17.83 $\pm$ 0.01 & 76.2 $\pm$ 2.5 & $-$0.80 $\pm$ 0.01 \\ 
$R$ & 2\,456\,828.1 $\pm$ 1.5   &   16.73 $\pm$ 0.01  & 2\,456\,908.2 $\pm$ 1.7 &  17.47 $\pm$ 0.01 & 80.1 $\pm$ 2.3 & $-$0.74 $\pm$ 0.01 \\ 
$I$ & 2\,456\,828.0 $\pm$ 1.1   &   16.66 $\pm$ 0.01  & 2\,456\,913.0 $\pm$ 1.4 &  17.12 $\pm$ 0.01 & 85.0 $\pm$ 1.8 & $-$0.46 $\pm$ 0.01 \\ 
\hline                                                                                       
   \multicolumn{7}{c}{AT~2017jfs} \\ \hline
$B$ & $<$ 2\,458\,118.8       &  $<$ 17.96       & slow decline        &  $>$ 19.54        & --              & exceeds $-$1.58     \\ 
$V$ & $\sim$ 2\,458\,113.7    & 17.34 $\pm$ 0.43 & 2\,458\,206.0 $\pm$ 10.3&  18.43 $\pm$ 0.04 & $\sim$ 92.3     & $-$1.09 $\pm$ 0.43 \\ 
$g$ & 2\,458\,115.3 $\pm$ 0.6 & 17.38 $\pm$ 0.03 & 2\,458\,169.5 $\pm$ 18.0&  19.08 $\pm$ 0.02 & 54.2 $\pm$ 18.0 & $-$1.70 $\pm$ 0.04 \\
$r$ & 2\,458\,116.0 $\pm$ 1.2 & 17.04 $\pm$ 0.05 & 2\,458\,208.0 $\pm$ 12.4&  17.67 $\pm$ 0.01 & 92.0 $\pm$ 12.5 & $-$0.63 $\pm$ 0.05 \\
$i$ & $<$ 2\,458\,118.8       &  $<$ 17.46       & 2\,458\,208.2 $\pm$ 12.8&  17.08 $\pm$ 0.02 & $\sim$ 92.2     & $\sim +$0.38       \\ \hline
\end{tabular}                                                                               
\end{table*}

The  photometric evolution of  NGC3437-2011OT1 is similar to that of NGC4490-2011OT1. 
The reference parameters of the blue peak light curve of NGC3437-2011OT1 reported in Table~\ref{lc_par} are those of the earliest multi-band observation. 
In fact, we assume the object reaches the blue peak on JD = 2\,455\,574.7, with $V=18.60\pm0.21$ mag, $B-V\approx0.5$ mag,  
$V-I\approx1.1$ mag, and $R-I\approx0.4$ mag. After peak, a luminosity decline is observed, lasting 24 d, when a local minimum is found at $V=19.20\pm0.33$ mag.  
Later on, the object experiences another luminosity rise to a red maximum, reached on JD = 2\,455\,671.6 (i.e., 97 days after the blue peak) in the $V$ band.  
We note that the time interval between the two $V$-band peaks in NGC3437-2011OT1 is shorter 
than that observed in NGC4490-2011OT1 (97 d vs. 140 d). At the red peak, the apparent magnitudes are $V=18.33\pm0.02$ mag,  
with $B-V\approx1$ mag, $V-I\approx1.3$ mag, and $R-I\approx0.8$ mag. 
After peak, the luminosity declines and in one of our latest detections (on JD = 2\,455\,738.5) it has the following $V$-band magnitude and colours: 
$V=19.97\pm0.12$ mag, $B-V \approx V-I \approx 1.85$ ($\pm0.2$ mag), and $R-I=0.8\pm0.2$ mag. 
Similar to NGC4490-2011OT1, the peak of the spectral energy distribution (SED) of NGC3437-2011OT1 at late phases shifts to longer wavelengths.

The red colour of UGC12307-2013OT1 at discovery suggests that this object was already at late phases. 
The rise time to maximum is slow (as inferred from the two earliest CRTS detections), and is slower  
than the post-peak decline. On the basis of these photometric considerations along with the striking similarity of the first UGC12307-2013OT1 spectrum with those of the  
other transients discussed in this paper obtained during the red maximum (see Sect. \ref{spectra_evol}), we propose that UGC12307-2013OT1 was discovered after the blue  
peak, during the rise to the red maximum. The peak is reached on JD $=2\,456\,481.9\pm9.0$ in the $V$ band. The earliest multi-band data were obtained on  
JD = 2\,456\,508.5 (hence, almost four weeks after the red maximum), with $V=19.34\pm0.06$ mag. At this epoch, we measure the 
following colours, corrected for Galactic reddening: $V-I=0.9\pm0.1$ mag and $R-I=0.3\pm0.1$ mag. The earliest epoch with $B$ band information is three days later,  
when the $B-V$ colour is $1.6\pm0.6$ mag. 
Then, a decline is observed in all bands, and last detection is reported on JD = 2\,456\,625.3, with the following magnitudes and colours: $V=23.66\pm0.35$ mag,  
$B-V=1.6\pm0.4$ mag, $V-I=1.3\pm0.4$ mag, and $R-I=0.6\pm0.3$ mag. Hence, also this object becomes redder with time. As we will see  
in Sect.  \ref{spectra_evol}, there is spectroscopic evidence for molecular band formation, which suggests possible dust condensation. Nonetheless,
all broadband light curves show a flattening at over 90 days after the red peak, which is the opposite than what expected when dust is forming. 
 
We also observed SN 1997bs at three epochs, two around the blue peak and one at late phase. Our early photometry does not add significant information to that
 published by \citet{van00}. If we account for a total reddening of $E(B-V)=0.21$ mag for SN 1997bs (in agreement with Van Dyk et al.), we obtain
the following colours at the blue peak: $B-V\approx0.35$ mag, $V-I\approx0.4$  mag and  
$R-I\approx0.25$ mag. Then the object declined to a sort of plateau \citep[similar to that shown in the $B$-band light curve of AT~2017jfs,][]{pasto19}, with reddening-corrected colours $V-I=1.1\pm0.1$ mag, and $R-I=0.5\pm0.1$ mag.  
Later, the object disappeared behind the Sun, and was recovered only five months later. For this reason, we cannot definitely assess whether this transient is a LRN or another type of gap transient.
However, when recovered from 232 d to 269 d after the blue peak, the average $V-I$ colour was around $3.2\pm0.1$ mag. Our $R$-band detection limit on JD = 2\,450\,816.8 (phase = 262 d) allows us to constrain the following colours at that epoch:  
$V-R <1.34$ mag and $R-I >1.41$ mag. Further late-measurements in 2004, 2009 and 2014 reported by \citet{ada15}, indicate the  object was then fainter than the magnitude of the progenitor reported by \citet{van00} ($V=22.86\pm0.16$ mag). A more recent reddening-corrected colour of the source at the position of SN 1997bs is estimated from HST images \citep{ada15}, providing  
$V-I=0.86\pm0.20$ mag in 2009, and $V-I=1.11\pm0.56$ mag in 2014 (under the rough assumption that the wide HST bands are similar to Johnson-Cousins filters). 
  
\begin{figure*}[!t] 
\centering
\includegraphics[width=17.0cm,angle=0]{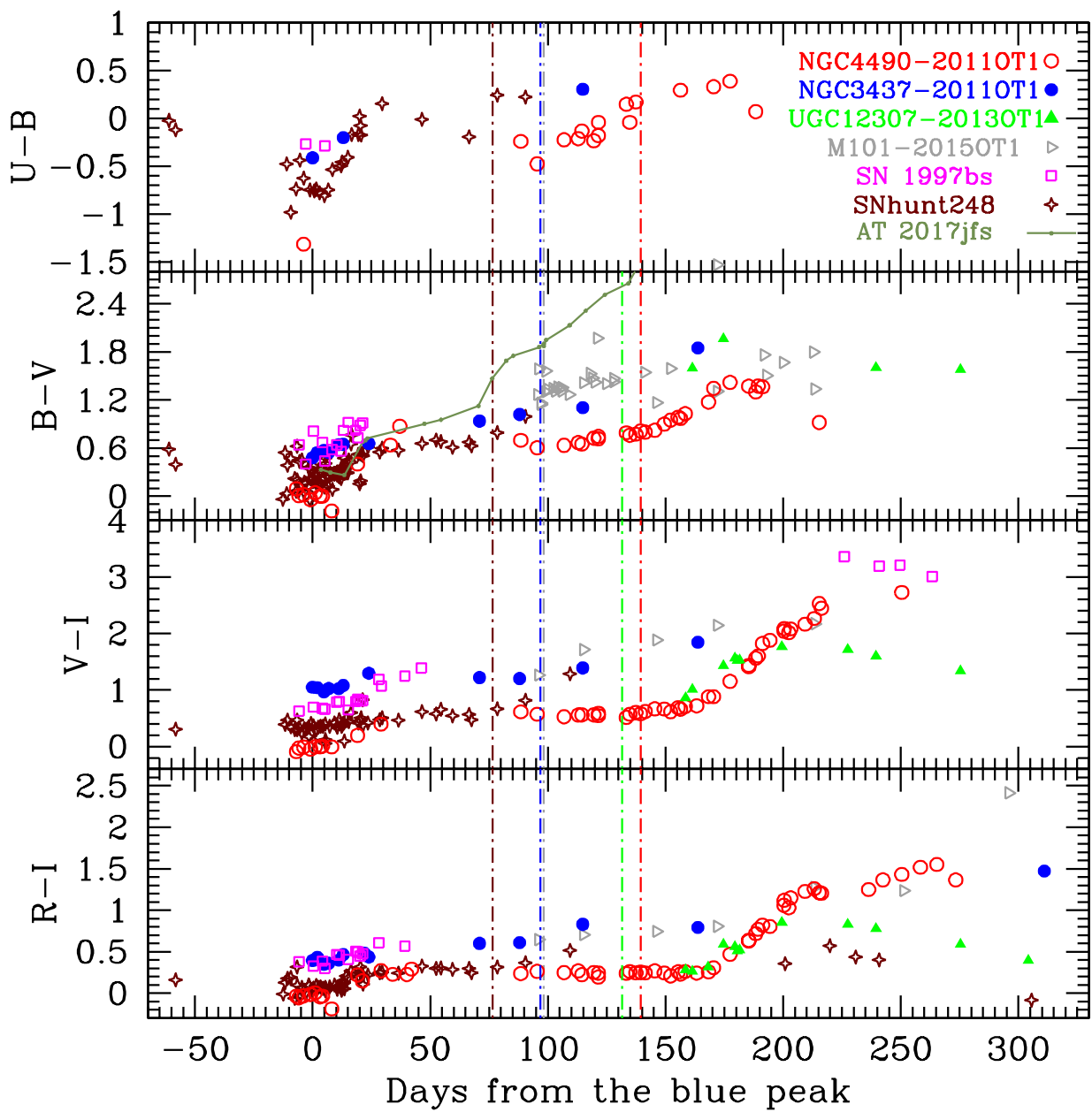}  
  \caption{Colour evolution for our sample of LRNe. The phase is in days from the first $V$ band maximum.  
For UGC12307-2013OT1, we tentatively adopted JD = 2\,456\,350 as epoch of the first peak. The dot-dashed vertical lines 
(plotted maintaining the same colour codes as the symbols) mark the epochs of the red maximum for the different objects.
In SN~1997bs, the red maximum was not observed. The $B-V$ colour curve of AT~2017jfs \protect\citep{pasto19} is also shown
as a comparison (solid line).\label{colorr}} 
\end{figure*} 

The light curve of M101-2015OT1 is extensively described in \citet{gor16} and \citet{bla16}, and the new data provided here do not change the
main outcomes. There are multiple detections of a stellar source before the main outburst, with the earliest detection being over 20 years before 
the discovery \citep[$B\approx22$ mag,][]{gor16}.
A subsequent detection in 2000, shows the object roughly at the same luminosity, and with the following colours: $B-V=0.0\pm0.3$ mag and $V-R=0.7\pm0.4$ mag \citep{bla16}.
During the following years, the source experienced only minor changes in brightness, although from about 5.5 years before discovery its luminosity began to grow, with the colours becoming
slightly redder \citep[$B-V=0.33\pm0.25$ mag and $V-R=0.28\pm0.24$ mag;][]{ger15}. At the time of discovery (on JD = 2\,456\,972.5), the object was much brighter, at $R\sim16.6$ mag.
This point was very likely in the proximity of the first light curve peak \citep{cao15}. Later, the light curve of  M101-2015OT1 showed a rapid decline to a minimum 1.6 mag fainter \citep{ger15}. 
About 100 days after the 
first maximum, the transient reached the second light curve peak (JD = 2\,457\,070.8) at $V=17.61\pm0.03$ mag. At this phase, the observed colours were $B-V\approx1.3$ mag,
$V-I\approx1.5$ mag, and $R-I\approx0.6$ mag. During the decline after the second maximum, the colours became even redder.
The new data presented in this paper are mostly pre-discovery observations of amateur astronomers. However, we also include unfiltered observations of the outburst 
(converted into $V$ or $R$ band photometry depending on the response curve of the CCDs used by the amateur astronomers). 
In particular, we report a detection of the first outburst on JD = 2\,456\,975.34 
at $R=15.78\pm0.13$ mag, suggesting that the Cao et al. detection was actually before the blue light curve peak. Later photometric points cover the red peak
evolution, and well match the light curve of \citet{bla16}.

Finally, the last object in our sample is SNhunt248, widely discussed in \citet{kan15} and \citet{mau15}. 
Here we provide additional data, with the most valuable ones obtained by the Palomar Transient Factory \citep[PTF,][]{law09,law14} survey, and made publicly available through 
the NASA/IPAC Infrared Science  Archive (IRSA).\footnote{{\it http://irsa.ipac.caltech.edu}}
For a detailed description of the light curve, we
address the reader to the above two papers, and to the parameters listed in Table \ref{lc_par}. Briefly, we note that SNhunt248 has a similar colour evolution as other objects of this sample
until the second post-outburst maximum (see Sect. \ref{photo_abs}), while the $R-I$ colour is bluer (up to $\sim1$ mag)  at very late phases (200-300 d after the first peak).  
The time span between the two light curve peaks ranges from 69 to 85 d, depending on the filter. 
In general, all LRNe reach the second peak later in the redder bands (see Table \ref{lc_par}).
We targeted SNhunt248 with the 1.82-m Copernico telescope of Mt. Ekar at very late phases (on March 2017, over 1000 d after the blue peak), but 
we only measured an upper limit ($R>22.3$ mag) at the position
of the transient. However, \citet{mau17} report multiple detections in HST images obtained in June 2015 at $F555W=25.46\pm0.03$ mag and 
$F555W-F814W$ (nearly $V-I$) $=0.9\pm0.1$ mag.

\subsection{Photometric comparison with similar transients, and pseudo-bolometric light curves}  \label{photo_abs} 
   
\begin{figure} 
\centering
\includegraphics[width=9.0cm,angle=0]{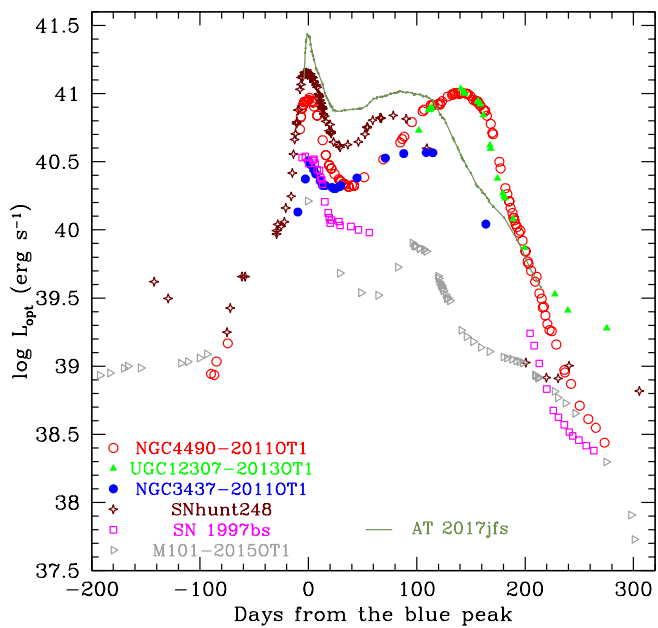}  
  \caption{Pseudo-bolometric (optical) light curves for six LRNe presented in this paper. The phase is in days from the first peak. AT~2017jfs data (solid line) are from \protect\citet{pasto19}. \label{bolom}} 
\end{figure} 

 The evolution of the $U-B, B-V, V-I$, and $R-I$ colours for the six objects of our sample, corrected for the reddening values as discussed in Sect. \ref{sample}, 
is shown in Fig. \ref{colorr}. In particular, for NGC4490-2011OT1, we now account also for the host galaxy reddening, adopting the total value $E(B-V)=0.32$ 
mag, as in \citet{smi16b}.
We note an overall similarity in the colour evolution of these objects at early phases, although with some subtle differences. 
M101-2015OT1 (Fig. \ref{colorr}) appears slightly redder than other LRNe, and this is possibly due to lower photospheric temperature
or an underestimate of 
the line-of-sight extinction. On the opposite side, at late epochs, UGC12307-2013OT1 and SNhunt248 have significantly bluer $V-I$ and $R-I$ 
colours than other LRNe.  
In this context, AT~2017jfs \citep{pasto19} is an outlier, as its $B-V$ rises rapidly to a much redder colour than other LRNe ($\sim$2.7 mag at 4 months past blue maximum).

Some diversity can also be noticed in the intrinsic luminosities of our LRN sample, which approximately
span one order of magnitude. In general, all events have peak absolute magnitudes $M_V<-12$ mag, although none of them 
reaches  $M_V=-15$ mag. We remark that RNe fainter than $\sim-12$ do exist, and have been observed in the Milky Way and in M~31, but their low
intrinsic luminosity makes their detection problematic outside the Local Group. As we will see in Sect. \ref{nature}, there is very likely 
a continuity in the RN and LRN luminosity distribution.

In order to best compare the light curves in our LRN sample, we compute the pseudo-bolometric light curves by integrating the fluxes 
in the well-sampled $BVRI$ bands (or, when observed, Sloan-$gri$). 
When a photometric point is not available at a given epoch for one of the filters, its contribution is estimated through an interpolation 
of the available data or extrapolating the last available epoch accounting for the colour information. 
The fluxes at the filter effective wavelengths, corrected for extinction, provide the SED at each epoch, which is 
integrated using the trapezoidal rule. No flux contribution is assumed outside the extremes of the integration regions.
We note, however, that the UV flux is likely significant at the epoch of the first peak, and the IR band contribution may increase with time,
becoming predominant at late phases. 
The observed flux is finally converted to luminosity using the values for the distances 
of the transients discussed in Sect. \ref{sample}. The resulting pseudo-bolometric light curves are shown in Fig. \ref{bolom}.
For comparison, the pseudo-bolometric light curve of LRN AT~2017jfs \citep{pasto19} is also reported.

Although there are some differences in the photometric evolution of our transient sample, a double-peaked light curve is clearly observable in most cases. 
When a double peak is not observed, this is possibly due to an incomplete coverage, or the red peak is shallow resembling a sort of
plateau (like in SN~1997bs). The pseudo-bolometric luminosity of the two peaks 
in NGC4490-2011OT1 is similar, with $L_{opt}\sim10^{41}$ erg s$^{-1}$. The blue peak of SNhunt248 is slightly more luminous than that of NGC4490-2011OT1,
peaking at $L_{opt}\sim1.4\times10^{41}$ erg s$^{-1}$, but it is fainter than AT~2017jfs \citep{pasto19} that peaks at $L_{opt}\sim3\times10^{41}$ erg s$^{-1}$.
However, the former reaches a red light curve maximum one-month earlier than NGC4490-2011OT1, at a luminosity of $7\times10^{40}$ erg s$^{-1}$. 
The pseudo-bolometric light curve of NGC3437-2011OT1 has a similar shape as NGC4490-2011OT1, but it is slightly fainter, 
with both peaks having $L_{opt}\sim3$-$4\times10^{40}$ erg s$^{-1}$. 
The other three objects, UGC12307-2013OT1, SN~1997bs and M101-2015OT1, have an incomplete photometric coverage. However, a flattening in their light curves 
and/or spectroscopic considerations (e.g., the identification of molecular  bands in the late spectra of UGC12307-2013OT1, see Sect. \ref{mole}) support 
their proximity to LRNe. For UGC12307-2013OT1, light curve monitoring covers only the red peak, 
 which reaches a luminosity of about $10^{41}$ erg s$^{-1}$. SN~1997bs has a blue peak with $L_{opt}\sim3.3\times10^{40}$ erg s$^{-1}$. 
Then, instead of a clear red peak, its pseudo-bolometric light curve flattens to about $10^{40}$ erg s$^{-1}$ (but the photometric coverage 
is incomplete in this phase). In contrast, M101-2015OT1 is much fainter than
all  other transients: the single detection during its blue peak allows us to constrain its luminosity to  
$\geq1.4\times10^{40}$ erg s$^{-1}$, while the red peak reaches $L_{opt}=6.6\times10^{39}$ erg s$^{-1}$.

\subsection{Searching for previous outbursts in archival data} \label{historic} 
     
\begin{figure*}[!t]
\centering
\includegraphics[width=14.0cm,angle=270]{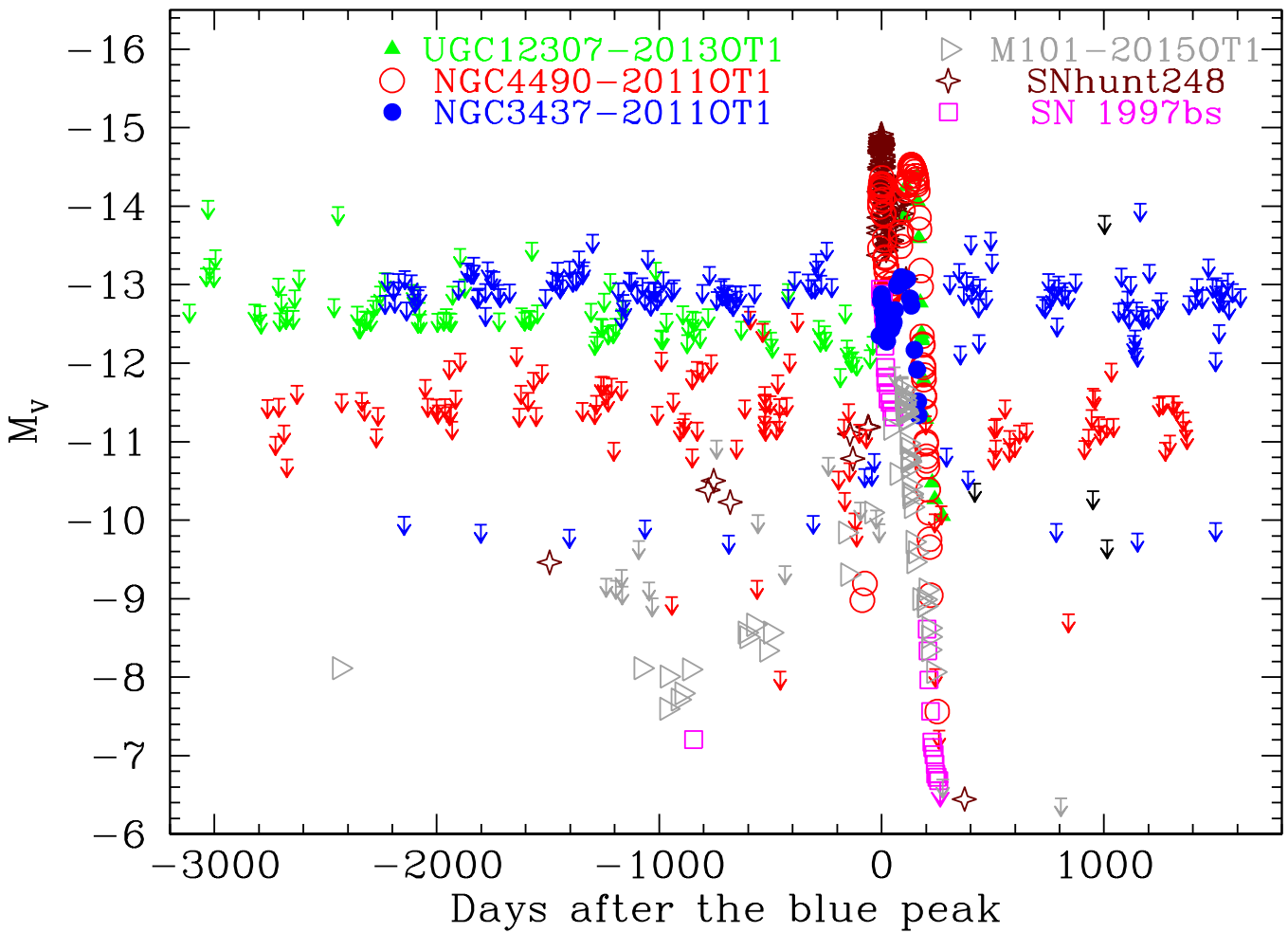}  
  \caption{Long-term photometric monitoring (specifically, $V$-band absolute magnitudes) for extra-Galactic LRNe discussed in this paper,
covering about a decade of their evolution. 
Pre- and post-discovery detection limits are indicated with ``down-arrow'' symbols.  
The phase is in days from the first (blue) peak in the $V$ band. The absolute light curves of LRN outbursts are compared in detail in Sect. \ref{nature}.\label{lc_abs2}} 
\end{figure*} 

The LRN sites were frequently monitored in past years by a number of amateur astronomers and professional sky surveys. 
In particular, we collected a large amount of images from the Catalina Sky Survey (CSS), and from PTF (obtained via NASA/IPAC IRSA). Other data were 
collected from the Isaac Newton Telescope Archive.\footnote{{\it http://casu.ast.cam.ac.uk/casuadc/ingarch/query}, maintained by the  Cambridge 
Astronomical Survey Unit (CASU), which  is part of the Institute of Astronomy, Cambridge University.} Finally, data were provided by a 
number of amateur astronomers, and sparse pre-outburst images were retrieved through other public archives. 

We inspected these images to detect signatures of previous outbursts. 
Historical detections and upper limits ($V$-band absolute magnitudes) at the expected positions of the transients
are reported in Fig. \ref{lc_abs2}. NGC4490-2011OT1 was detected about 1.5 months before the blue peak at a magnitude significantly brighter than that 
of the quiescent progenitor recovered in HST images and discussed by \citet{fra11} and \citet{smi16b}. Faint detections are measured in archival images 
(obtained from about 20 years to $\sim$1 year before the outbursts) in two cases (see Fig. \ref{lc_abs2}): M101-2015OT1 \citep[with $M_V$ in the range from $-7.5$ to $-9.5$ mag, see][]{bla16,gor16} and 
SNhunt248 \citep[with $M_V\approx-10.5$ mag, and $M_R$ in the range from $-9.5$ to $-11$ mag,][]{kan15}. 

The remaining three LRNe were not detected prior to their discoveries.
We note, however, that only NGC4490-2011OT1 has sparse detection limits down to about $-8$ or $-9$ mag, while the detection limits for both UGC12307-2013OT1 and 
NGC3437-2011OT1 are typically less stringent ($M_V>-12.5$ mag and  $-$13 mag, respectively). In order to increase the depth of the detection limits, 
for NGC3437-2011OT1 we have created some deep seasonal stacks, obtaining more stringent limits down to $M_V\approx-9.8$ mag. 
Therefore, with the available archival data, it is unlikely that the objects discussed in this paper experienced major outbursts in the past few years before their discoveries. 

\begin{table*} 
\centering
  \caption{Log of spectroscopic observations of NGC4490-2011OT1, NGC3437-2011OT1 and UGC12307-2013OT1. \label{tab_spec}} 
  \begin{tabular}{cccccc} 
  \hline  \hline 
   Date &  JD          &  Phase &  Instrumental configuration & Range  & Resolution  \\ 
  (dd/mm/yy) & (+2\,400\,000)   & (days)$^\ddag$  &  & (nm) &  (nm) \\ 
 \hline 
   \multicolumn{6}{c}{NGC4490-2011OT1} \\ \hline 
18/08/11 & 55\,792.39 & -5.8 & TNG + LRS + LRR & 510--1010 & 1.4 \\ 
19/08/11 & 55\,793.38 & -4.8 & TNG + LRS + LRB & 330--810 & 1.6 \\ 
20/08/11 & 55\,794.41 & -3.8 & NOT + ALFOSC + gm4 & 340--910 & 1.8 \\ 
28/08/11 & 55\,802.38 & +4.2 & TNG + LRS + LRB & 330--810 & 1.1 \\ 
02/09/11 & 55\,807.37 & +9.2 & WHT + ISIS + R300B/R158R & 320--1030 & 0.5/0.6 \\ 
21/11/11 & 55\,886.72 & +88.5 & Ekar1.82 + AFOSC + gm4 & 350--820 & 1.2 \\ 
28/11/11 & 55\,893.62 & +95.4 & Ekar1.82 + AFOSC + gm4 & 350--820 & 1.4 \\ 
08/12/11 & 55\,903.72 & +105.5 & NOT + ALFOSC + gm7/gm8 & 380--835 & 0.4/0.4 \\ 
21/12/11 & 55\,916.77 & +118.6 & WHT + ISIS + R300B/R158R & 300--1000 & 0.5/0.6 \\ 
22/12/11 & 55\,917.75 & +119.6 & NOT + ALFOSC + gm8 & 580--840 & 0.5 \\ 
21/01/12 & 55\,947.80 & +149.6 & OHP1.93 + CARELEC + 300T/mm & 400--730 & 0.5 \\ 
23/01/12 & 55\,950.42 & +152.2 & Ekar1.82 + AFOSC + gm4 & 370--820 & 1.2 \\ 
30/01/12 & 55\,956.54 & +158.3 & Ekar1.82 + AFOSC + gm4 & 360--820 & 1.2 \\ 
18/02/12 & 55\,975.68 & +177.5 & Ekar1.82 + AFOSC + gm4 & 415--810 & 2.6 \\ 
01/03/12 & 55\,987.64 & +189.4 & NOT + ALFOSC + gm4 & 340--900 & 1.4 \\ 
29/03/12 & 56\,015.56 & +217.4 & Ekar1.82 + AFOSC + gm4 & 355--820 & 1.2 \\ 
30/03/12 & 56\,016.60 & +218.4 & CAHA2.2 + CAFOS + g200 & 385--1020 & 1.3 \\ 
23/04/12 & 56\,040.61 & +242.4 & CAHA2.2 + CAFOS + g200 & 410--1020 & 1.3 \\ 
15/05/12 & 56\,062.68 & +264.5 & LBT + MODS + G670L + GG495 & 490--1000 & 0.6 \\ \hline 
   \multicolumn{6}{c}{NGC3437-2011OT1} \\  \hline 
17/01/11 & 55\,578.60 & +3.9 & CAHA2.2 + CAFOS + g200 & 375--960 & 1.3 \\ 
24/01/11 & 55\,585.85 & +11.2 & NTT + EFOSC2 + gr.13 & 365--930 & 2.8 \\ 
12/02/11 & 55\,604.77 & +30.1 & NTT + EFOSC2 + gr.13 & 580--925 &  2.8 \\ 
04/05/11 & 55\,685.54 & +110.8 & WHT + ISIS + R300B/R158R & 310--1030 & 0.4/0.6 \\ \hline 
   \multicolumn{6}{c}{UGC12307-2013OT1} \\ \hline 
01/08/13 & 56\,505.60 & +24.0$^\ast$ & Ekar1.82 + AFOSC + gm4 & 365--820 & 1.4 \\  
07/08/13 & 56\,511.56 & +30.0$^\ast$ & Ekar1.82 + AFOSC + gm4 & 365--820 & 1.4 \\ 
26/08/13 & 56\,530.74 & +49.1$^\ast$ & SOAR + Goodman + RALC 300 + GG385 & 365--895 & 0.9 \\ 
24/10/13 & 56\,589.53 & +107.9$^\ast$ & GTC + OSIRIS + R500R &  490--930 & 1.5 \\ 
\hline                                                                                         
\end{tabular}   
 
\tablefoot{$^\ddag$  Days from the blue maximum (in the $V$ band);                                                                     
$^\ast$  Days from the red maximum (in the $V$ band).}
\end{table*} 

\section[]{Spectroscopic data}  \label{spectra_redu} 
 
Our spectral sequences of NGC4490-2011OT1, NGC3437-2011OT1, and UGC12307-2013OT1 were obtained using a number of telescopes available to our collaboration, including 
the 3.58m Telescopio Nazionale Galileo (TNG) equipped with LRS, the 2.56m Nordic Optical Telescope (NOT) with ALFOSC, the 4.2m William Herschel  
Telescope (WHT) with ISIS, and the 10.4m Gran Telescopio Canarias (GTC) + OSIRIS. All of them are hosted at the  Roque de los Muchachos Observatory  
(La Palma, Canary Islands, Spain).  Additional spectra were taken with the 1.82m Copernico Telescope plus AFOSC of the INAF - Osservatorio Astronomico di Asiago 
(Ekar1.82; Mt. Ekar, Asiago, Italy),  
the 2.2m reflector telescope of the Calar Alto Observatory with CAFOS (CAHA2.2; Almeria, Spain), the 3.58m New Technology Telescope (NTT) + EFOSC2  
(ESO - La Silla Observatory, Chile), the 1.93m telescope of the Observatoire de Haute-Provence plus CARELEC (OHP1.93; Alpes-de-Haute-Provence d\'epartement, France),  
the 2 $\times$ 8.4m Large Binocular Telescope (LBT) with MODS of the Mount Graham International Observatory (Arizona, US), and the 4.1m Southern Astrophysical Research (SOAR)  
telescope plus the Goodman Spectrograph of the Cerro Tololo Inter-American Observatory (Cerro Pach\'on, Chile). 

The spectra were reduced following standard procedures in IRAF. We first corrected the 2-D spectra for bias, overscan and flat-field. 
Then, the 1-D spectra of the targets were extracted and calibrated in wavelength using arc lamp spectra 
obtained during the same night and with the same instrumental configuration as our targets. The flux calibration was performed using spectra of  
spectro-photometric standard stars. To check the accuracy of the flux-calibration of the LRN spectra, 
we compared the spectro-photometric data derived from our spectra with the broadband photometry obtained during the same night and, in case of discrepancy, 
a correction factor was applied to the spectra. Occasionally, when a wavelength-dependent flux loss was found, a linear flux 
correction was applied to the spectrum. The spectra of standard stars were also used to 
remove the strongest telluric absorption bands (O$_2$ and H$_2$O) from the LRN spectra. 
We remark that an incomplete removal of these telluric bands may occasionally affect the profile of individual spectral features. 

\begin{figure*}[!t] 
\centering
{\includegraphics[width=13.4cm,angle=270]{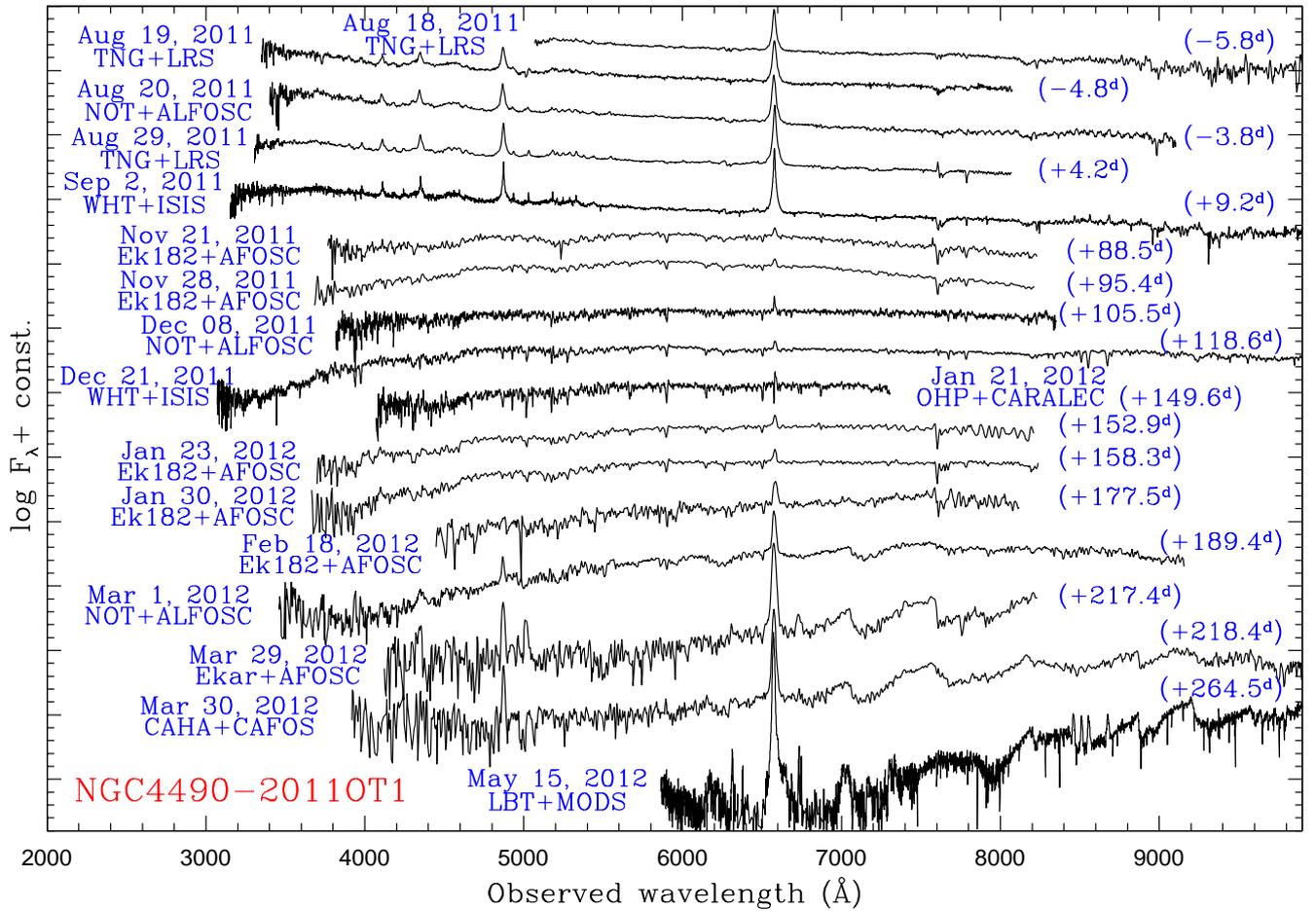}  
\includegraphics[width=8.8cm,angle=0]{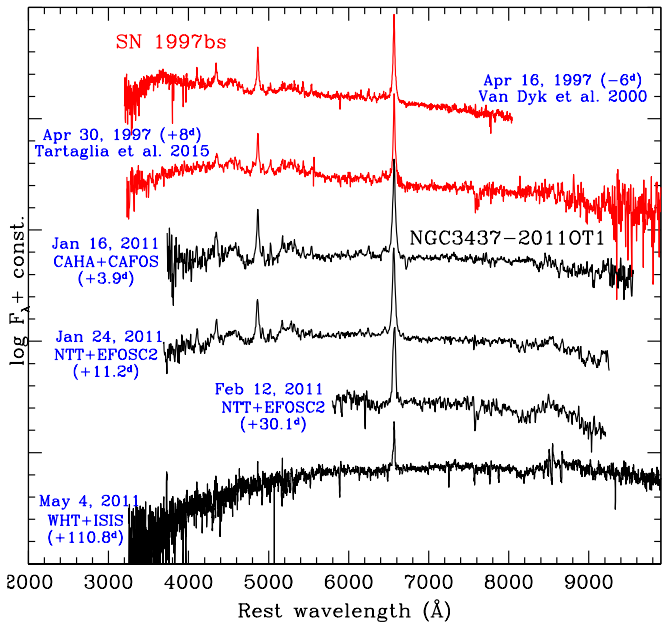}  
\includegraphics[width=8.8cm,angle=0]{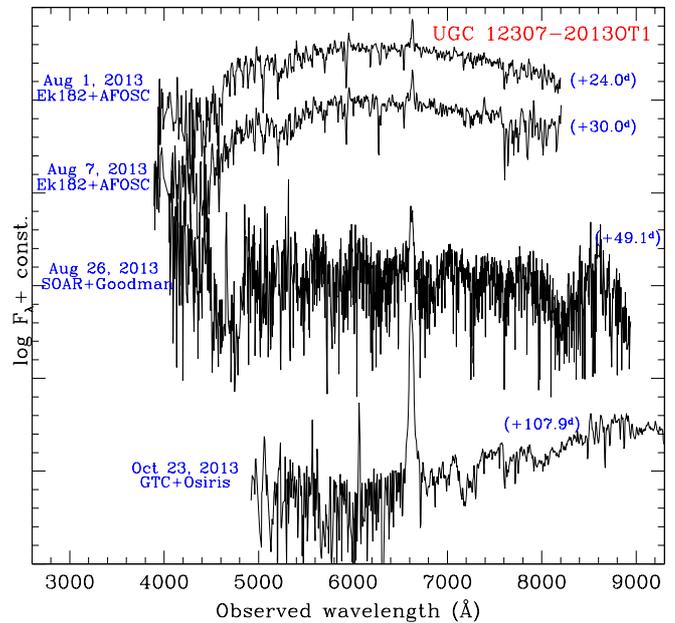} } 
  \caption{Top: Spectral evolution of NGC4490-2011OT1. Bottom-left: spectra of NGC3437-2011OT1 shown with those of SN~2007bs 
available in the literature \citep{van00,tar15}. Bottom-right: spectral evolution of UGC12307-2013OT1 during and after the red light-curve peak. A low S/N
SOAR spectrum of UGC 12307-2013OT1 obtained on 2013 August 26 is shown for completeness. \label{spec}} 
\end{figure*} 

\begin{figure*}[!t]
\centering
\includegraphics[width=9.1cm,angle=0]{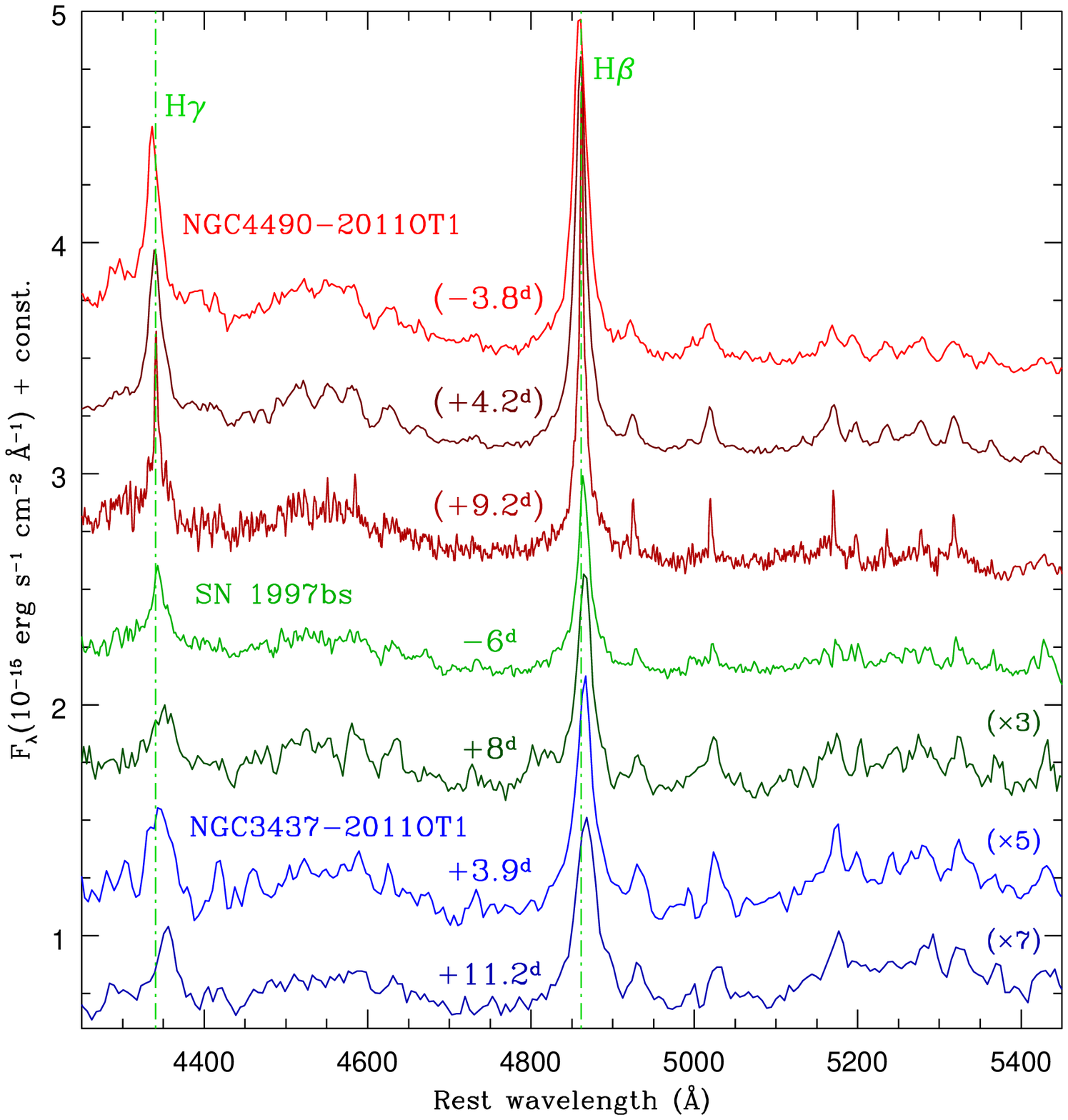} 
\includegraphics[width=9.1cm,angle=0]{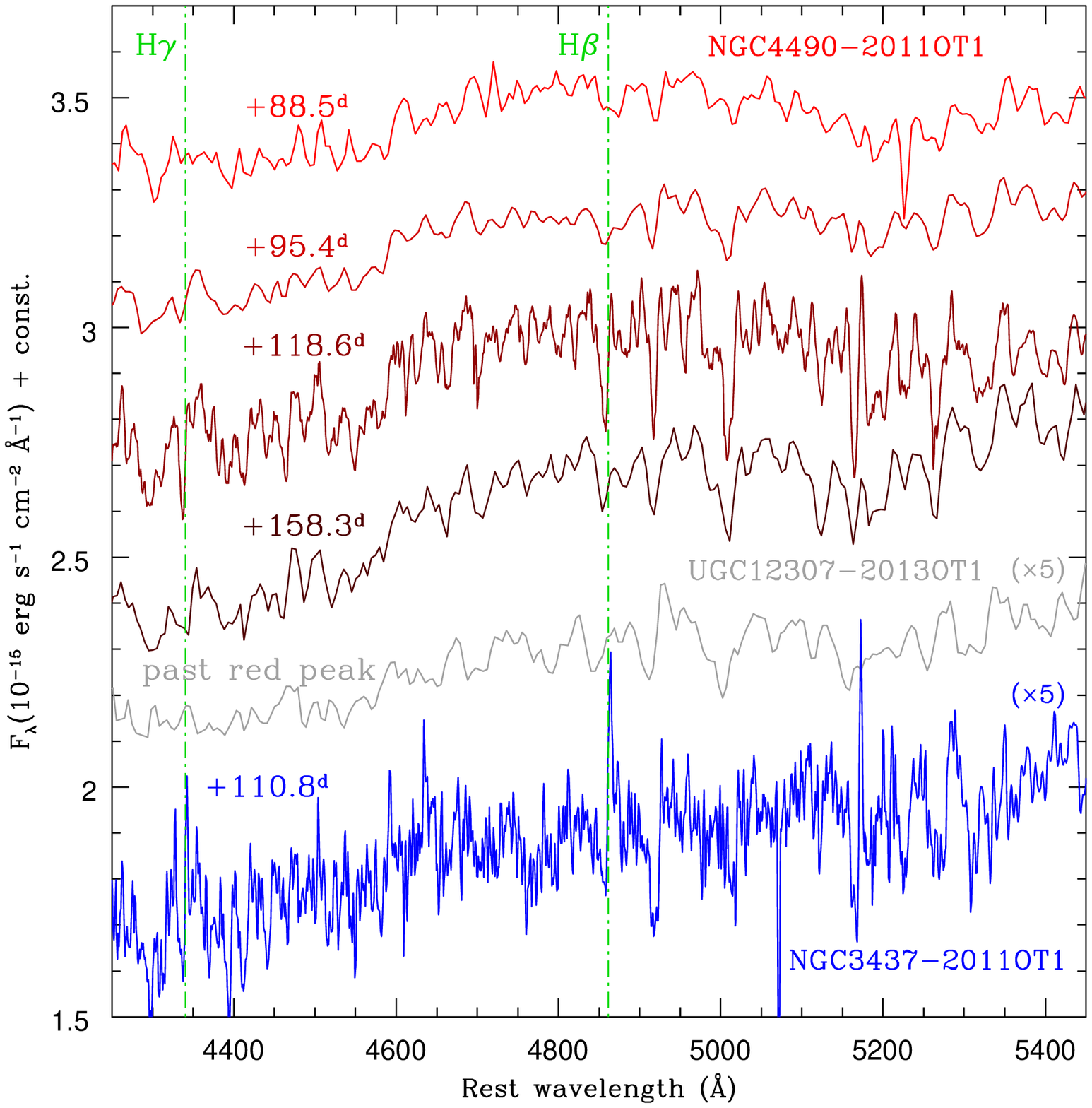}
  \caption{Detail of region 4250-5450~\AA~for selected LRN spectra presented in this paper.
Left: early-time LRN spectra, obtained during the first luminosity peak. Right: LRN spectra during the second, red peak. 
The vertical dot-dashed lines mark the rest wavelengths of H$\beta$ and H$\gamma$. \label{spec_Hbeta}} 
\end{figure*}

Basic information on the spectra is given in Table \ref{tab_spec}, while the final spectra are shown in Fig. \ref{spec}. 
A blow-up of the region between 4250 and 5450~\AA~is shown for a sub-sample of LRN spectra in Fig. \ref{spec_Hbeta}.
While H$\gamma$, H$\beta$ and numerous Fe~II multiplets are  visible in emission during the blue light-curve peak (left panel)
the same features are mostly in absorption during the broad red peak (right panel). The spectral evolution of our LRN sample will be
illustrated in detail in Sect. \ref{spectra_evol}.

\begin{figure*}
\centering
\includegraphics[width=17.0cm,angle=0]{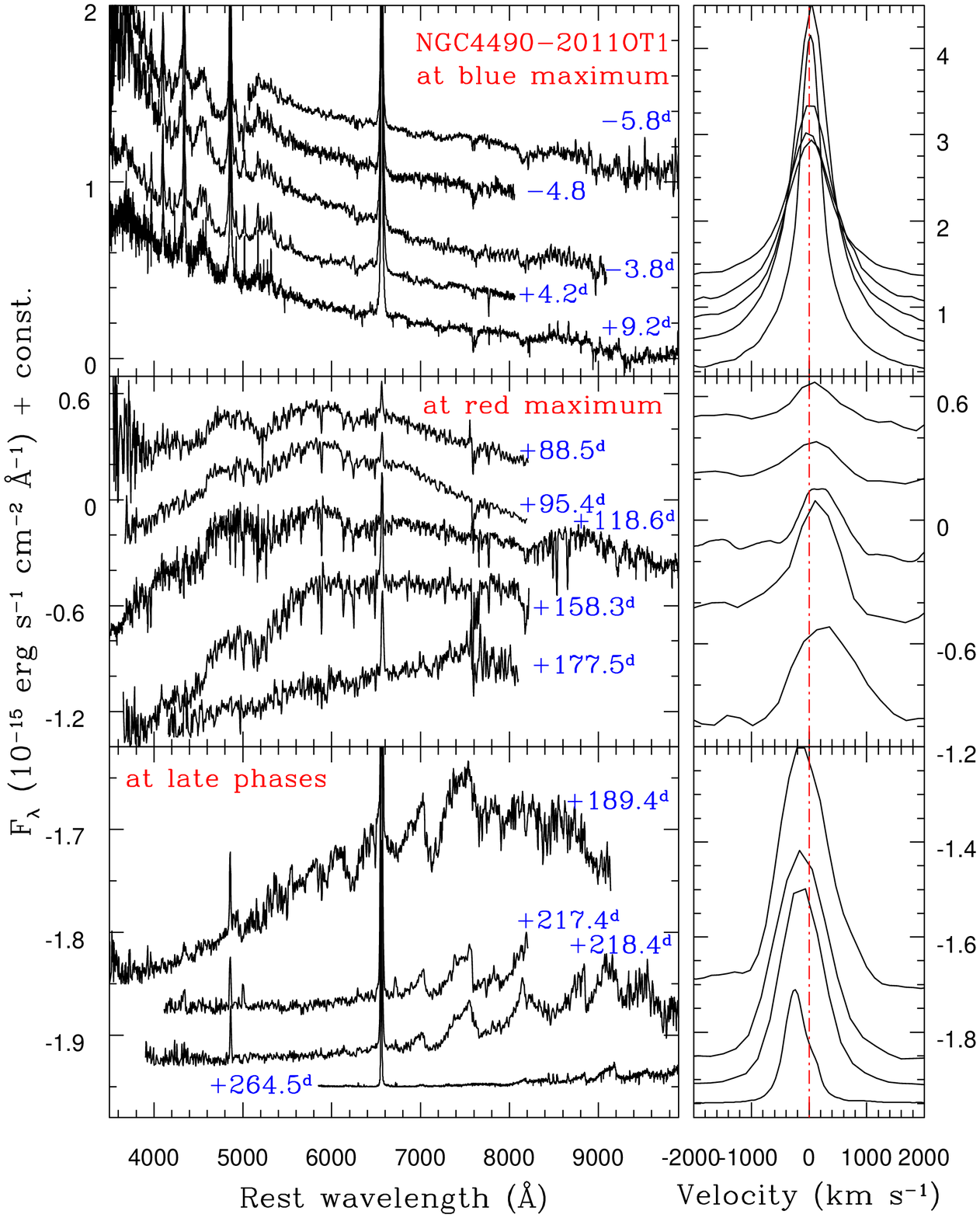}  
  \caption{Left panels: spectral evolution of NGC4490-2011OT1, with only higher S/N spectra shown. 
Right panels: evolution of H$\alpha$ in velocity space. The spectra obtained at the blue peak are shown in the top panels, those
obtained at the red peak are in the middle panels, and the late spectra are in the bottom panels. The spectra were reported to the rest frame applying
a Doppler correction (the redshift $z = 0.002265$ is determined from a nearby H~II region, and was corrected accounting for the heliocentric velocity); 
no reddening correction was applied to the spectra. \label{N4490_spevol2}} 
\end{figure*}

\subsection{Spectroscopic evolution of individual objects} \label{spectra_evol}

The spectroscopic follow-up of NGC4490-2011OT1 covers all crucial phases of its evolution (Fig. \ref{spec}, top panel). 
Early spectra (from 2011 August 18 to September 2, obtained at the time of the blue light curve peak; Fig. \ref{N4490_spevol2}, top-left panel)  
resemble those of other SN impostors \citep[e.g.,][]{van00,pasto10} and Type~IIn SNe. These spectra are characterized by a blue continuum with superposed prominent 
narrow lines of the Balmer series. The H features are in emission and have a Lorentzian profile, with a full-width at half-maximum (FWHM) velocity v$_{FWHM}\sim400$-$500$ km s$^{-1}$ and wings extending up to about 3200 km s$^{-1}$.
The flux measured for H$\alpha$ is of the order of $10^{-13}$ erg s$^{-1}$ cm$^{-2}$.  
In addition, a number of lines of Fe~II, O~I, and Ca~II with P~Cygni profiles are detected, and become more prominent with time. In this period, the continuum temperature, estimated through a blackbody fit,  
drops from about 18000 to 14000~K.

\begin{figure*}
\centering
\includegraphics[width=13.6cm,angle=270]{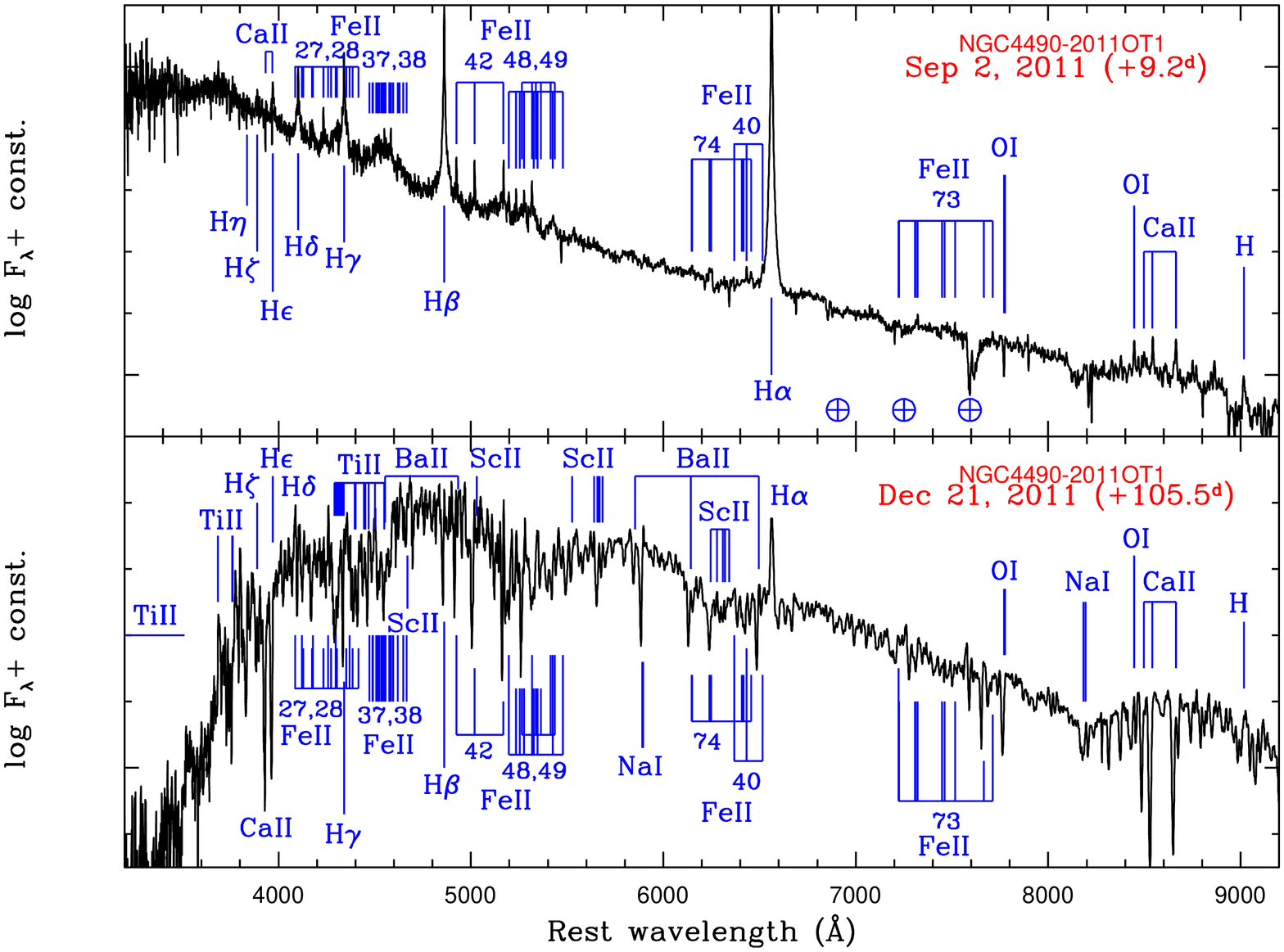}  
  \caption{Line identification in two spectra of NGC4490-2011OT1, at the first (top) and  the second peaks (bottom). \label{lineid}} 
\end{figure*}

\begin{figure*} 
\centering
\includegraphics[width=17.0cm,angle=0]{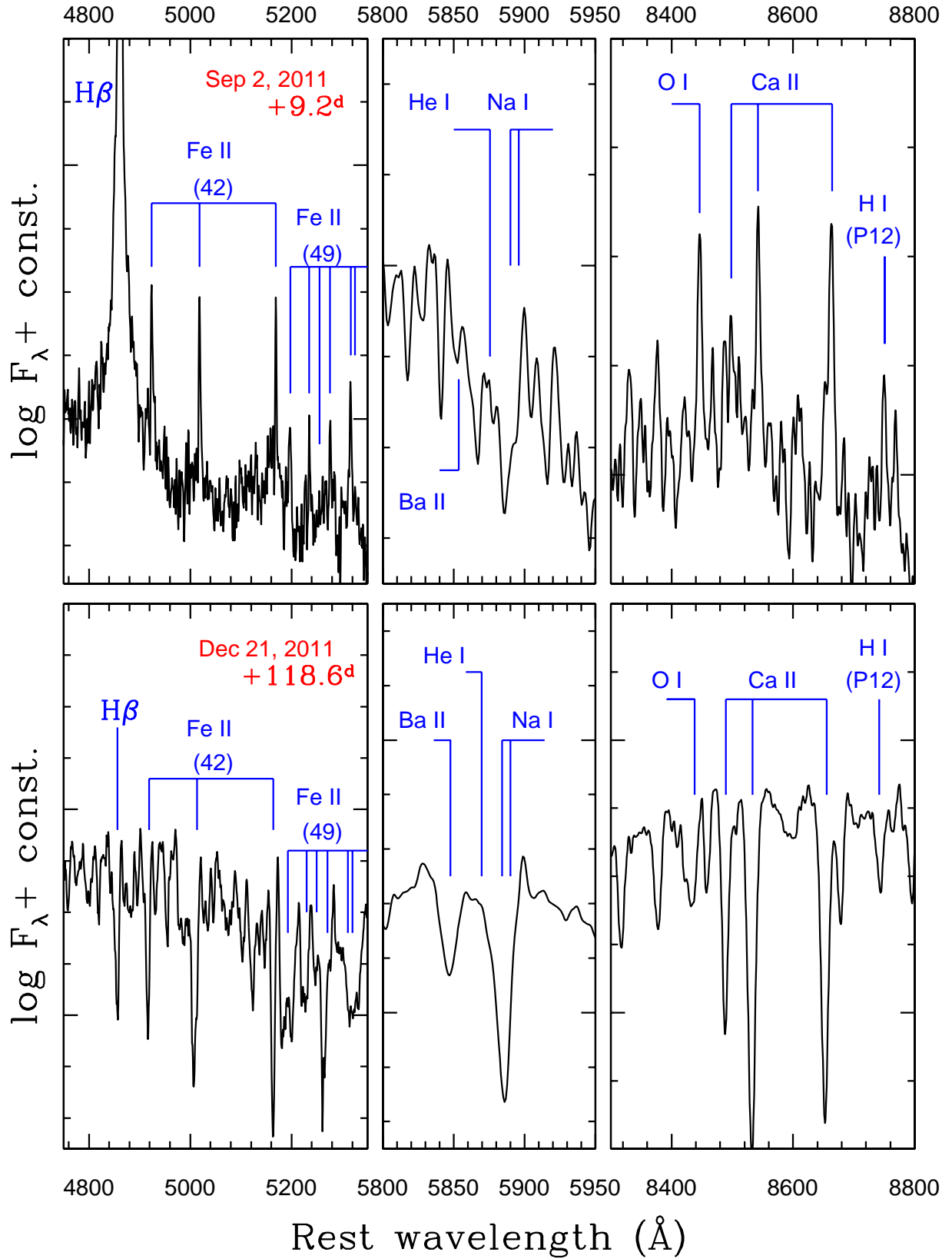}  
  \caption{Blow-up of selected spectral windows  in two spectra of NGC4490-2011OT1, at first peak (top panels) and during second maximum (bottom panels). The regions at 4750-5350 \AA~(left panels), 5800-5950 \AA~(middle panels) and 8300-8800 \AA~(right panels) are shown. The positions marked in the bottom panels are those of the P Cygni minima, obtained assuming 
an average expansion velocity of 300 km s$^{-1}$ (see text). The position of He~I $\lambda$5876 is also marked, although that feature is not clearly detected.
\label{linezoom}} 
\end{figure*}

Later spectra (from 2011 November 21 to 2012 January 30; roughly covering the early evolution around the second peak, Fig. \ref{N4490_spevol2}, middle-left panel) become much redder, and the line profiles drastically change:  
at these epochs, a forest of metal lines in absorption is observed. In particular, many lines of Fe~II are observed, with  the multiplet 42 being very strong.
Ti~II lines are likely responsible for the flux deficit at 
blue wavelengths.\footnote{Although not securely identified, transitions due to V~II may also affect the spectral continuum below about 3600~\AA. } 
Along with these metal lines, we now identify also Sc~II, Ba~II, and Na~I. The O~I ($\lambda$$\lambda$7772-7775 and $\lambda$8446), and Ca~II features (both H$\&$K and the NIR triplet) are now seen in absorption. 
He lines are not detected at any phase. With time,
the Balmer lines become much weaker,  and are dominated by the absorption component. Only H$\alpha$ still has a moderately prominent emission component, 
whose flux decreases by  one order of magnitude with reference to the early epochs. 
The wind velocity, as inferred from the minimum of isolated Fe~II lines, is about 300-350 km s$^{-1}$, while from H$\alpha$ (that has now a Gaussian profile) we estimate  
v$_{FWHM}$ $\sim$ 200-300 km s$^{-1}$ from our highest resolution spectra.\footnote{We note, however, that the narrow features are marginally resolved in our best resolution spectra, so these estimates
are only indicative.} The continuum temperature declines from about 7000~K to 4500~K. 
Comprehensive line identification in two spectra of NGC4490-2011OT1 obtained in the proximity of the first and the second light curve peaks is shown in Fig. \ref{lineid} (top and bottom panels, respectively), while a detailed view of the two spectra in selected spectral regions is shown in Fig. \ref{linezoom}.
We note that \citet{smi16b} extensively  monitored the object in spectroscopy during the second light curve peak. From their highest resolution spectrum (obtained on January 12), they 
infer a velocity of 280 km s$^{-1}$ from the middle of the blueshifted H$\alpha$ absorption component (consistent with our estimates), with a maximum velocity of 650 km s$^{-1}$
from the blue edge of the absorption feature.

\begin{figure} 
\centering
\includegraphics[width=8.4cm,angle=0]{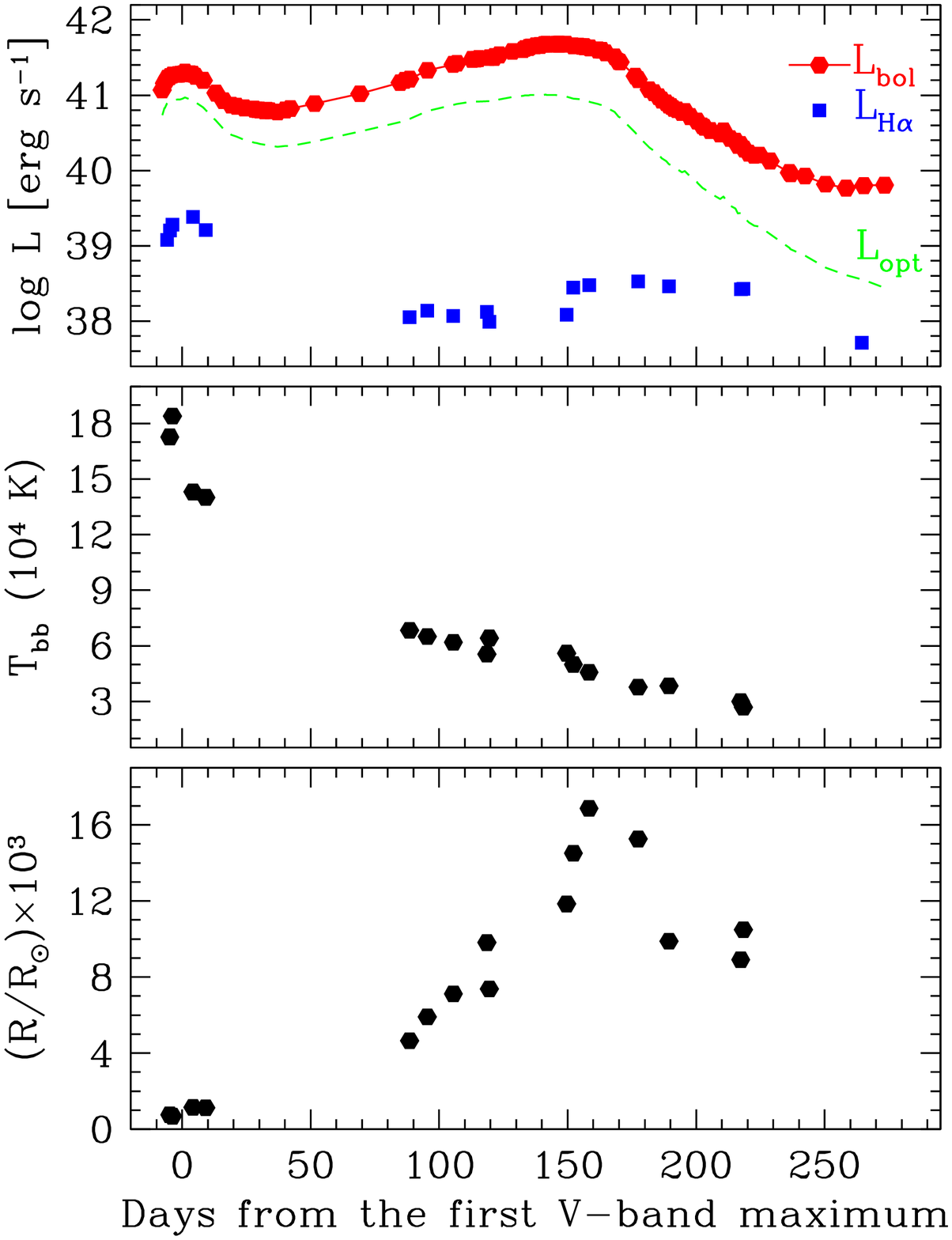}  
  \caption{Evolution of the bolometric, optical ($BVRI$) and H$\alpha$ luminosities (top panel), continuum temperature (middle panel) and photospheric radius (bottom panel) in our spectra of NGC4490-2011OT1.  
A conservative error of 500 K is estimated for the blackbody continuum temperature. \label{radius}} 
\end{figure}

At later phases, after 2012 February 18 and during the light curve decline following the second peak, the narrow Balmer lines in emission become quite prominent, with  
$v_{FWHM}$ increasing to about 700-900 km s$^{-1}$. In our latest spectrum (2012 May 15), the Ca~II NIR triplet along with O~I $\lambda$8446 are  also clearly detected in emission.
At the same time, the very red continuum ($T\approx$ 3000-4000 K) is characterized by broad molecular absorption bands (Fig. \ref{N4490_spevol2}, bottom-left panel), as will be discussed in Sect. \ref{mole}. The overall spectral evolution of NGC4490-2009OT1 closely resemple that of LRN AT017jfs \citep{pasto19}.
 
The complete evolution of the H$\alpha$ line profile (in velocity space) is shown in the right panels of Fig. \ref{N4490_spevol2}. In particular, the middle-left panel emphasizes the weakness of H$\alpha$ during the second peak. We also note that the H$\alpha$ emission peak is slightly blueshifted in the late spectra of NGC4490-2011OT1 (bottom-right panel of Fig. \ref{N4490_spevol2}). This is usually interpreted as a signature of dust condensation.

Using the information available for the bolometric light curve\footnote{For this purpose, the bolometric light curve of NGC4490-2011OT1 was computed from the quasi-bolometric light curve presented in Sect. \ref{photo_abs} (calculated from the $B$ to the $I$ bands), applying a correction 
that accounts for the $U$-band and IR contributions. The correction was estimated using the bolometric data of M101-2015OT1.} and the continuum temperature, 
we can compute the radius evolution of the emitting photosphere, which is 700-1000 R$_\odot$ at the time of the first peak. It rises to $\sim$5000-6000 R$_\odot$ at around 100 d after the first maximum, and peaks to almost 
17000 $R_\odot$  ($R\approx1.17\times10^{15}$ cm) soon after the second (red) light curve maximum. Later on (70 days after), the radius steadily declines by a factor $\sim$2. 
The overall evolution of the bolometric and H$\alpha$ line luminosities, the continuum temperature and the radius of the emitting region in NGC4490-2011OT1 are shown in Fig. \ref{radius}. 
We note that the temperatures inferred for NGC4490-2011OT1 at the time of the red maximum are comparable to those estimated by \citet{bla16} for M101-2015OT1. 
The radius of the emitting photosphere at the red maximum is a factor 3 higher in NGC4490-2011OT1 than in M101-2015OT1. 
All of this favours a more massive progenitor for NGC4490-2011OT1. Another interesting comparison is with SNhunt248 \citep{kan15}. In this case, while the radius of SNhunt248 at the time of 
the blue peak is one order of magnitude larger than NGC4490-2011OT1, at the time of the red peak the radii are very similar.

The spectral evolution of NGC3437-2011OT1 (Fig. \ref{spec}, bottom-left panel) is similar to that of NGC4490-2011OT1. The three early spectra (from January to February 2011) are 
representative of the blue peak phase, with  H$\alpha$ having a Lorentzian profile. The spectrum of NGC3437-2011OT1 taken on 2011 May 4 is very similar to those of NGC4490-2011OT1 at the second maximum, with a red continuum, 
a low-contrast H$\alpha$ in emission with a Gaussian profile, and a number of metal lines visible in absorption. The spectra of NGC3437-2011OT1 in Fig. \ref{spec} are shown along with those of SN~1997bs available in the literature 
\citep{van00,tar15} obtained at the time of blue light curve peak. In both objects, an indicative $v_{FWHM}\sim700$ km s$^{-1}$ is estimated for H$\alpha$. We also note that none of these objects show the [Ca~II] doublet emission which usually characterizes other types of gap transients, including the intermediate-luminosity red transients \citep[ILRTs,][]{bot09,kas11} and S~Dor-type LBV outbursts \cite[e.g.,][and references therein]{wal17}.
 
The first two spectra of UGC12307-2013OT1 (taken on 2013 August 1 and 8) cover the evolution of the transient during the second (red) light curve maximum\footnote{Very little information can be inferred from 
the third spectrum (2013 August 26) because of its poor S/N, and it is shown only for completeness.} (see Fig. \ref{spec}, bottom-right panel). For the narrow H$\alpha$ emission,
we measure  $v_{FWHM}\sim$ 800-1000 km s$^{-1}$.
The last spectrum of UGC12307-2013OT1,  obtained on 2013 October 24  is reminiscent of the very late spectra of NGC4490-2011OT1 shown in the top panel, with a very red continuum, a strong H$\alpha$ in emission, 
and broad molecular bands in absorption (see Sect. \ref{mole}).

\subsection{Detection of molecules and dust formation} \label{mole} 
 
\begin{figure} 
\centering
\includegraphics[width=8.4cm,angle=0]{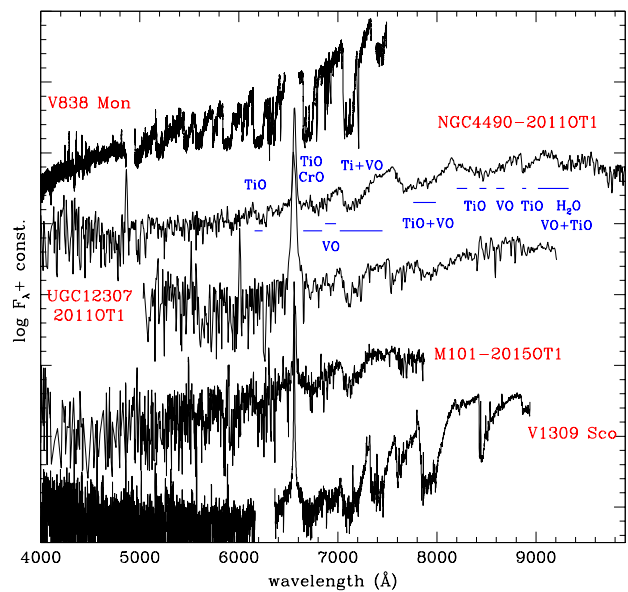}  
  \caption{Detection of molecules in late spectra of NGC4490-2011OT1, UGC12307-2013OT1 and M101-2015OT1 \protect\citep{bla16}, and comparison with late spectra of RNe V838~Mon \protect\citep{kam09}  
 and V1309~Sco \protect\citep{mas10}. H$\alpha$ and H$\beta$ are excised in the spectrum of V838~Mon to emphasize the 
broad molecular absorption bands.
\label{sp_molecules}} 
\end{figure}

Late-time spectra of NGC4490-2011OT1, UGC12307-2013OT1, and M101-2015OT1 \citep[][]{bla16} are extremely red, and show broad absorption bands, along with the usual H$\alpha$ emission (Fig. \ref{sp_molecules}). These absorptions are identified with molecular 
bands \citep[see, e.g.,][]{pasto19}, and are frequently observed in RNe and in late-type stars. Late spectra of the two objects mentioned above are compared in  Fig. \ref{sp_molecules}
with late spectra of the RNe V838 Mon and V1309 Sco. 
The prominent H$\alpha$ and H$\beta$ features have been cut in the spectrum of  V838 Mon to emphasize the absorption bands. Following \citet{kam09}, we identify
a number of features due to VO and TiO. However, also H$_2$O and CrO are very likely present. 

We note that RNe usually experience a metamorphosis from a bluer spectral type at 
the epoch of the light curve peak to an M-type at late phases, and all their spectra transition from being continuum-dominated to a configuration characterized by broad
and strong molecular absorption features \citep[see, e.g., V4332 Sgr,][]{mar99}. \citet{bar14} suggested that the spectrum
of V4332 Sgr observed over a decade after the outburst, showing simultaneously a late M-type continuum, molecular bands and emission features of metals,
was due the contribution of a third, red companion that was dynamically destroying the expanding envelope produced in the merging event.

The detection of molecular bands in the optical spectra is an argument frequently used to support the dust formation at late phases. 
In fact,  a strong IR flux excess was measured by \citet{smi16b} in Spitzer images of NGC4490-2011OT1 obtained about three years after the outburst.
From their analysis, Smith et al. determined the presence of cool (725 K) circumstellar dust, at $\sim80$ AU and an emitting dust mass of at least $10^{-9}$ M$_\odot$. 
This dust shell may have formed in the ejected material after the eruptive event, or was possibly already present in the circumstellar environment. 
The strong IR excess can be more likely explained as dust heated by a survived stellar source (e.g., the outcome of a stellar merger), otherwise it may be the signature of an IR echo \citep{smi16b}. 
Another object of this family showing a clear NIR excess about one year after the outburst is SNhunt248. \citet{mau17} estimated the presence of dust 
(650 K to 1450 K) at the location of SNhunt248. Using similar arguments as \citet{smi16b}, \citet{mau17} favoured a scenario with the dust being heated by a surviving
star rather than an IR echo. A somewhat different interpretation was given by \citet{eva03} for V838 Mon, from the inspection of a NIR spectrum obtained 
almost 10 months after discovery. The stellar radius exceeding 800 R$_\odot$ and the extremely red continuum, with T$_{eff} \lesssim$ 2300 K as constrained by the strength 
of some molecular bands (in particular, H$_2$O and CO) and the absence of CH$_4$, were considered as an evidence for the presence of a very cool supergiant
(a late L-type star), without necessarily linking these observables with dust signatures.

As a support for the newly formed dust scenario in  NGC4490-2011OT1, we note that the large photospheric radius estimated at the time of the second (red) peak 
(17-20 $\times 10^3$  R$_\odot$, which is about 80-90 AU; see mid panel in Fig. \ref{radius}) suggests that pre-existing dust at $\sim$80 AU would have been 
swept away by the fast expanding  material ejected during the outburst. In addition, the late-time blueshift of the H$\alpha$ emission peak is a further argument
that supports the formation of new dust.

Thus, the presence of previously existing or new dust may be a common property in double-peaked transients, including known RNe \citep[e.g.,][and references therein]{ban15,ext16}.
The origin of the dust is controversial also for RNe, and plausible scenarios include interstellar \citep[e.g.,][]{tyl04,cra05,tyl05,kam11} vs. circumstellar dust \citep{bon03,van04,bon07,wis08}, or even a combination of both 
\citep[see][]{ban06}. \citet{tyl16} found that dust was present before the outburst of V1309 Sco, but new dust also formed soon after the outburst in the expanding envelope.
Due to the presence of dusty environments, these objects are expected to contribute to the rich population of IR transients recently discovered in nearby galaxies 
by the SPitzer InfraRed Intensive Transients Survey \citep[SPIRITS;][]{kas17}.
We will discuss the implications in Sect. \ref{discussion}.

\section[]{On the nature of double-peaked SN impostors} \label{discussion}

As reported previous sections, the six LRNe studied in this paper form 
a small family of transients with well-defined observational properties, that can be summarized as follows.

\begin{itemize}

\item Their light curves are reminiscent of those of RNe, with an initial blue light curve maximum, followed by a broader red peak which, in some cases (e.g., SN~1997bs), is 
stretched to become a sort of plateau. However, LRNe are significantly more luminous than RNe, ranging between $M_V = -12$ and $-15$ mag at maximum.
\item From the inspection of archival images, the precursors of this family of transients do not show further major outbursts (with luminosity comparable with those of the blue and red peaks) 
in the years prior to their discovery.
\item Early spectra (during the blue peak) are similar to those of Type IIn SNe, with a blue continuum and prominent Balmer lines in emission. During the red light-curve peak, the spectra become
similar to those of mid to late spectral-type stars, with a forest of narrow absorption metal features and Balmer lines being barely visible. 
\item Finally, at very late phases (from $\sim$ 4 to 6 months after the blue maximum), their optical spectra display evident absorption features due to molecular bands, and - at least in some cases - there are indications (in the IR domain) 
of a dusty circumstellar medium (CSM).
\item The LRN properties at very late phases, as detailed for NGC~4490-2011OT1 (Fig. \ref{radius}), suggest that a luminous 
stellar-like source is still visible, having a very extended ($R\sim10^4$ R$_\odot$) optically thick photosphere which is radiating $L\approx6\times10^{39}$ erg s$^{-1}$ 
(hence log $L/L_\odot\sim6.2$). This very red source is powered by a combination of radiated energy from the survived stellar merger (which has still
a burning core), plus residual thermal energy released during the merging event, and perhaps interaction between colliding shells.
\item As we will see in Sect. \ref{progenitors}, when the progenitor stars are observed in a pseudo-quiescent stage in deep HST archival images, none of them is very luminous, or
has a very large mass. 
These recovered progenitor stars (or stellar systems) are preferentially moderate luminosity, relatively massive and hot supergiants.

\end{itemize}

In the forthcoming sections, we will summarize what is known on the progenitor stars of LRNe (Sect. \ref{progenitors}), 
we will characterize them in the context of RNe (Sect. \ref{nature}) and will offer interpretations for their
observables, providing insights on plausible end-of-life scenarios (Sect. \ref{two_peaks} and Sect. \ref{11ht}).

\subsection[]{Progenitors of luminous double-peaked transients}  \label{progenitors} 

So far, direct information on the progenitor stars is available for four of the transients discussed in this paper.
The stellar precursor of NGC4490-2011OT1 was recovered in a putative quiescent stage  by \citet{smi16b}
only in one band (in 1994 HST images), at $M_V \approx -$6.4 mag \citep[consistent with the 
absolute magnitude determined by][]{fra11}. Two non-detections 
from Spitzer images obtained in 2004 (assuming little or no photometric variability in ten years) help in constraining the spectral type. 
Nonetheless, a precise colour estimate is impossible because of the single band detection and the large uncertainty in the line-of-sight reddening to NGC4490-2011OT1.
\citet{smi16b} discussed three reddening scenarios: 1) The luminosity of the star is attenuated by Milky Way reddening only \citep[$E(B-V)=0.02$ mag;][]{sch11}. Adopting this low reddening
value, the putative progenitor is a yellow star with effective temperature $T_{eff}=7500$ K and a luminosity of about $10^4$ L$_\odot$, implying an F-type moderate-mass supergiant star of
around 10 M$_\odot$. 
2) Assuming $E(B-V)=0.32$ mag, \citet{smi16b} obtained $T_{eff}=15000$ K and $L\sim10^5$ L$_\odot$. 
In this case, the progenitor would be a more massive mid-B supergiant. Assuming a single-star model, the resulting progenitor mass is
about 20-30 M$_\odot$, still consistent with a relatively low mass LBV. 
3) Finally, adopting as total reddening the upper limit of $E(B-V)=0.67$ mag\footnote{This extinction value was estimated for the star nearest to the location of NGC4490-2011OT1.}, 
\citet{smi16b} obtained a very hot ($T_{eff}\geq50000$ K) and luminous ($L\sim10^6$ L$_\odot$)
progenitor, which would be an extremely massive O-type main sequence or a Wolf-Rayet star. 
Smith et al. suggested that the non-detection of a source at the position of  NGC4490-2011OT1 in 2004 deep Spitzer observations allowed us to safely rule out a late spectral type
for the progenitor of this LRN. On the other hand, a high reddening scenario is not plausible because this would imply the progenitor to be a very hot, luminous O-type or Wolf-Rayet star of
around 80-100 M$_\odot$ (to explain its enormous luminosity), difficult to reconcile with the remote location of the transient in NGC~4490.
The above arguments led \citet{smi16b} to favour scenario number 2. If the lower mass binary companion does not give a significant contribution to the total luminosity, the progenitor is more 
likely to be an intermediate-mass yellow supergiant to a moderate-mass LBV. However, a binary system with a significant contribution to the global luminosity from the companion star 
cannot be ruled out. This would lead to a different mass and spectral type for the progenitor of NGC4490-2011OT1.

An object having more stringent information on the quiescent progenitor star is M101-2015OT1. 
\citet{bla16} argued that the progenitor was a binary system with a primary F-type yellow supergiant 
($T_{eff}\approx7000$ K), with a luminosity of $8.7\times10^4$ L$_\odot$, hence with $M\approx16$-$18$ M$_\odot$,
and a modest-mass secondary companion. \citet{gor16}, on the basis of its proximity with nearby OB-supergiant associations, proposed a somewhat bluer progenitor.
While a common envelope was surely ejected by the binary progenitor of M101-2015OT1, the post-outburst fate is uncertain, as the outcome can be either a merger or a surviving 
close binary \citep{bla16}. 
In the latter scenario, we would expect to detect signs of photometric variability in future IR photometric 
observations. 

\begin{table*}[!t] 
\centering
  \caption{Main parameters of known merger candidates. References for the distance, reddening estimates and light curves, respectively,
are identified by numbers in the last column. \label{tab_LRN}} 
  \tiny
  \begin{tabular}{ccccccccc} 
  \hline   \hline
            \noalign{\smallskip} 
Milky Way & Outburst Period       &  $d$        & $E_{tot}(B-V)$ & $M_{V,star}$ & $M_{V,prec}$ & $M_{V,1peak}$ & $M_{V,2peak}$ & Source    \\
  Transients                 &                          &   (kpc)          &   (mag)       &    (mag)   &    (mag)   &     (mag)   &    (mag)    &  codes     \\ \hline 
CK Vul$^\P$        & Jun 1670 to Jun 1671     &    0.7           & 2.2          & $>-$5.4     &      -     &   $-$8.4   &   $-$8.8     & 1;2;2  \\ 
V1148 Sgr$^{\bigstar}$  & $\sim$1943               &    2.7$^\ominus$           & 1.3           &     -      &       -    &   $-$9.5    &     -       & 3;4;5  \\ 
V4332 Sgr$^\P$ $^{\bigstar}$    & 1994                     &  $>$5.5          & 0.55         & $<$2.65       &      -     &     -      &   $<-$7.01    & 6;6;7-9   \\ 
V838 Mon$^\P$ $^{\bigstar}$ $^\square$ & Jan 2002 to May 2002     &    6.1           & 0.85          & $-$1.06    & $-$6.68    & $-$9.79     &  $-$9.46    & 10;11;12-16 \\
OGLE 2002-BLG-360$^\nabla$ & 2002 to 2006             &    8.2           & 2.00$^\dag$    & $-$0.80    & $-$1.21    & $-$4.10     &  $-$3.09    & 17;18;17 \\ 
V1309 Sco$^\P$ $^{\bigstar}$    & Jun 2008 to Mar 2010     &    3.5           & 0.55          & 3.80       & $>-$0.93   & $-$6.56     &  $-$5.02   & 19;20;19-21  \\  \hline 
Extragalactic & Outburst Period   &  $d$        & $E_{tot}(B-V)$ & $M_{V,star}$ & $M_{V,prec}$ & $M_{V,1peak}$ & $M_{V,2peak}$ & Source    \\
 Transients                  &                          &   (Mpc)          &   (mag)       &    (mag)   &    (mag)   &     (mag)   &    (mag)    &  codes     \\ \hline
M31-RV            & Jun 1988 to Oct 1988     &    0.78          & 0.12          & $-$5.04    &     -      &  -          & $-$8.69     & 23;22;22 \\
M31-LRN2015       & Jan 2015 to May 2015     &    0.78          & 0.35          & $-$2.13    &     -      & $-$10.16    & $-$8.93     & 23;23;23-25 \\ 
M101-2015OT1      & $<$Feb 2015 to Mar 2016  &    6.4           & 0.01          & $-$7.19    & $-$10.10   & $-$12.70    & $-$11.46    & 26;4;27-29 \\
NGC4490-2011OT1   & May 2011 to May 2012     &    9.6           & 0.32          & $-$7.32    & $-$9.18    & $-$14.35    & $-$14.54    & 30;31;23$\&$30 \\ 
NGC3437-2011OT1   & Jan 2011 to Jun 2011     &   20.9           & 0.02          & $>-$9.98   & $>-$10.83  & $-$13.06    & $-$13.33    & 32;4;29 \\ 
UGC12307-2013OT1  & $<$Jun 2013 to Nov 2013  &   39.7           & 0.22          & $>-$11.88  &     -      &      -      & $-$15.03    & 29;4;29 \\
SN~1997bs (NGC~3627) & Apr 1997 to Jan 1998  &    9.2           & 0.21          & $-$7.61    &     -      & $-$13.34    & $-$11.51    & 29;33;29$\&$33-35 \\ 
SNhunt248 (NGC~5806) & Apr 2014 to May 2015  &   22.5           & 0.04          & $-$8.99    & $-$11.18   & $-$14.87    & $-$14.07    & 36;37;37-38 \\
AT~2017jfs        & Dec 2017 to Jul 2018     &   34.7           & 0.02          & $>-$11.26  &     -      & $-$15.46    & $-$14.38    & 39 \\ 
\hline                                                                                       
        \noalign{\smallskip}
\end{tabular}                                                                               

\tablefoot{
1  = \protect\citet{haj07}; 2  = \protect\citet{sha85}; 3 = \protect\citet{min95};  4 = \protect\citet{sch11};
5  = \protect\citet{may49}; 6  = \protect\citet{tyl15}; 7 =  \protect\citet{mar99}; 8 = \protect\citet[][and references therein]{gor07}; 
9  = \protect\citet{kim06}; 10 = \protect\citet{spa08}; 11 = \protect\citet{afs07}; 12 =  \protect\citet{mun02}; 
13 = \protect\citet{gor02}; 14 = \protect\citet{kim02}; 15 = \protect\citet{cra03}; 16 = \protect\citet{cra05}; 
17 = \protect\citet{tyl13}; 18 = \protect\citet{koc14}; 19 = \protect\citet{tyl11}; 20 = \protect\citet{mas10}; 
21 = \protect\citet{wal12}; 22 = \protect\citet{bos04}; 23 = \protect\citet{kur15}; 24 = \protect\citet{wil15}; 
25 = From Astronomer's Telegrams; 26 = \protect\citet{sha11};27 = \protect\citet[][and references therein]{gor16};  
28 = \protect\citet{bla16}; 29 = this paper; 30 = \protect\citet{smi16b}; 31 =  \protect\citet{pasto08};
32 = \protect\citet{sor14}; 33 = \protect\citet{van00}; 34 = \protect\citet{li02}; 35 = \protect\citet{ada15};
36 = \protect\citet{tul09}; 37 =  \protect\citet{kan15}; 38 =  \protect\citet{mau15};
39 = \protect\citet{pasto19}.\\
Additional information: $^\P$  \protect\citet{ozd16} (see their tables 2 and 3) give the following alternative distance and reddening estimates for four
Galactic RNe: $d=4.48\pm0.24$ kpc and $E(B-V)=0.75\pm0.05$ mag for CK~Vul; $d=1.14\pm1.00$ kpc and $E(B-V)=0.32\pm0.10$ mag for V4332~Sgr;
 $d=2.5\pm0.4$ kpc and $E(B-V)=0.55\pm0.05$ mag for V1309~Sco; $d\geq7$ kpc and $E(B-V)=0.87$ mag for V838~Mon. 
$^{\bigstar}$ As a consistency check, we also estimated the distance of Galactic RNe using the Gaia DR2 parallaxes \protect\citep{gaia1,gaia2}, and obtained the following values:
$\omega=0.4306\pm0.0591$ mas and $d=2.33_{-0.37}^{+0.84}$ kpc for V1148~Sgr; $\omega=0.0017\pm0.2798$ mas and $d=3.85_{-1.57}^{+4.65}$ kpc for
V4332~Sgr; and $\omega=-0.0014\pm0.1051$  mas and  $d=6.06_{-2.02}^{+3.33}$ kpc for V838~Mon. All of them are consistent (within the errors) with those reported in the table.
$^\ominus$ We assume that V1148~Sgr is at the distance of the globular cluster NGC~6553.
$^\square$ \protect\citet{mun05} estimated a similar reddening $E(B-V)=0.87\pm0.02$ mag, and a somewhat larger distance ($d\sim10$ kpc) to V838~Mon. 
$^\dag$ Following \citet{koc14}, we adopt a total reddening comprising both a Milky Way component \citep[$E(B-V)_{MS}=1$ mag, with $R_V=2.5$, following][]{nat13}, 
and a circumstellar dust component ($E(B-V)_{CSD}=1$ mag, with a standard reddening law $R_V=3.1$).
$^\nabla$ As only the $I$-band light curve is well monitored at all phases, the $V$-band peak magnitudes were computed from the $I$-band magnitudes, adopting
the available $V-I$ colour information.
}
\end{table*} 
    
The pre-outburst history of SNhunt248 is somewhat different, as there is evidence of significant photometric oscillations (of the order of 1 mag) in the pre-outburst phase \citep{kan15}.
Adopting only the Milky Way reddening contribution, from observations obtained about 2 years before the outburst, Kankare et al. find $T_{eff}\approx6700$ K and $L\approx10^6$ L$_\odot$ 
for the progenitor. 
\citet{mau15} computed $T_{eff}\approx6500$ K and $L\approx4\times10^5$ L$_\odot$ for the stellar precursor detected in images obtained over 9 years before the outburst,
which are consistent with the above estimates accounting for the past variability history.
In both estimates, the progenitor is consistent with being a luminous yellow (late F-type to early G-type) hypergiant of $M\approx30$ M$_\odot$. A super-Eddington outburst 
of an LBV-like star or close stellar encounters with a binary companion are alternative explanations for the observed outburst  \citep{mau15}. However, the above mass value
should be taken with a bit of caution, as the magnitude estimates of the progenitor star (or stellar system) were possibly obtained in a non-quiescent stage.

Finally, progenitor mass estimates are available also for SN~1997bs. \citet{van99} found a $M_V \approx-7.4$ mag source in 1994 December HST images. As there is a single-band detection, 
no information is provided on the intrinsic colour and variability, although the authors favoured a super-outburst of an LBV as the most likely explanation for the outburst.
\citet{van12} estimated  $T_{eff}>6300$ K (hence a spectral type hotter than F6) and a mass exceeding 20 M$_\odot$ for the progenitor of SN~1997bs;
\citet{ada15} favoured a marginally lower mass star, of about 20 M$_\odot$, rather than a massive LBV.
Post-eruption HST imaging obtained in March 2001 revealed the source to be still marginally visible, about 3 mag fainter than the 1994 detection (and with a relatively blue colour, 
$V-I <0.8$ mag), initially leading \citet{li02} to doubt that the star had survived the outburst. However, \citet{van12} argued that the formation of large dust 
grains produces obscuration without a significant reddening. This would explain the relatively blue colour of the post-outburst source, hence favouring the survival of the star. On the other hand,
 \citet{ada15} reached a different conclusion, and argued that SN~1997bs was a faint SN explosion rather than a SN impostor, although they could not rule out a surviving star embedded 
in a dusty massive shell. Clearly, the controversial interpretation of the nature of SN~1997bs does not allow us to firmly constrain its progenitor.

Despite the large uncertainty on the information inferred for the LRN progenitors, a usual interpretation favours blue to yellow spectral type stars, 
although likely in a wide mass range. It is plausible that the erupting star is the primary member of a close, interacting binary system, although an outburst of a more massive single star 
 cannot be definitely ruled out. The evolution of the binary system leading to double-peaked events will be further discussed in Sect. \ref{two_peaks}.

\subsection[]{A continuum of properties from RNe to LRNe?}  \label{nature} 
 
\begin{figure*}[!t]
\centering
\includegraphics[width=15.6cm,angle=0]{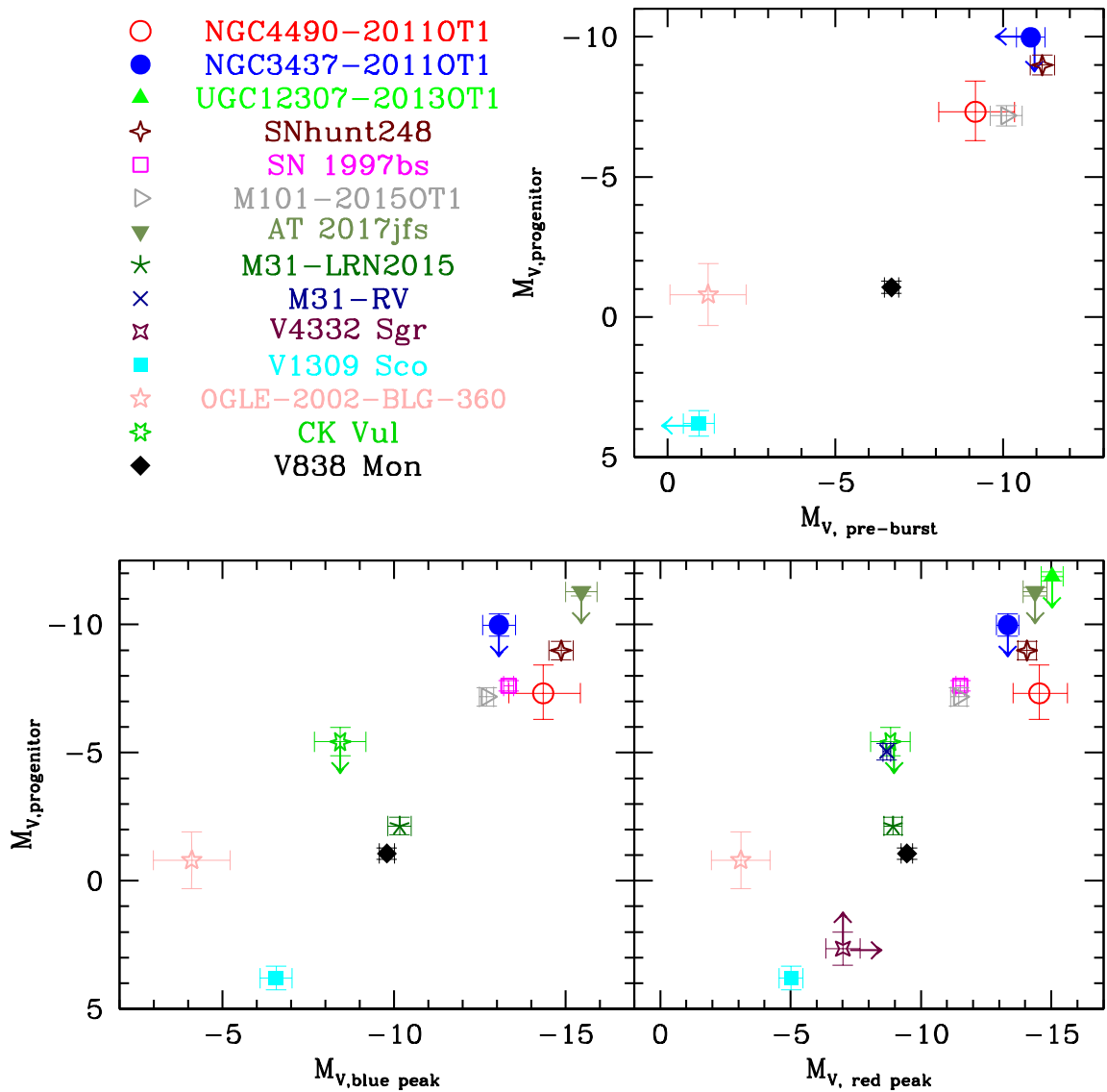}   
  \caption{Absolute $V$-band magnitudes of putatively quiescent progenitors of merger candidates vs. absolute $V$ band magnitudes of brightest pre-outburst detection (top-right), 
blue peak (bottom-left) and red peak (bottom-right). The absolute magnitudes of the progenitor and the light-curve 
red peak of V4332~Sgr are indicated as lower limits, as only a lower limit is known for its distance (Table \ref{tab_LRN}). \label{correlation_mergers}} 
\end{figure*} 

Known Galactic RNe from the sub-luminous V1309~Sco \citep{mas10,tyl11} to much brighter events such as V4332~Sgr \citep[which erupted in 1994,][]{mar99} 
and the famous V838~Mon \citep[e.g.,][]{mun02}, as well as extra-Galactic transients such as two well-studied RNe in M31 (M31-RV and M31-LRN2015). 
Although a number of authors proposed that RNe were peculiar thermonuclear-driven novae \citep[e.g.,][]{sha10},
they are usually interpreted as the result of the common envelope ejection and/or gravitational merging process in low 
to moderate mass systems \citep[e.g.][]{sok03,iva13,tyl16,mac17}. This will be discussed in Sect. \ref{two_peaks}.

\citet{smi16b} emphasized the similarity of NGC4490-2011OT1 with RNe. 
Here we compare the observational properties of RNe and LRNe to eventually correlate their physical parameters and shed light on the nature of their progenitors.
In particular, we measure the faintest pre-outburst detection of the progenitor (or its detection limit), from which we derive the bona fide absolute magnitude $M_V$ of the quiescent 
progenitors. Then, we estimate $M_V$ of the minor pre-outburst brightening, and those of the blue and the red peaks. The photometric parameters for all proposed merger 
candidates are reported in Table \ref{tab_LRN}.

We note that the LRN (or RN) labelling was proposed in the past for two gap transients, M85-2006OT1 and V1148~Sgr.
The classification of M85-2006OT1 as a peculiar RN was first proposed by  \citet{kul07}. However, this was questioned by \citet{pasto07} and \citet{tho09}, who noted 
similarities with SN~2008S-like transients and faint core-collapse SNe \citep[see also][]{koc14}. In particular, observational arguments can be used to disfavour
the RN/LRN classification.  Firstly, the spectra of M85-2006OT1 show a quite prominent [Ca~II] $\lambda\lambda$~7291,7324 feature, which is typical 
of ILRTs, such as SN 2008S \citep{bot09}. In addition, molecular bands have not been detected in the latest spectrum. The above
spectroscopic properties support its classification as an ILRT \citep[][and references therein]{kas11}.
Hence, we will not further discuss this transient in the context of the merger candidates.\\
V1148~Sgr is another possible RN, although this claim is not supported by an accurate photometric monitoring. 
However, the spectrum described by \citet{may49} is that of a K-type star, with strong H$\&$K absorption lines along with possible TiO bands. 
A few days later, a prominent H$\alpha$ was clearly detected. The description of the spectra closely matches the spectral appearance of double-peaked 
events, during their transition from the red light curve peak to very late phases (Fig. \ref{sp_molecules}). For sake of completeness, the modest information available 
for this object is reported in Table \ref{tab_LRN}.

In Fig. \ref{correlation_mergers}, we show the  $V$-band absolute magnitudes of the quiescent progenitors of the objects of Table \ref{tab_LRN} vs.
the absolute magnitudes of the pre-outburst brightening (top-right panel), the blue peak (bottom-left panel) and the red peak (bottom-right panel).
Although this analysis may be affected by selection biases (e.g., RNe have not been detected in the Local Group), 
progenitor and outburst luminosities seems to be somewhat correlated, with more luminous quiescent progenitors are producing 
brightest outbursts.\footnote{A promising correlation between peak luminosity of the outburst and the
wind outflow velocity was found by \citet{pej16} and \citet{mau17}.} This qualitative trend suggests that
all outbursts of Table \ref{tab_LRN} may have been triggered by a similar mechanism. 
 
\begin{figure*} 
\centering
{\includegraphics[width=8.6cm,angle=0]{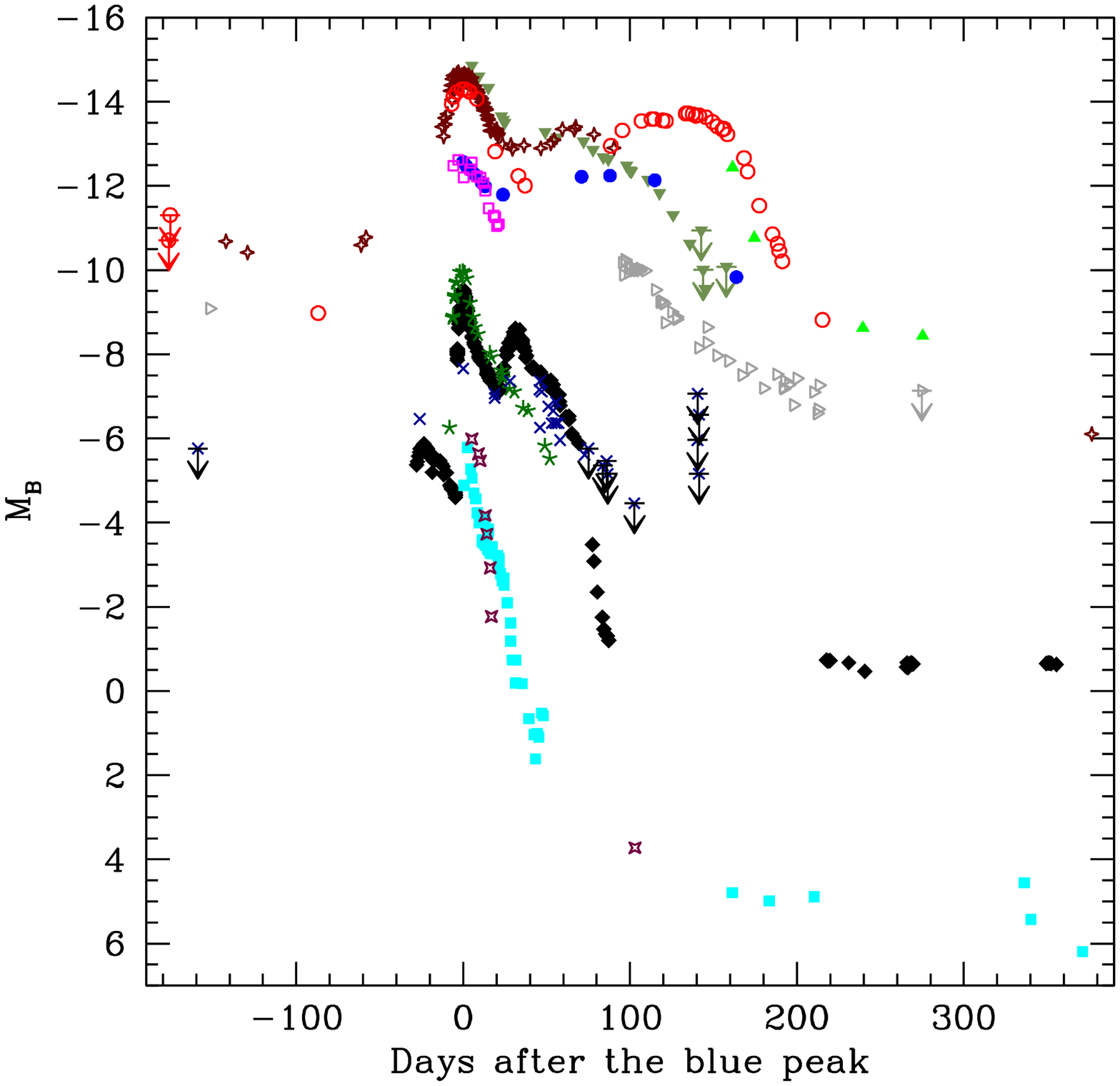}  
\includegraphics[width=8.6cm,angle=0]{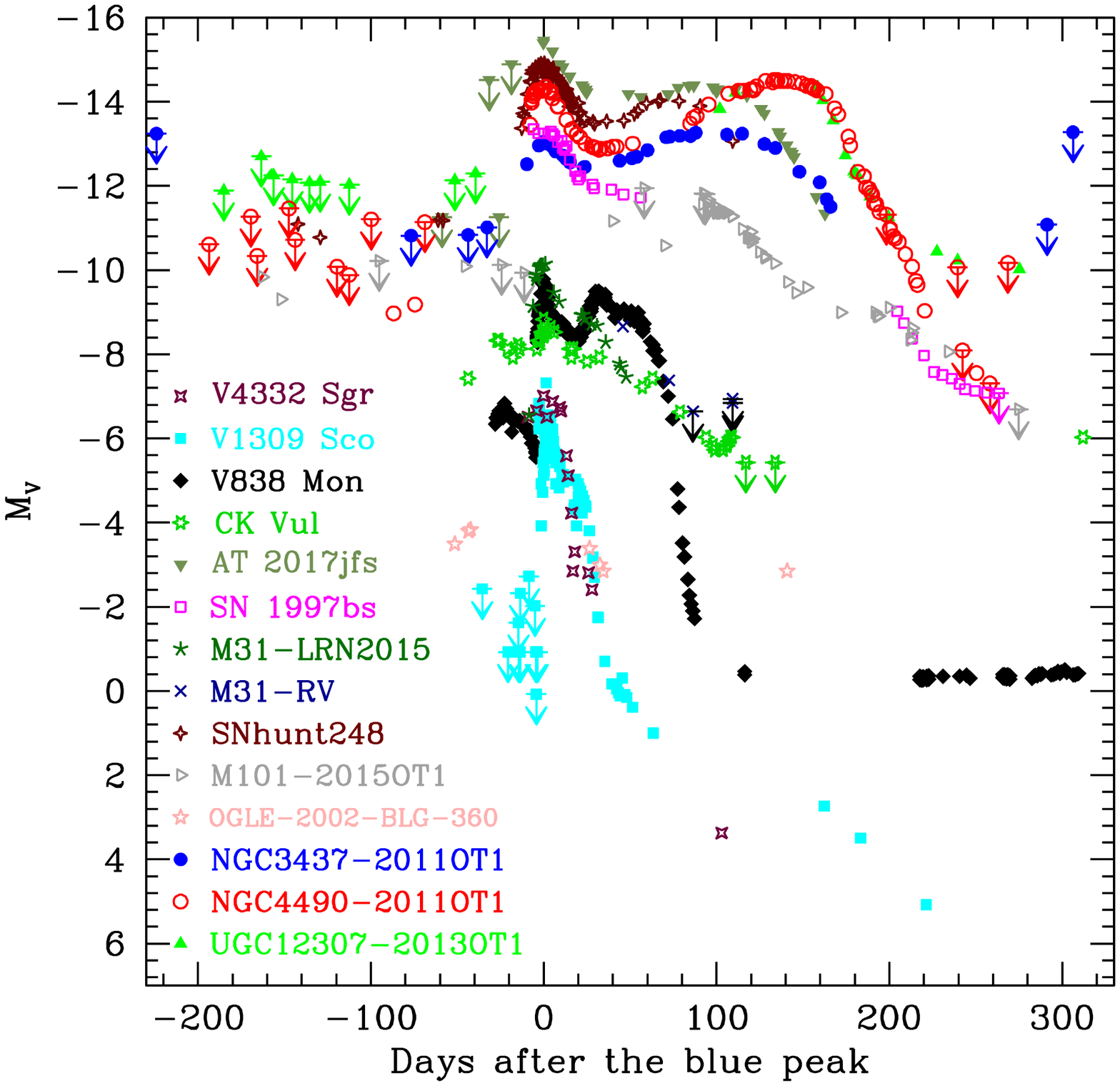}  
\includegraphics[width=8.6cm,angle=0]{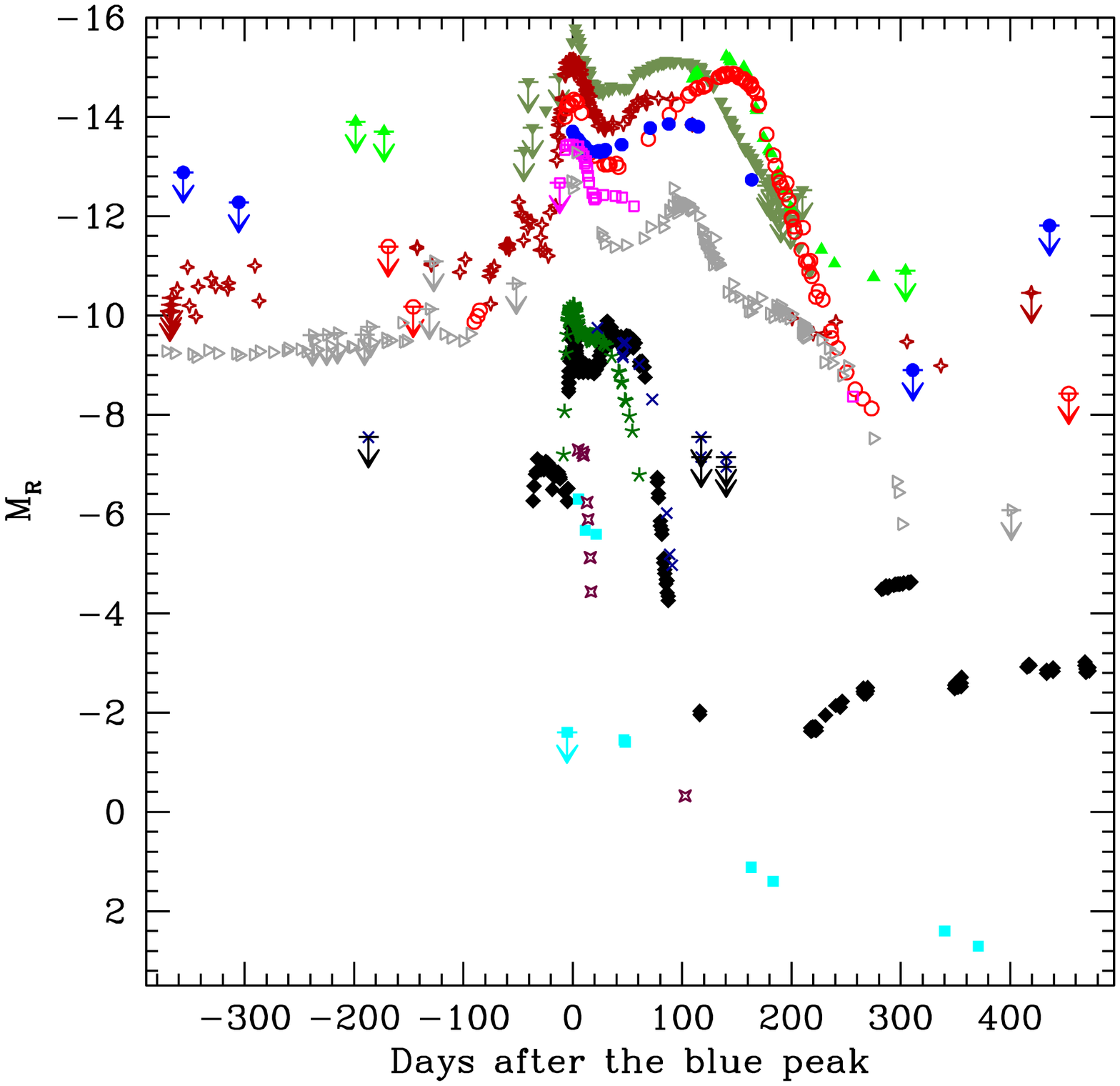}  
\includegraphics[width=8.6cm,angle=0]{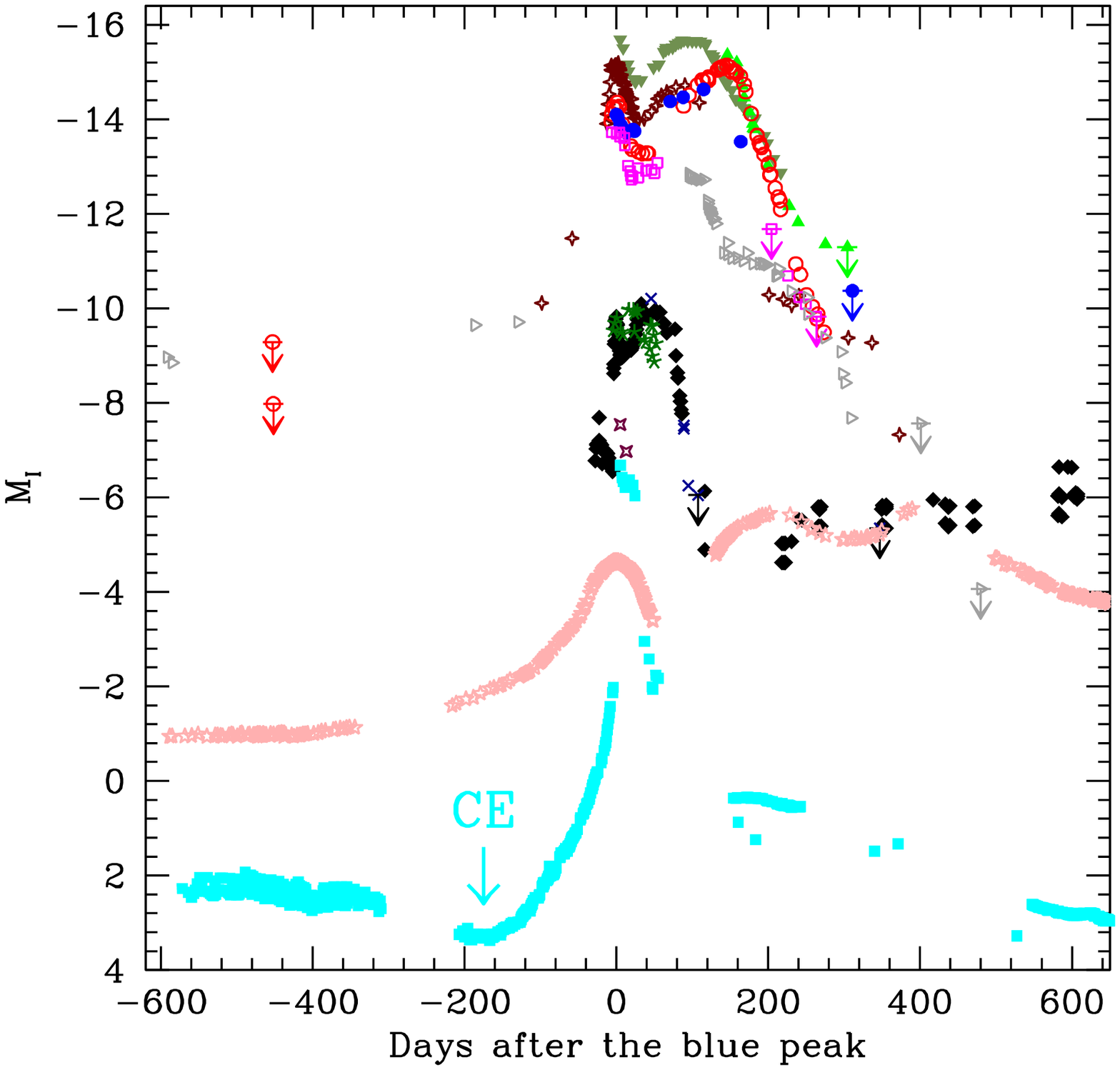}} 
  \caption{$B$ (top-left), $V$ (top-right), $R$ (bottom-left), and $I$-band (bottom-right) absolute light curves of merger candidates listed in Table \ref{tab_LRN}. 
The approximate time of the common envelope ejection  in the I-band light curve of V1309~Sco is also marked. \label{lc_mergers}} 
\end{figure*} 

The available $B, V, R$ and $I$ absolute light curves of our RN/LRN sample are compared in Fig. \ref{lc_mergers}.  
Within the framework of the common envelope scenario, then the intrinsic photometric heterogeneity of the objects likely depends on the time of the common envelope
ejection, the total mass of the binary system \citep{koc14}, and the mass ratio between the two stellar components.
Assuming that the mass and luminosity of the quiescent progenitor system are correlated, the qualitative comparison in Fig. \ref{correlation_mergers} 
also supports a major result of \citet{koc14}, with higher mass systems producing more luminous outbursts.
V4332 Sgr, OGLE-2002-BLG-360, and V1309 Sco are expected to arise from systems formed by low-mass (1-2 M$_\odot$) stars, 
while the two RNe in M31 are likely associated to more massive binaries \citep[3-6 M$_\odot$, e.g.,][]{lip17,mac17},\footnote{A best-fit 
model obtained by \citet{lip17} using an adapted version of the radiative hydrodynamical STELLA code \citep{bli06}, 
reproduced the light curve of M31-LRN 2015 through a merging event, in which the total merger's mass was only 3 M$_\odot$.} 
and V838~Mon was produced by an even more massive system, with a primary of 5-10 M$_\odot$ \citep{tyl05}.
According to this view, the progenitor systems of the objects discussed in this paper are even more extreme, with the total mass likely 
being between 15 and 30 M$_\odot$ \citep{smi16b}.

\subsection[]{Common envelope ejection and merger scenario for double-peaked transients} \label{two_peaks}

\subsubsection[]{Current understanding of the red nova phenomenon} \label{two_peaks_LRN}

Although thermonuclear runaway \citep[e.g.,][for M31-RV]{ibe92}, or post-asymptotic giant branch 
(AGB) He-shell flash scenarios \citep{mun02,kim02} have been proposed in the past for RNe \citep[see, also,][]{van04}, there are growing
evidences that these gap transients originate from binary stellar systems. In particular, while the initial decrease
of the photospheric temperature is expected near to the outburst maxima, no thermonuclear models comfortably explain 
the very late-time temperature evolution of RNe towards cooler and cooler temperatures \citep{sok03}.

Most massive stars exchange material with a binary companion during their life \citep[about 70\%, according to][]{sana12}, and this strongly affects the evolutionary 
path of the two components. This interaction may, in fact, change the relative masses and the spins of the two stars, and trigger mixing processes in the inner stellar 
layers, affecting their chemical evolution. In addition, a significant fraction of massive stars \citep[up to one-fourth,][]{sana12} are expected to merge.
Hence, merging events are very likely common \citep{koc14}. For example, stellar mergers are a natural explanation for peculiar stars such as the blue stragglers
\citep[e.g.,][]{per09,nao14}. This has been recently proposed by \citet{fer19} as the outcome of V1309~Sco. 
Other studies suggested that a large fraction of massive stars are the outcome of past merging events \citep{sel14}.
\citet{ofe08} set relatively high rates of stellar mergers in the Milky Way, providing a lower limit of 0.019 yr$^{-1}$, later revised by \citet{koc14} who estimated
a rate of about 0.5 yr$^{-1}$ for events brighter than $M_V=-3$ mag, and fading to $\sim0.1$ yr$^{-1}$ for objects brighter than $M_V=-7$ mag. 
For this reason, it is not surprising that we are finding a number of merger candidates in the 
Local Group. 

Two Galactic objects gave fundamental insight to our knowledge on RNe: V838 Mon and V1309 Sco.

Firstly, V838~Mon is one of the best-studied Galactic variables in the past decade. The variable lies in a young environment, 
with numerous hot stars, rich in gas and dust \citep{afs07}. A B-type companion was revealed by spectroscopy, and its wide separation
makes implausible an active role in the V838~Mon outburst \citep[see,][and references therein]{bar17}.
The luminosity of the progenitor system of V838~Mon remained approximately constant at a visual brightness of about 15.6 mag for over 
half a century, until early 1994\footnote{The S/N of these old observations on photographic plates and the accuracy of the light curve calibration 
do not allow us to support (neither to rule out) a photometric modulation during the monitoring period.} \citep{gor04}.
Unfortunately, while after January 2002 the outburst was well monitored, the field was not targeted from March 1994 to December 2001,
and the evolution of the source during this critical phase was not observed. Although a core stellar instability 
cannot be definitely ruled out, the merging model is favoured to explain the structured light curve of V838~Mon \citep[e.g.,][]{sok03}. 
This scenario is also supported by new observational constraints from similar objects (see below).
Thus, the progenitor stellar system of V838~Mon would be formed by three stars. Along with the detached B-type companion mentioned above, 
a close binary system formed by another blue star of a few M$_\odot$ plus a lower-mass (a few $\times10^{-1}$ M$_\odot$) companion
likely produced the outburst.\\ 
When the primary inflated, filling its Roche lobe, 
mass flowed to the other star, with the system becoming dynamically unstable. Unfortunately, this phase was not directly observed
in V838~Mon. However, with the loss of angular momentum, the orbital distance is expected to diminish, and a series of minor 
photometric oscillations superposed on a longer-term moderate brightening have likely occurred. The modulation of the light curve would be the consequence of subsequent 
passages of the secondary star to the periastron \citep{tyl05}. This unobserved approaching phase was followed by a major encounter 
which partially disrupted the companion star and produced a luminous pre-outburst event reaching 10th mag in January 2002 \citep{mun02}. This model 
would explain the entire photometric evolution of V838~Mon. \\
Following \citet{tyl05,tyl06}, the primary star violently ejected the envelope producing an initial, low-luminosity  
light curve peak. Then, the secondary star accreted onto the primary component within the extended, common envelope. The complex 
merging process, along with ejecta-shell collisions, likely gave rise to the double-peaked light curve.
 This peculiar light curve shape is actually observed both in the Galactic comparison objects, and in more luminous extra-Galactic 
transients presented in this paper. The final outcome of this process was the production of a late-type star. 

Secondly, V1309~Sco is another key object. It provided unequivocal proofs of the final merging event. 
This RN was monitored for a long period (and with high-cadence observations) before the outburst
\citep[][see their figure 1]{tyl11}, allowing us to directly observe the courtship dance of the two stars leading to the final coalescence. 
Its pre-outburst photometric evolution initially showed a slow trend of increasing luminosity (approximately from 2002 to mid-2007),
which was probably due to increased mass loss that moved the photospheric radius \citep{pej16b} outwards. The slow 
(but moderate) luminosity increase was characterized by frequent oscillations superposed to the main brightening trend. 
The period of these oscillations decreased with time following an exponential trend. 
This evolution was due to the inspiraling orbital motion \citep{tyl11} reducing the distance of the 
two stars,  and an anisotropical obscuration of the binary from one direction \citep{pej16}.
During this phase, the mass-loss rate of the system increased progressively, settling onto a value of about 
$10^{-4}$ M$_\odot$ yr$^{-1}$ \citep{pej14}. Then, the light curve of V1309~Sco reached a minimum in late 2007 to early 2008,
in coincidence with the disappearance of its short timescale variability \citep[e.g.,][]{pej17}.
This can be explained with the expanding envelope progressively obscurating the binary system, with only the gas outflow remaining visible 
 \citep{pej16}, or with the formation of a dusty excretion disk \citep{tyl11,nic13}. This feature in the V1309~Sco light curve is marked  
 with a down-arrow in Fig. \ref{lc_mergers} (bottom-right panel).
Later on, a steep luminosity rise was observed from March to late August 2008, without the unequivocal photometric 
modulations due to the relative motion of the two stars. This brightening was very likely
due to an optically thick gas outflow from the primary star which enwrapped the secondary companion and produced a 
common envelope \citep{tyl11,pej14}. During this 5-months period, the mass-loss increased from about $10^{-3}$ 
to a few $\times10^{-1}$ M$_\odot$ yr$^{-1}$  \citep[see][their figure 5]{pej17}. Then, the huge brightening of $\sim3.5$ mag in less than
two weeks is likely due to the final coalescence of the secondary star's core onto the primary \citep[see, also,][]{nan14}. 
A similar photometric evolution of the outburst, although less densely sampled, was also observed in the RNe V4332~Sgr \citep{kim06} and M31-LRN2015 \citep{dong15,mac17}.
In general, simulations of merger events show that the runaway orbital decay with escalating mass ejection followed by
the stellar coalescence may produce a luminous transient flare with a duration similar to that of the binary orbital period \citep{mcl18}.\\
We note that, while the arguments provided to support the merger scenario for  V1309~Sco are robust, the physical mechanisms leading to
the rapid stellar coalescence are still debated. The increasing mass loss rate from the primary through the outer Langrangian point (L$_2$) 
may trigger the decrease of angular momentum \citep{pej17}. In this case, the formation of a circum-binary disc is expected. 
Alternatively, \citet{nan14} propose the Darwin instability, with dissipation of orbital momentum to the rotation of the stellar components, 
as a primary responsible of the merging event.  Finally, the presence of a third companion can lead a binary system to lose momentum, 
as suggested above for V838~Mon. However, the presence of a third star has been ruled out for V1309~Sco.

After the early blue peak, a plateau or a redder secondary peak is observed in RNe (see Fig. \ref{lc_mergers}), followed by a steep luminosity decline,
possibly accompanied by dust formation. The optical luminosity at very late phases can be significantly fainter than
the integrated luminosity of the original binary system, and shows no evidence of modulation in the light curve \citep{nic13}. 
As mentioned in Sect. \ref{mole},
signatures of dust formation are also observed in the late-time evolution of the luminous double-peaked events discussed in this
paper \citep[][see also Sect. \ref{two_peaks_NGC4490}]{bla16,smi16b}.

Although the binary evolution described above is globally consistent with the observables of RNe before and after the main outburst, the process
powering the light curve of the outburst with the characteristic double-peak is still debated. 
\citet{bar14} suggested that the most luminous, blue peak was produced by the strong, shock-induced photospheric heating triggered 
by the rapid coalescence process, while the broader red peak was due to the slow thermal energy release following the adiabatic
expansion of the envelope.
A reasonable explanation for the double peak is also provided by \citet{lip17} for M31-LRN2015. With the expansion, the temperature
decreases and a cooling front eventually recombines the hydrogen envelope. This may produce a plateau-like photometric evolution 
after the blue peak or a shallow, longer-duration red peak.
The presence of multiple H shells and a more complex density profile of the gas may complicate the scenario, generating a 
light curve with a more pronounced red maximum or even multiple peaks. However, as pointed out by \citet{mac17}, the H
recombination alone cannot explain the first, luminous and blue light curve peak.

\citet{met17} modelled the whole RN evolution, proposing that the first peak was due to the release of thermal
emission from hot gas, which is basically freely expanding in the polar direction.\footnote{A polar gas outflow in RNe is supported
by spectro-polarimetric observations \citep[e.g.,][for V4332~Sgr]{kam13}.} The observed properties are somewhat similar to 
those produced by the cooling gas in core-collapse SNe, in particular the early blue peak of the light curve. 
The subsequent (quite heterogeneous) light curve evolution of RNe can be explained, according to \citet{met17}, in terms of
interaction of the ejecta with lower-velocity equatorial wind, with the shock-heated material producing the second peak.
The gas shell swept up by the shock front would generate a cool dense shell, consistent with the red colours observed after the 
second peak, and would be the ideal site for the rapid dust condensation observed in most of these transients.
This explanation hence comfortably accomodates all the observables of both RNe and (very likely) their higher-luminosity counterparts.

The aspherical nature of RNe, inferred mostly by previous polarimetric and spectro-polarimetric observations \citep[see, e.g.,][]{kam13}, 
has been recently confirmed by submillimeter observations of V4332~Sgr and V1309~Sco \citep{kam18}, that have unequivocally 
revealed the existence of bipolar molecular outflows resulting from the merger event.

\subsubsection[]{Massive merger scenario for LRNe} \label{two_peaks_NGC4490}

Events brighter than V838~Mon ($M_V\sim-10$ mag) are rare, and have a Galactic rate of only 0.03 yr$^{-1}$ \citep{koc14}, hence we expect 
that luminous transient events, like those presented in this paper, are extremely rare. In fact, \citet{koc14} estimate integral event rates of
$\sim1$ to $5\times10^{-3}$ yr$^{-1}$ for events with $M_V\sim-15$ and $-13$ mag, respectively.
Despite their rarity, the physical mechanisms producing luminous double-peaked outbursts are very likely the same as RNe. From an observational point of view,
LRNe are in fact scaled-up clones of RNe (Sect. \ref{nature}). In addition, signatures of the slow pre-outburst brightening that might be attributed to the orbital shrinkage have been observed 
in a few objects presented in this paper (see Sect. \ref{two_peaks_LRN}). We speculate that these brightenings can be related to the common envelope phase,
in analogy with RNe.

\begin{itemize}
\item In NGC4490-2011OT1, a modest brightening was observed in 2011 June-July, about 3 months before the blue peak. In this phase, 
the object was about 5 mag fainter than at the blue peak, but 3.5-4 mag brighter than the 1994 HST detection reported by 
\citet{fra11} and \citet{smi16b}. 
\item In SNhunt248, the pre-outburst phase was monitored with sparse observations over about 15 years before the blue peak. During 
this period, its $R$-band magnitude was variable, ranging from about 22.2 to 20.5 mag. The object significantly brightened
 $\sim1.5$ months before the blue peak, but remained $\sim3.4$ mag fainter than the peak (Fig. \ref{lc}, bottom-right).
\item In M101-2015OT1, the Sloan $r$-band light curve remained below 21 mag until (at least) early 2008. From August 2009 to August 2014 
the object showed a minor erratic photometric variability superposed to a general trend of slow luminosity increase (from $r\sim20.5$ 
to 19.6 mag). The brightest magnitude of the pre-outburst phase was reached about three months before the blue peak, like in 
NGC4490-2011OT1.
\end{itemize}

\begin{figure*} 
\centering
\includegraphics[width=13.5cm,angle=270]{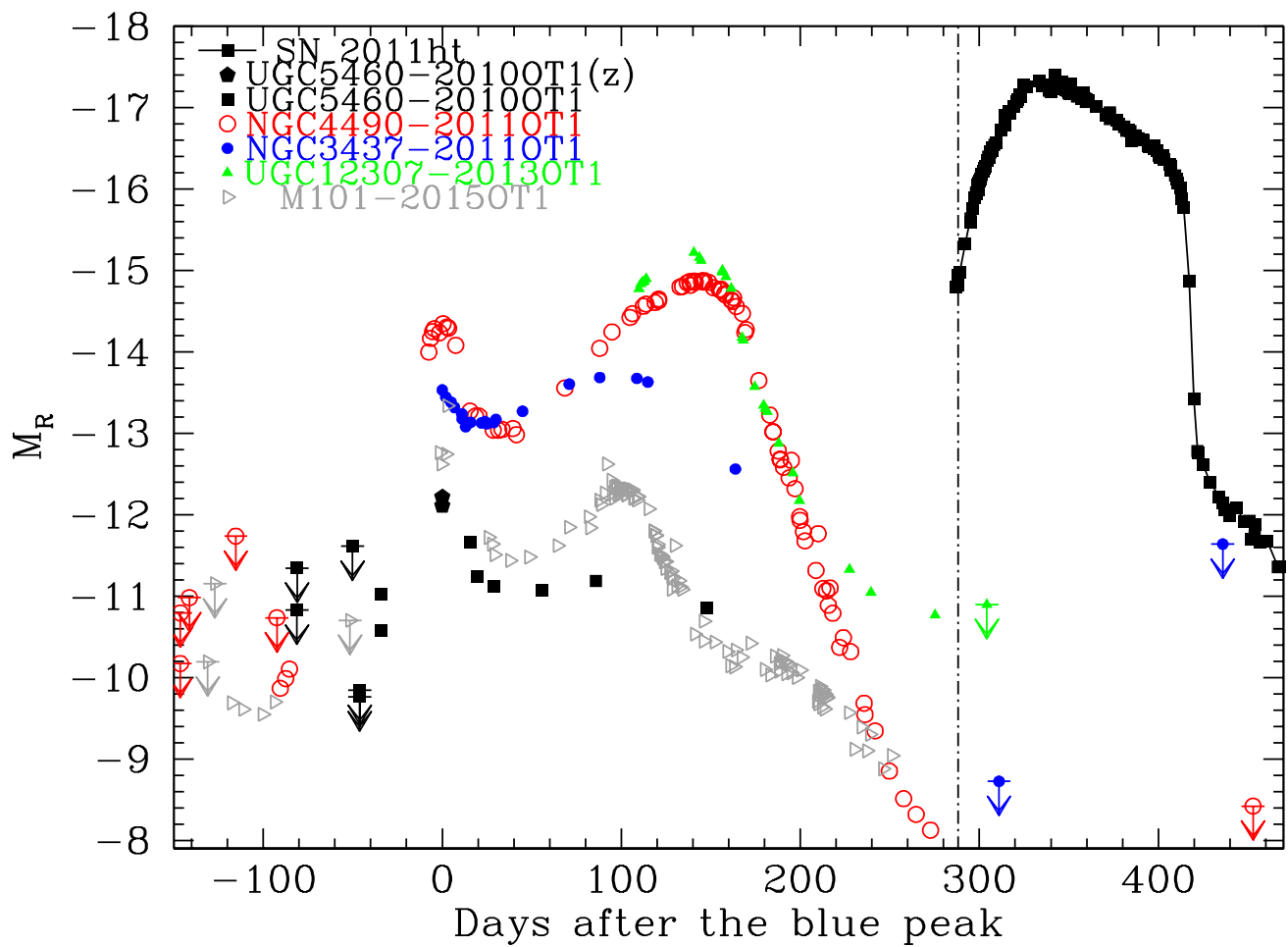}  
  \caption{Comparison of absolute light curves between a sample of LRNe
 and UGC5460-2010OT1, precursor of SN~2011ht. The light curve of the SN is also shown \protect\citep[from][and Pastorello et al., in preparation]{mau13}. 
The phase is in days from the $R$-band blue peak. Only significant 
detection limits are shown for the different objects. The dot-dashed vertical line marks the epoch of the earliest available spectrum of SN~2011ht. \label{lc_abs11ht}} 
\end{figure*} 

Recent papers have proposed that LRNe are the consequence of common envelope ejections 
in relatively massive contact binary systems  \citep[e.g.,][]{bla16}, likely followed by the stellar coalescence \citep{smi16b,mau17,lip17}.
Following this interpretation, we suggest that all transients analysed in this paper underwent a similar fate. The main differences
with RNe are in the values of the parameters involved, with NGC4490-2011OT-like transients having faster outflows, longer duration outbursts and higher
luminosities. Furthermore, LRNe are likely produced by more massive binaries \citep{koc14} than RNe. The typical range 
of systemic mass of the faintest RNe is in fact $\sim$1-5 M$_\odot$, that of the intermediate-luminosity V838~Mon is 5-10 M$_\odot$ 
\citep{mun02,tyl05}, and in LRNe masses are likely up to a few tens of M$_\odot$ \citep[see Table 1 of][and references therein]{met17}.

At the highest-mass boundary, \citet{smi11b} proposed that the fluctuations visible in the 19th century light curve of the Giant Eruption (GE) of $\eta$~Car
were produced by periastron encounters of two stellar components \citep[see, also,][]{mau17}. The high luminosity of the GE, along 
with the high mass  \citep[over  10 M$_\odot$, according to][]{smi03}
ejected during the event, are consistent with a very large systemic mass for $\eta$~Car. The GE of $\eta$ Car
and the formation of the Homunculus Nebula have also been explained with a merger event of a massive (90~M$_\odot$) close binary, 
triggered by the gravitational interaction of a 30~M$_\odot$ companion \citep{por16}.

Alternative explanations for double-peaked transients, explored for instance by \citet{sok03}, invoke other types of eruptive events from stars in a wide range of 
masses (AGBs to blue-to-yellow supergiants and hypergiants) or interacting systems where the common envelope ejections may prevent the merging event. However,
these can not be easily reconciled with the peculiar light curve shape and the spectral evolution towards a late-type star. These similar observational properties link transients 
spanning an enormous range of luminosities (and, hence, masses), and are more comfortably explained with a common envelope ejection. 
Whether the final outcome of the binary interaction process in LRNe is a merger, or alternatively the system rearranges into a new stable binary configuration 
is still debated, and only late-time photometric observations - especially in the IR domain - may solve this puzzle.

\subsection[]{Post-LRN evolution: a link with SN 2011ht-like events?} \label{11ht} 
 
In this section, we explore a possible evolutionary channel for LRNe, inspired by their photometric similarity 
with the outburst (named UGC5460-2010OT1) observed a few months before 
the explosion of the interacting SN~2011ht, and discussed by \citet{fra13}. The evolution of UGC5460-2010OT1/SN~2011ht will be
studied in detail in a forthcoming paper (Pastorello et al., in preparation).
 
For UGC~5460, we adopt a distance $d=17.9$ Mpc (from the HyperLeda Virgo-corrected recessional 
velocity and a standard cosmology), hence $\mu=31.26\pm0.15$ mag. The interstellar extinction within
the host galaxy is negligible, as expected from the location of the object in the outskirts of UGC~5460,
hence the modest total reddening is only due to the Milky Way contribution, $E(B-V)=0.01$ mag \citep{fra13}. The $R$-band absolute light 
curve comparison of UGC5460-2010OT1 (plus SN~2011ht)  with a sub-sample of the objects discussed in Sect. \ref{photo_abs}
is shown in Fig. \ref{lc_abs11ht}. The data of UGC5460-2010OT1 are from \protect\citet{fra13} and \protect\citet{ofe14}.

\begin{figure} 
\centering
\includegraphics[width=9.0cm,angle=0]{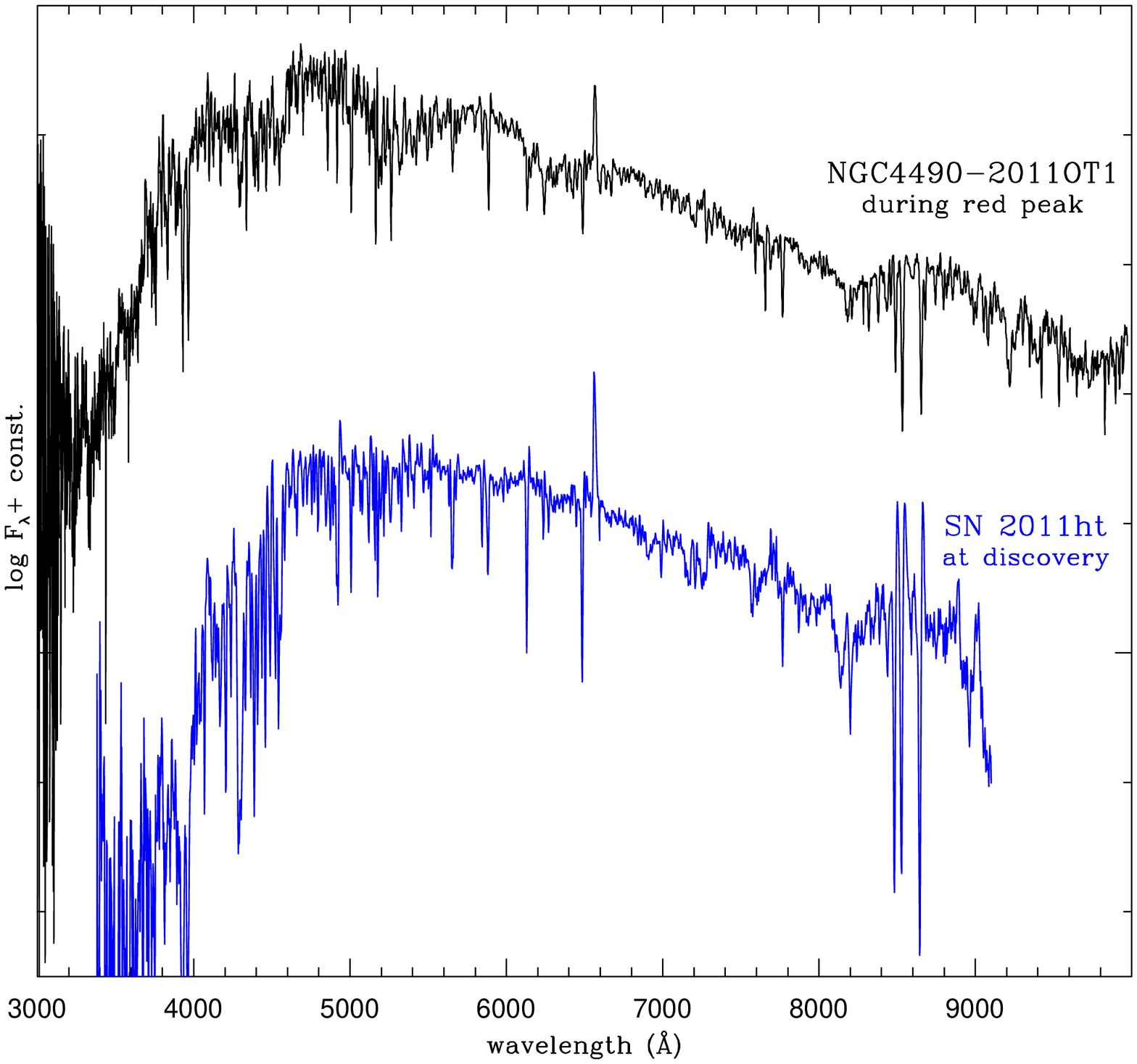}  
  \caption{Comparison of a spectrum of LRN NGC4490-2009OT1 taken on 2011 December 21 
(hence around red light curve peak) with
classification spectrum of SN~2011ht obtained on 2011 September 30  \protect\citep{pasto11}. The SN~2011ht spectrum 
was taken at the early SN luminosity rise  ($\sim -$54 d from the V-band 
maximum), and before it showed the classical Type IIn spectral features. \label{spec11ht}} 
\end{figure}

The  UGC5460-2010OT1 outburst started about 300 days before the discovery of SN~2011ht.  
Although slightly fainter than the LRNe discussed in this paper, this event shares many photometric  
analogies with them. In particular, although UGC5460-2010OT1 was poorly monitored and lacks
precise colour information, we note an appealing similarity of its light curve
with LRNe.
  Unfortunately, no spectrum was obtained during the UGC5460-2010OT1 observational campaign to be compared
with the spectra of our sample. The first spectrum of the source in UGC~5460 (whose epoch is marked with a vertical dot-dashed line
in Fig. \ref{lc_abs11ht}) was obtained at time of the putative SN~explosion \citep{pasto11}, during the early SN rising phase.\footnote{This
remarkable spectrum, obtained on 2011 September 30, is to our knowledge the earliest available in the literature, 
25 days before that shown by \protect\citet{mau13}.}
We note the similarity of this spectrum of SN~2011ht with those of the LRN NGC4490-2009OT1 obtained at the red maximum 
(Fig. \ref{spec11ht}), suggesting similar properties of their circumstellar environments. In particular, 
the line velocities inferred from the position of the blueshifted absorption of H$\alpha$ in the two spectra of Fig. \ref{spec11ht} 
are comparable,\footnote{From the highest S/N spectrum of NGC4490-2009OT1 obtained  by \protect\citet{smi16a} on 2012 January 19 (which is very close to the epoch
of the red light curve peak), the core velocity of the blueshifted narrow absorption of H$\alpha$
is about 400 km s$^{-1}$, extended to 650 km s$^{-1}$ for the blue edge.
This value is consistent with the core velocity of the blueshifted absorption measured in our Echelle 
spectra of SN~2011ht (Pastorello et al. in preparation).} indicating that the wind outflows from the progenitor star (or
progenitor's binary system) occurred at nearly identical rates for the two transients. 

The circumstellar cocoon was likely produced by the UGC5460-2010OT1 event, and was embedding the progenitor star at the time of the SN~2011ht discovery.
The presence of high density surrounding material initially produced a spectrum similar to that of an intermediate-type hypergiant. It evolved 
towards that of a more classical SN IIn after a few days, when the SN~2011ht ejecta started the interaction with the CSM. 
Whether SN~2011ht was a genuine (but weak) interacting SN explosion, or it was the outcome of gas shell collisions 
produced in non-terminal outbursts is still an open issue. However, some clues favour the terminal SN explosion for SN~2011ht, including
the late-time luminosity decline marginally consistent with the $^{56}$Co decay rate \citep[][see also below]{mau13,smi13}.

Proving a connection of UGC5460-2010OT1/SN~2011ht with LRNe would have important implications to our understanding of these gap transients. 
Hence, two scenarios are proposed to explain the UGC5460-2010OT1/SN~2011ht sequence of outbursts:

{\bf 1. UGC5460-2010OT1 as an eruption of a single massive star -} 
The UGC5460-2010OT1 event was interpreted by a number of authors \citep[e.g.,][]{mau13,smi13,fra13} 
as a major instability of a massive star approaching the core-collapse. The chain of events
is either consistent with the outburst of a super-AGB star followed by an electron-capture SN explosion
\citep{bar71,nom84,wan09,pum09}, or that of a more massive star \citep[with $M>25$ M$_\odot$,][]{mau13} 
which ends its life as a fall-back SN \citep{woo95,zam98,fry99}.
In both cases, the outcome is expected to be a $^{56}$Ni-poor ($10^{-3}$ to $10^{-2}$ M$_\odot$) core-collapse SN, whose ejecta
interact with H-rich CSM, consistent with the observables of SN 2011ht \citep[see, e.g.,][]{smi13,mor14,chu16}.
However, the above interpretations were questioned by \citet{hum12}, who
stated that the 2010 event was the initial manifestation
of an eruptive phase lasting many months, then producing the shell-shell collision event
known as SN~2011ht. Hence, according to the later interpretation, SN~2011ht was a non-terminal
event. We note, however, that the existence of a remarkably homogeneous family of SN~2011ht-like events 
(sometimes dubbed SNe IIn-P)\footnote{This subclass of SNe IIn, first defined by \citet{mau13}, is 
characterized by narrow P~Cygni line spectra and optical light curves resembling those of SNe II-P, 
especially in the red-to-NIR bands. The group, along with SN~2011ht \citep{rom12,mau13} includes: 
SN~1994W \citep{sol98,chu04}, SN~2005cl \citep{kie12}, and SN~2009kn \citep{kan12}.} 
would favour a faint, $^{56}$Ni-poor SN scenario  for all of them \citep{mau13}.

\begin{figure*}
\centering
\includegraphics[width=14.0cm,angle=0]{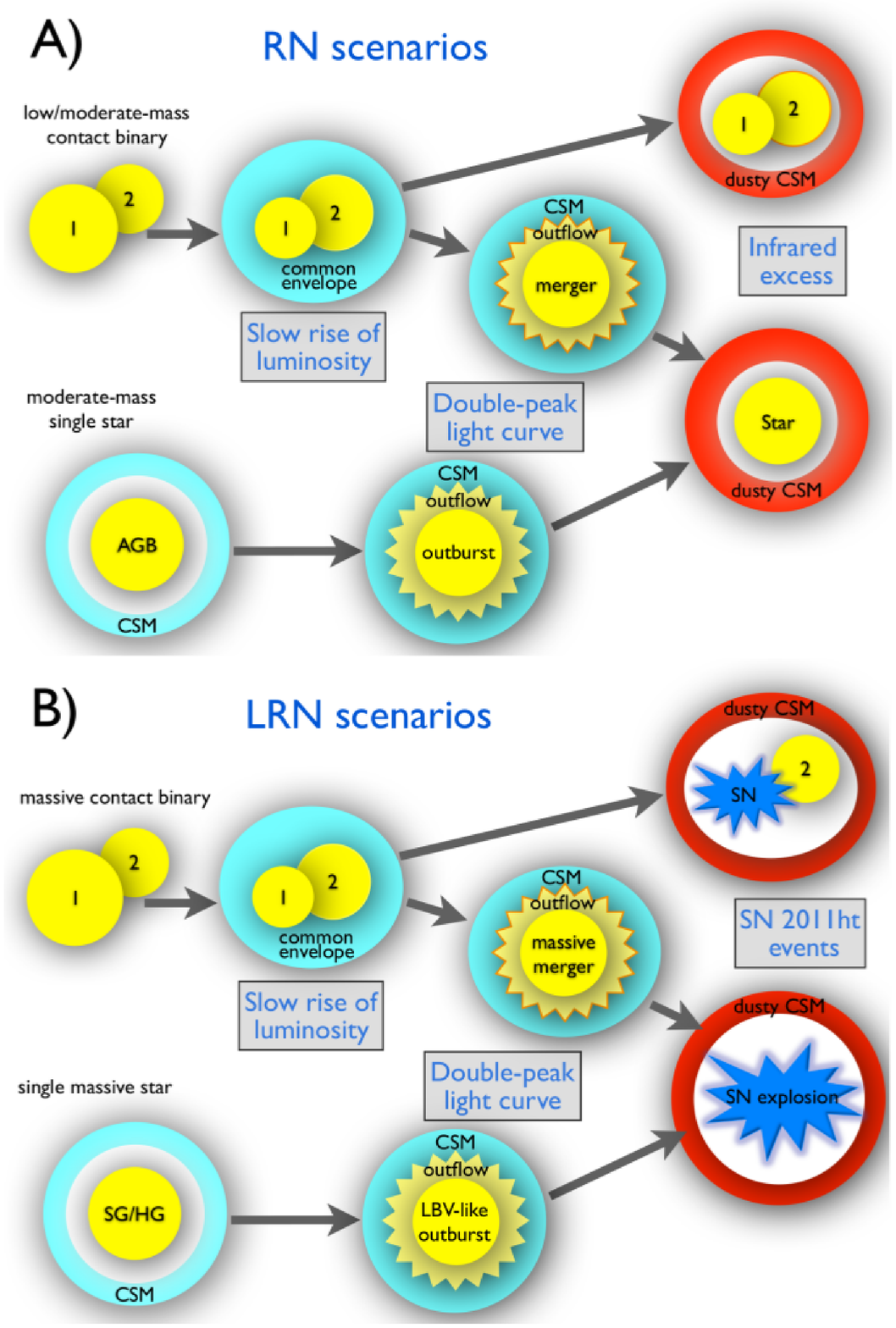}
  \caption{Sketch showing possible evolutionary paths for RNe (top) and LRNe (bottom).
Plausible scenarios for double-peak transients include eruptive phases of the evolution 
of moderate to high mass stars (with a shell ejection impacting pre-existing CSM), a common envelope ejection 
in a binary system formed by non-degenerate stars, and eventually (though not necessarily) a merging event 
with gas outflow interacting with a circumstellar shell. 
If the primary is an evolved high-mass star, a supergiant (SG) or an hypergiant (HG) with a $M_{MS}>7-8$ M$_\odot$, or the resulting 
merger is massive enough, the final outcome may be a core-collapse SN explosion, likely similar to 
SN~2011ht. \label{camminoevolutivo}} 
\end{figure*} 

{\bf 2. UGC5460-2010OT1 as a common envelope or a merging event in a binary system -} 
As an alternative to the single-star eruption for UGC5460-2010OT1 (followed by a possible SN explosion), one may propose a common
envelope ejection scenario involving a massive binary system, eventually (although not necessarily) followed by a merging event.
This is analogous to the scenario suggested for NGC4490-2011OT1 and LRNe,
with the main difference being that the merger producing UGC5460-2010OT1 is followed a few months later 
by a much more luminous outburst,  SN~2011ht. Again,  SN~2011ht can be interpreted as a shell-shell collision event
or as a terminal core-collapse SN.
According to the latter interpretation, the primary star of the binary progenitor system of UGC5460-2010OT1 would 
be an evolved massive star likely exploding as a SN soon after the binary interaction outburst or the coalescence.\footnote{The
evolutionary channel here discussed is somewhat similar to that proposed by a number of authors for the Type II SN~1987A 
\protect\citep[see][and references therein]{mor18}, although occurring in different time scales, as in that case the merging event occurred 20\,000 years ago \protect\citep{mor07}.} 
A physical explanation for the UGC5460-2010OT1 outburst as a LRN-like event would have important implications, 
and UGC5460-2010OT1/SN~2011ht would become a reference object in predicting the latest evolution of massive binary systems.

Only a continued post-outburst monitoring will shed light to the final fate of LRN progenitors and, most importantly, if they will 
encounter a sequence of outbursts similar to those of the progenitor of UGC5460-2010OT1/SN~2011ht. A sketch outlining  
the scenarios described above
for  double-peak transients and their possible evolutionary paths is in Fig. \ref{camminoevolutivo}. Whether these predictions
are realistic should be tested through detailed data modelling, post-outbursts follow-up and surveying known pre-merger candidates 
(Sect. \ref{future}).

\subsection[]{Chronicle of an embrace foretold} \label{future}

We have seen that a growing number of merger candidates  are discovered and studied 
after the coalescence outburst. However, very little is known about their precursor systems. 
Contact eclipsing binaries such as TY Pup have been proposed as RN progenitor candidates \citep{sar18}.
In this context, we should mention a couple of stellar systems which might merge in very short time scales (a few years).

KIC~9832227 is a complex stellar system formed by a contact binary with total mass of about 1.7 M$_\odot$, with a mass ratio $q\sim0.23$, 
plus a possible very low-mass third component of $\sim0.1$  M$_\odot$ \citep{mol17b}. Calculations based on light curve information have predicted with a surprisingly 
high precision that the contact binary in KIC~9832227 should merge on year $2022.2\pm0.6$ \citep{mol15,mol17a,mol17b}. The coalescence 
involving low-mass stars is expected to generate a low-luminosity RN event, similar to  V1309~Sco \citep{mas10,tyl11}. 
We note, however, that the  coalescence prediction of Molnar et al. has been  questioned by a recent study \citep{soc18}.

On the other extreme of the mass/luminosity function, an approaching merging has been proposed for another stellar system, VFTS~352 in 30~Dor (in the Large Magellanic Cloud).
It is formed by two O-type stars, and the two components 
have nearly identical masses \citep[28-29 M$_\odot$ each;][]{alm17}. The most likely outcome for this system is a stellar 
merger producing a transient more luminous than a RN \citep{sok06}.
This is very similar to the scenario proposed by \citet{smi16a} for the LRN NGC4490-2011OT1.

The different scenarios discussed in Sect. \ref{11ht} can also be proposed for VFTS~352. 
If the chemical evolution of the binary remains homogeneous, the two stellar components 
would remain blue, luminous and compact, avoiding the coalescence but maintaining a fast rotation rate \citep{sel09}. 
The two stars of the system are expected to individually evolve as WRs, finally exploding as stripped-envelope SNe, 
possibly generating a double black hole system \citep{alm17}. 

Alternatively, the outcome of the coalescence of the two  VFTS~352 components might also be an initially fast-rotating 
massive star in a low-metallicity environment. The resulting star is expected to end its life producing a SN, possibly associated with a long 
gamma-ray burst \citep[L-GRB,][]{yoo05,woo06}; but see discussion in \citet{alm17}. However, if the merger embedded in a dense
CSM also retains a significant fraction of its H envelope, this would produce a normal core-collapse SN without a L-GRB.
In this case, the ejecta would interact with the H-rich CSM, likely producing an event similar to  SN~2011ht 
\citep{mau13}.

Although none of the NGC4490-2011OT1-like objects presented in this paper has been followed so far by a SN explosion,
long-duration, post-outburst photometric monitoring is necessary to verify if they will eventually produce an ejecta-CSM 
interacting SN like SN~2011ht in relatively short time scales.

\section{Conclusions} 
 
We have presented new optical observations of a sample of LRNe showing a characteristic double-peaked light curve.
Their intrinsic colours are very blue during the first peak, typically ranging from $B-V\approx0$ to 0.8 mag.
The second peak has a much redder colour, with $B-V\approx0.6$ to 1.5 mag.
The time span between the blue and red peaks depends on the filter considered, and is different among 
the objects of our sample, ranging from about 2.5 to over 4 months in the $V$ band.
During the blue peak, the spectra show a blue continuum and prominent Balmer lines in emission, with typical
$v_{FWHM}$ of a few hundreds km s$^{-1}$. During the red peak, the spectra become remarkably similar to those of a late-G to K star, 
with a redder continuum and a forest of metal lines, while H features are only barely detected.
At late phases,  molecular bands are shown by the optical spectra of at least four LRNe (NGC4490-2011OT1,
UGC12307-2013OT1, M101-2015OT1, and AT~2017jfs), in analogy to that observed in some Galactic RNe  (V838~Mon, V4332~Sgr, and V1309~Sco).
In contrast with the lower-mass stellar systems producing RNe, the progenitors of LRNe are likely massive binaries, although 
both RNe and LRNe are probably the consequence of a common envelope ejection plus a stellar merging event
\citep{smi16b,bla16,mau17}.

On the other hand, we have also found remarkable analogies between LRNe and UGC5460-2010OT1, the outburst precursor of the Type IIn-P SN~2011ht:
the double-peaked light curve,  and the first spectrum of SN~2011ht closely resembles that of LRNe at the epoch of the red light curve peak.
A pre-SN eruption followed by a core-collapse SN was a reasonable explanation proposed by a number 
of authors \cite[e.g.,][]{mau13}, although others suggested that the whole cycle of UGC5460-2010OT1/SN~2011ht
variability was the display of a long-duration eruptive phase of a still-living massive star \citep{hum12}, with the SN IIn-P observables being 
consistent with circumstellar shell-shell interaction \citep[as suggested by][for the Type IIn-P SN~1994W]{des09}. 
We speculate that this striking similarity may connect UGC5460-2010OT1 with LRNe, hence to a pre-SN merging event.

A massive star resulting from a merging event is predicted to explode with some delay with respect the lifetime of a single star with the
same mass \citep{zap17}. For this reason, studying the stellar population in galaxies hosting LRNe is necessary to test the robustness of the merger channel
proposed in this paper for massive stars. Combining an increased sample of LRNe with  high-cadence photometric monitoring in the optical and IR domains,
good resolution spectroscopy, and detailed studies of their environments is crucial to clarify the nature of LRNe, hence predicting their fate.

 \begin{acknowledgements}

We dedicate this work to our friend Alex Dimai. His enthusiasm and competence will remain as a precious gift for those who had the priviledge of working with him.

We thank the referee Elena Barsukova for the insightful comments that improved the manuscript.

We are grateful to Marco Fiaschi for his observations at the Asiago Telescopes, Stefano Valenti and Mattias Ergon for their observations at ESO-La Silla, 
and Avet Harutyunyan for his support with the observations with TNG. We also acknowledge M. MacLeod and O. Pejcha for helpful discussions, and N. Blagorodnova
for providing the data of M101-2015OT1, and for useful suggestions.
SB, LT, PO, MT, MTB are partially supported by the PRIN-INAF 2017 
{\sl ``Towards the SKA and CTA era: discovery, localisation and physics of transient sources''} (PI M. Giroletti).
NER acknowledges support from the Spanish MICINN grant ESP2017-82674-R and FEDER funds.
YC is supported by the China Scholarship Council. KM is supported by STFC through an Ernest Rutherford Fellowship.
ST is supported by TRR33 {\sl``The Dark Universe''} of the German Research Foundation.
SJS acknowledges funding from ERC Grant 291222 and STFC grant Grant Ref: ST/P000312/1.
SGS, AJD, and the CRTS survey have been supported by the NSF grants AST-1313422, AST-1413600, and AST-1749235.
KM acknowledges support from H2020 through an ERC Starting Grant (758638).
Support for GP is provided by the Ministry of Economy, Development, and Tourism's Millennium Science Initiative 
through grant IC120009, awarded to The Millennium Institute of Astrophysics, MAS.\\

This article is based on observations made with the following facilities: 
the Italian Telescopio Nazionale Galileo operated on the island of La Palma by the Fundaci\'on Galileo Galilei of the INAF (Istituto Nazionale di Astrofisica) at the Spanish Observatorio del Roque de los Muchachos of the Instituto de Astrof\'isica de Canarias; 
the Nordic Optical Telescope, operated by the Nordic Optical Telescope Scientific Association at the Observatorio del Roque de los Muchachos, La Palma, Spain, of the Instituto de Astrof\'isica de Canarias; 
the William Herschel Telescope and the Isaac Newton Telescope, which are operated on the island of La Palma by the Isaac Newton Group of Telescopes in the Spanish Observatorio del Roque de los Muchachos of the Instituto de Astrof\'isica de Canarias; 
the Gran Telescopio Canarias, installed at the Spanish Observatorio del Roque de los Muchachos of the Instituto de Astrof\'isica de Canarias, in the island of La Palma; 
the Calar Alto 2.2m Telescope of the Centro Astron\'omico Hispano-Alem\'an, Almer\'ia, Spain; 
the 1.93m OHP telescope of the Observatoire de Haute-Provence; 
the 3.56m New Technology Telescope and the Dutch 0.9-m telescopes at ESO-La Silla; 
the Copernico and the Schmidt telescopes (Asiago, Italy) of the INAF - Osservatorio Astronomico di Padova; 
the 8.4m Large Binocular Telescope at Mt. Graham (Arizona, USA);
the Southern Astrophysical Research (SOAR) telescope, which is a joint project of the Minist\'erio da Ci\^encia, Tecnologia, Inova\c{c}\~aos e Comunica\c{c}\~oes do Brasil (MCTIC/LNA), the U.S. National Optical Astronomy Observatory (NOAO), the University of North Carolina at Chapel Hill (UNC), and Michigan State University (MSU);
the Liverpool Telescope which is operated on the island of La Palma by Liverpool John Moores University in the Spanish Observatorio del Roque de los Muchachos of the Instituto de Astrof\'isica de Canarias with financial support from the UK Science and Technology Facilities Council; 
the 2m Faulkes North Telescope of the Las Cumbres Observatory Global Telescope Network (LCOGTN).

This paper used data obtained with the MODS spectrographs built with funding from NSF grant AST-9987045 and the NSF Telescope System
Instrumentation Program (TSIP), with additional funds from the Ohio Board of Regents and the Ohio State University Office of Research.
The LBT is an international collaboration among institutions in the United States, Italy and Germany. LBT Corporation partners are: 
The University of Arizona on behalf of the Arizona university system; Istituto Nazionale di Astrofisica, Italy; LBT Beteiligungsgesellschaft, 
Germany, representing the Max-Planck Society, the Astrophysical Institute Potsdam, and Heidelberg University; The Ohio State University, 
and The Research Corporation, on behalf of The University of Notre Dame, University of Minnesota and University of Virginia.\\

This paper is also based on observations collected at the European Organisation for Astronomical Research in the Southern Hemisphere under ESO 
programmes 184.D-1140 and 184.D-1151.\\

This research has made use of the NASA/ IPAC Infrared Science Archive, which is operated by the Jet Propulsion Laboratory, California Institute of Technology, under contract with the National Aeronautics and Space Administration. This research has also made use of the NASA/IPAC Extragalactic Database (NED) which is operated by the Jet Propulsion Laboratory, California Institute of Technology, under contract with the National Aeronautics and Space Administration. We acknowledge the usage of the HyperLeda database ({\it http://leda.univ-lyon1.fr}).
\end{acknowledgements}

\begin{appendix}
\section{Supplementary figure}


\begin{figure}[!t]
\centering
\includegraphics[width=9.0cm,angle=0]{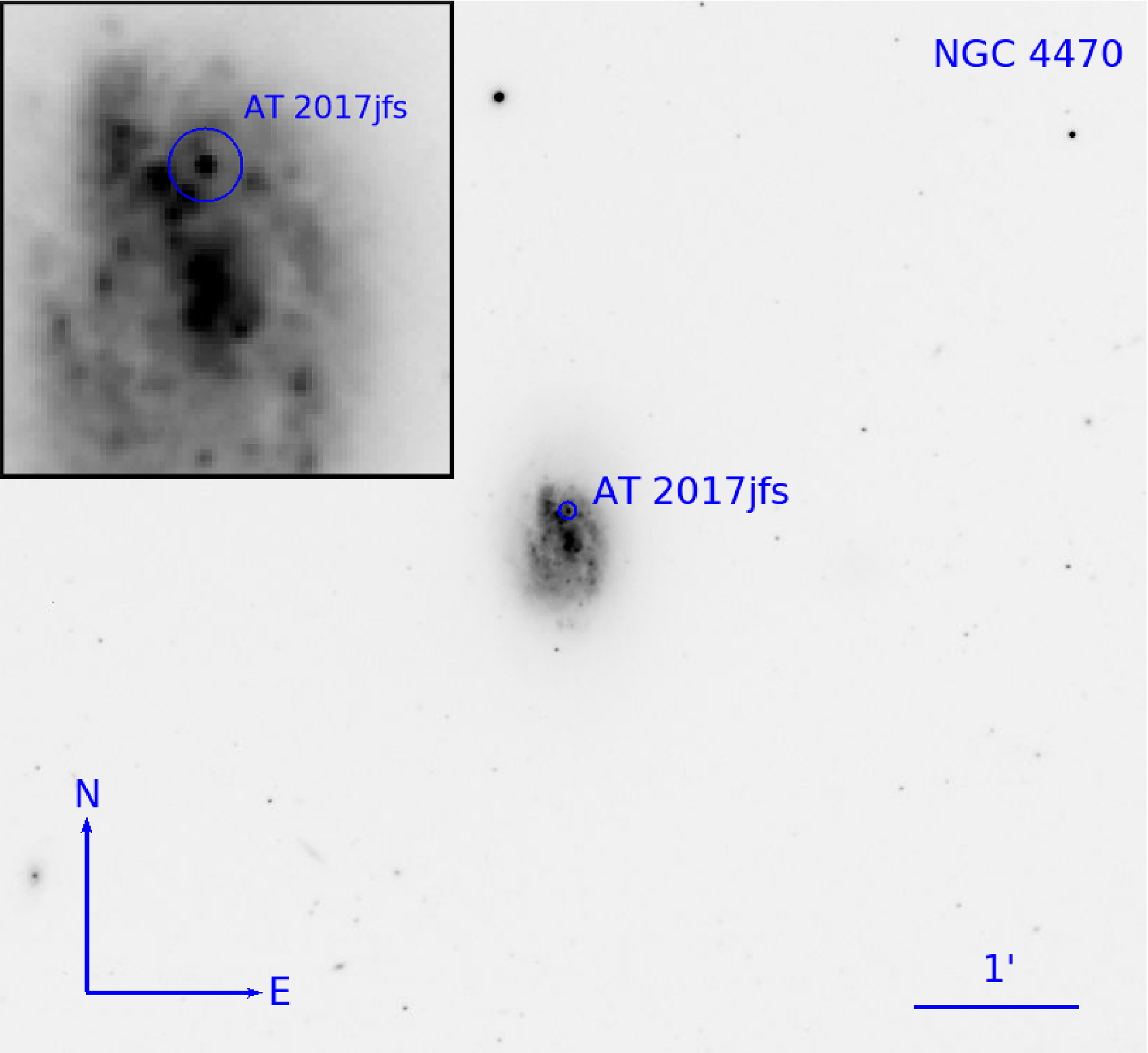}  
  \caption{Finder chart of reference LRN AT~2017jfs in NGC 4470, published in \citet{pasto19}. 
The Sloan $r$-band image was obtained on 2018 May 19 with the NOT telescope, equipped with ALFOSC.  \label{finder_17jfs}}
\end{figure}

\end{appendix}

\end{document}